\pdfoutput=1
\documentclass[11pt]{article}
\usepackage{cite}
 \usepackage {soul}
\usepackage{graphicx}
\usepackage{latexsym}   
\usepackage{mathrsfs}
\usepackage[scr=rsfs,cal=boondox]{mathalfa}
\usepackage[overload]{textcase}

\font\grande=cmr9.5 scaled \magstep4
\font\medio=cmr9.5 scaled \magstep2
\outer\def\beginsection#1\par{\medbreak\bigskip
      \message{#1}\leftline{\bf#1}\nobreak\medskip
\vskip-\parskip
      \noindent}

\begin{document}
\bibliographystyle{unsrt}

\titlepage

\vspace{1cm}
\begin{center}
{\grande Quantumness of relic gravitons}\\
\vspace{1.5cm}
Massimo Giovannini \footnote{e-mail address: massimo.giovannini@cern.ch}\\
\vspace{0.5cm}
{{\sl Department of Physics, CERN, 1211 Geneva 23, Switzerland }}\\
\vspace{0.5cm}
{{\sl INFN, Section of Milan-Bicocca, 20126 Milan, Italy}}\\
\vspace*{1cm}
\end{center}
\vskip 0.3cm
\centerline{\medio  Abstract}
\vskip 0.5cm
Since the relic gravitons are produced in entangled states of opposite (comoving) three-momenta, 
their distributions and their averaged multiplicities must determine the maximal frequency of the spectrum above which the created pairs are exponentially suppressed. The absolute upper bound on the maximal frequency derived in this manner coincides with the THz domain and does not rely on the details of the cosmological scenario. The THz limit also translates into a constraint of the order of $10^{-33}$ on the minimal chirp amplitudes that should be effectively reached by all classes of  hypothetical detectors aiming at the direct scrutiny of a signal in the frequency domain that encompasses the MHz and the GHz bands. The obtained high-frequency limit is deeply rooted in the quantumness of the produced gravitons whose multiparticle final sates are macroscopic but always non-classical. Since the unitary evolution preserves their coherence, the quantumness of the gravitons can be associated with an entanglement entropy that  is related with the loss of the complete information on the underlying quantum field. It turns out that the reduction of the density matrix in different bases leads  to the same von Neumann entropy whose integral over all the modes of the spectrum is dominated by the maximal frequency. Thanks to the THz bound the total integrated entropy of the gravitons can be comparable with the cosmic microwave background entropy but not larger. Besides the well known cosmological implications, we  then suggest that a potential detection of gravitons between the MHz and the THz may therefore represent a direct evidence of macroscopic quantum states associated with the gravitational field.
\noindent
\vspace{5mm}
\vfill

\newpage
\renewcommand{\theequation}{1.\arabic{equation}}
\setcounter{equation}{0}
\section{Introduction}
\label{sec1}
It is difficult to assess a potential signal without any prior knowledge of its origin and of its gross physical features. This observation applies, in particular, to the diffuse backgrounds of gravitational radiation whose amplitudes are often comparable with the potential noises appearing in the same spectral domain. It seems therefore useful to investigate in depth the relic signals that represent a well defined (and probably unique) candidate source for typical frequencies\footnote{We are going to employ systematically the standard prefixes of the international system of units so that $1 \mathrm{kHz} = 10^{3} \mathrm{Hz}$, $1\, \mathrm{aHz} = 10^{-18} \, \mathrm{Hz}$ and similarly for all the other relevant frequency domains.} exceeding the kHz region where wide-band detectors are currently operating. Indeed, since the mid 1970s the existence of a stochastic background of relic gravitons has been repeatedly suggested \cite{AA1,AA2,AA3,AA4} as a direct consequence of the breaking of Weyl invariance \cite{AA5,AA6} but after the formulation of the inflationary scenario it became gradually more evident that the conventional lore would predict a minute spectral energy density in the MHz region \cite{AA7,AA8,AA9}. This is ultimately the reason why the most stringent tests of the conventional lore could come, in the near future, from the largest scales \cite{AA10} where the  limits on the tensor to scalar ratio $r_{T}$ \cite{AA11,AA12} are in fact direct probes of the spectral energy density in the aHz region \cite{AA13}. 

As already mentioned in the previous paragraph, the highest  frequencies currently scrutinized by ground-based detectors fall in the audio band (between few Hz and $10$ kHz) where the operating interferometers improved their bounds all along the last score year \cite{BB0a,BB0b}. While twenty years ago the direct constraints on the spectral energy density in critical units\footnote{The spectral energy density in critical units is  denoted hereunder by $\Omega_{gw}(\nu,\tau)$ and it depends both on the comoving frequency $\nu$ and on the conformal time coordinate $\tau$. It should be clear that $\Omega_{gw}(\nu,\tau)$ {\em does not} coincide with the energy density of the gravitons in critical units. Moreover $h_{0}$ is the Hubble rate expressed in units of $100\,\mathrm{Hz}\, \mathrm{km}/\mathrm{Mpc}$ and since $h_{0}^2$ appears in the denominator of $\Omega_{gw}(\nu, \tau_{0})$ it is a common practice to discuss directly  $h_{0}^2 \Omega_{gw}(\nu, \tau_{0})$ which does not depend on the specific value of $h_{0}$.} ($\Omega_{gw}(\nu,\tau_{0})$ in what follows) were ${\mathcal O}(10^{-3})$ they are today in the range ${\mathcal O}(10^{-9})$ \cite{BB1,BB2}. The other potential probes of diffuse backgrounds in  the nHz window (we recall $1\,\, \mathrm{nHz} = 10^{-9} \,\, \mathrm{Hz}$) are the pulsar timing arrays \cite{BB3,BB4}, even if the competing experimental collaborations still make different statements about the correlation properties of the observed signal which should eventually comply (at least in the case of relic gravitons) with the Hellings-Downs curve \cite{BB5} (see also \cite{BB5a} and discussions therein). Finally, as already mentioned, in the aHz region (we recall $1\,\, \mathrm{aHz} = 10^{-18}\,\, \mathrm{Hz}$) the direct measurements of the temperature and polarization anisotropies set relevant limits on the tensor-to-scalar ratio $r_{T}$ so that, according to current data, $r_{T} < 0.06$ or even $r_{T} <0.03$ \cite{AA11,AA12}.

Although cosmic gravitons are produced from any variation of the space-time curvature, when a conventional stage of inflationary expansion is followed by a radiation-dominated epoch $h_{0}^2\,\Omega_{gw}(\nu,\tau_{0})$ is quasi-flat for $\nu > 100 \, \mathrm{aHz}$. Between few aHz and $100$ aHz  $h_{0}^2\,\Omega_{gw}(\nu,\tau_{0})$ scales as $\nu^{-2}$ and this regime encompasses the wavelengths that reentered the Hubble radius after matter-radiation equality. From the nHz domain to the audio band (between few Hz and $10$ kHz) the spectral energy density of inflationary origin is, at most, ${\mathcal O}(10^{-17})$ and the deviations from scale-invariance  always lead to decreasing spectral slope in $h_{0}^2\,\Omega_{gw}(\nu,\tau_{0})$, at least in the context of the conventional scenarios. This happens, for instance, in the single-field case where, thanks to the consistency relations, the tensor spectral index $n_{T}$ is notoriously related to the tensor to scalar ratio $r_{T}$ as $n_{T} \simeq - r_{T}/8$. Since $r_{T}$ is currently assessed from the analysis of the temperature and polarization anisotropies of the Cosmic Microwave Background (CMB) \cite{AA11,AA12} $n_{T}$ cannot be positive. There are finally known sources of damping that further reduce the inflationary result (see  for a recent review) and, most notably, the free-streaming of neutrinos \cite{BB6a,BB6,BB7,BB8,BB9} for frequencies below the nHz.

Although in the conventional lore $h_{0}^2 \Omega_{gw}(\nu,\tau_{0})$ is minute at high frequencies,  it is comparatively larger in the aHz region where a direct detection is not excluded. While  the bounds on $r_{T}$ are getting progressively more stringent \cite{AA11,AA12,AA13} their detection would be a potential test of the quantization of gravity  since the production of relic gravitons is absent in the limit $\hbar \to 0$ \cite{BB10}. A similar logic is at heart of this paper where, however we preferentially consider the high-frequency band, i.e. between the MHz and the GHz. Indeed, as we move from the audio band to the MHz and GHz regions the potential signals associated with the relic gravitons can even be $8$ or $9$ orders of magnitude larger than the spectral energy density computed in the case of the concordance paradigm \cite{ST1,ST2,ST3,ST4,ST5,ST6,ST7,ST7a}. At high-frequencies a background of relic gravitons is also expected from bouncing scenarios \cite{AA13} originally envisaged by Tolman \cite{TTT1} and  Lema\^itre \cite{TTT2}. Bouncing models often appear various contexts encompassing the cosmological scenarios inspired by string theory and quantum gravity.  On a physical ground it is practical to distinguish  between the  bounces of the scale factor and the curvature bounces \cite{TTT3}: for the bounces of the scale factor the extrinsic (Hubble) curvature locally vanishes while in the case of the  curvature bounces it is the time derivative of the Hubble rate that goes to zero. In both cases a potential signal is expected 
in the high-frequency domain.

In this paper we shall demonstrate that the relic gravitons appear in a macroscopic quantum state whose statistical properties ultimately determine the maximal frequency of the spectrum and the related entanglement entropy which is characteristic of the process of particle production \cite{ST8,ST9,ST9a,ST9b,ST10,ST11}. While this observation does not rely on the details of the cosmological evolution,  the gist of the argument is that  gravitons are produced in entangled pairs of opposite three-momenta and therefore the averaged multiplicity of the created quanta is exponentially suppressed above the maximal frequency $\nu_{max}$ corresponding to the probability of producing  a single pair of gravitons. This observation is considered here as an operational definition of the maximal frequency of the spectrum; in other words the multiplicity distribution of the gravitons determines $\nu_{max}$ and this ultimately implies that the maximal frequency cannot exceed the THz region.

The layout of this paper is, in short, the following.
A general argument leading to the quantum mechanical bound on the maximal frequency is presented in section \ref{sec2} and the obtained conclusions are corroborated in section \ref{sec3} by a series of classical considerations where $\nu_{max}$ is interpreted as the inverse of the minimally amplified wavelengths that exited the Hubble radius at the end of inflation and reentered immediately after. The quantum bound of section \ref{sec2} agrees then with the classical determinations of the maximal frequency in a variety of cosmological scenarios. The constraint on $\nu_{max}$ also implies an upper limit on the chirp amplitudes expected below the THz: from this observation it is plausible to set the specific goals for hypothetical detectors aiming at the direct detection of quantum gravitons in the spectral domain ranging between the MHz and the GHz bands.  
The THz limit is ultimately based on the unitarity of the evolution and this is why in sections \ref{sec4} and \ref{sec5} we show that the multiparticle final states of the relic gravitons are inherently quantum mechanical. This aspect is analyzed in greater detail with particular attention to the region $\nu \ll \nu_{max}$ where the large multiplicities of the created gravitons would naively suggest a classical description. We show instead that unitarity is never lost so that the limit of large averaged multiplicities corresponds to a macroscopic quantum state possibly characterized by a large entanglement entropy which is computed in section \ref{sec5}.  Since the unitary evolution preserves the coherence of the gravitons their quantumness can be quantified in terms of their entanglement entropy which is caused by the loss of the complete information on the underlying quantum field. According to the THz constraint the total entanglement entropy of the gravitons must not exceed the entropy associated with the cosmic microwave background. To avoid digressions in the bulk of the paper, various relevant results have been relegated to the appendices \ref{APPA}, \ref{APPB} and \ref{APPC}. 

\renewcommand{\theequation}{2.\arabic{equation}}
\setcounter{equation}{0}
\section{Quantum production of relic gravitons}
\label{sec2}
According to the current wisdom the large-scale inhomogeneities have a quantum mechanical origin \cite{CL1} as postulated by Sakharov \cite{SAK} almost two decades before the formulation of conventional inflationary scenarios. The same mechanism leading to low-frequency gravitons 
(potentially affecting the large-scale observations through the tensor to scalar ratio \cite{AA11,AA12,AA13}) also produces high-frequency pairs. In time-dependent backgrounds the classical fluctuations are given once forever (on a given space-like hypersurface). The quantum fluctuations keep on reappearing all the time and become the dominant source of inhomogeneity only when the classical inhomogeneities are suppressed. Since the  quantum fluctuations cannot dominate by fiat \cite{SAK} the classical inhomogeneities must be asymptotically suppressed and this aspect can be scrutinized by following the evolution of the spatial gradients \cite{CL2,CL2a}. In conventional inflationary models classical gradients are indeed suppressed \cite{CL3} provided the accelerated stage is sufficiently long and a similar conclusion follows for a stage of accelerated contraction \cite{CL4}. In this section the quantum fluctuations are the dominant source of inhomogeneity as it happens when an inflationary stage of  non-minimal\footnote{If the duration of inflation is minimal (or close to minimal) the classical fluctuations (which were super-horizon sized at the onset of inflation) are not necessarily suppressed by the inflationary stage and may have computable large scale effects. The minimal duration of inflation does not affect however the shortest scales of the problem which are the ones relevant for the derivation of the bound discussed in this section. } duration (i.e. larger than ${\mathcal O}(60)$ $e$-folds) precedes a phase of ordinary decelerated expansion. 
The purpose of this section is to show that the averaged multiplicity of the produced gravitons and their multiplicity distribution are sufficient to derive a general constraint on the maximal frequency which cannot exceed the THz domain. 
It is relevant to stress, in this respect, that we are considering here quantum fields propagating 
in a curved space-time background \cite{PARKTH,BIRREL,PTOMS}. A similar approach is employed in quantum optics \cite{MANDL,HBT4} where a classical (pump) field interacts with a nonlinear material to create pairs of photons. The same strategy 
is at the heart of the derivation of the Hawking radiation and of the inflationary estimates of the 
scalar power spectra of curvature inhomogeneities \cite{PTOMS}. While the resulting theory is 
non-renormalizable, it makes sense as an effective field theory and it tyopically fails for curvature scales that are nearly Planckian. In this spirit, one the purposes of this section is to
show that the momenta of the gravitons (as well as the corresponding frequencies) cannot exceed an ultra-violet cut-off that depends on the (maximal) rate of variation of the geometry. In the quantum perspective pursued here this effective cut-off is model-independent and although the technical aspects of the effective description are not crucial for the present ends, it is useful to mention them with some care; this is why, in the final part of this section, a quick reminder of the effective field theories  
 of the inflationary scenarios (and on their corresponding inhomogeneities) has been added in a self-contained perspective.

\subsection{The averaged multiplicity}
In the Heisenberg description the production of particles from the vacuum is described by the following unitary transformation \cite{AA5,AA6} (see also \cite{PARKTH,BIRREL,PTOMS}) which is closely analog to the ones arising in the context of the quantum theory of parametric amplification \cite{louisell,mollow1,mollow2,RGS0,HBTG1}:
 \begin{eqnarray}
\widehat{a}_{\vec{p}}(\tau) &=& u_{p}(\tau)\,\, \widehat{b}_{\vec{p}}\,- \,
v_{p}(\tau)\,\, \widehat{b}_{-\vec{p}}^{\,\dagger},  
\label{FFF1}\\
\widehat{a}_{-\vec{p}}^{\,\dagger}(\tau) &=& u_{p}^{\ast}(\tau) \,\,\widehat{b}_{-\vec{p}}^{\,\dagger}\,  - \,v_{p}^{\ast}(\tau)\,\, \widehat{b}_{\vec{p}},
\label{FFF2}
\end{eqnarray}
where $\tau$ denotes throughout the conformal time coordinate and the time-dependent (complex) functions $u_{p}(\tau)$ and $v_{p}(\tau)$ are constrained by the unitary evolution that must preserve the commutation relations between the two different sets of creation and annihilation operators. It is relevant to stress that Eqs. (\ref{FFF1})--(\ref{FFF2}) have been in fact derived in section \ref{sec4} (see, in particular,
Eqs. (\ref{DNT6})--(\ref{DNT7})). For the argument presented here, however, the  details 
(thoroughly analyzed in section \ref{sec4}) are not essential; this is why, incidentally, the polarization indices have been suppressed. For the purposes of the argument discussed here it is sufficient to view Eqs. (\ref{FFF1})--(\ref{FFF2}) as a unitary transformation between the initial and the final stages of particle production; similar unitary transformations 
appear in condensed matter theory and in quantum optics. In the language of Eqs. (\ref{FFF1})--(\ref{FFF2})
$\widehat{b}_{-\vec{p}}^{\dagger}$ creates a quantum of momentum $-\vec{p}$ and the dagger denotes 
the standard Hermitian conjugate. In the continuous mode representation the commutation relations 
between the operators appearing in the initial stage are $[\widehat{b}_{\vec{p}},\, \widehat{b}_{\vec{k}}^{\,\dagger}] = \delta^{(3)}(\vec{k} - \vec{p})$; the commutation relations at a generic time $\tau$ are 
also given by $[\widehat{a}_{\vec{p}},\, \widehat{a}_{\vec{k}}^{\,\dagger}] = \delta^{(3)}(\vec{k} - \vec{p})$
since the evolution is unitary; this means that the two complex functions  $u_{p}(\tau)$ and $v_{p}(\tau)$
must be subjected to the following unitarity condition:
\begin{equation}
\bigl| u_{p}(\tau) \bigr|^2 - \bigl|v_{p}(\tau) \bigr|^2 = 1. 
\label{FFF3}
\end{equation}
As already mentioned above, Eqs. (\ref{FFF1})--(\ref{FFF2}) and (\ref{FFF3}) have been written, for the sake of simplicity,  in the case of a single tensor polarization and describe the production of graviton pairs with opposite three-momenta. Different physical scenarios may modify the values of the maximal frequency (or maximal wavenumber) of the spectrum\footnote{As we shall see in section \ref{sec3} when a standard inflationary stage precedes the radiation-dominated epoch we would have $\nu_{max} = {\mathcal O}(200)$ MHz but the value of $\nu_{max}$ can exceed $200$ MHz if the postinflationary evolution is dominated by a plasma whose equation of state is stiffer than radiation \cite{ST1,ST2,ST3,ST7,ST7a}. In what follows we are not going to assume a specific value of $\nu_{max}$ and the purpose is  
rather to find a general bound on the maximal frequency. The present 
discussion improves and complements the analysis of Ref. \cite{ST7a}.} but from the quantum mechanical viewpoint the maximal frequency simply corresponds {\em to the production of a single graviton pair}. The averaged multiplicity for $\nu \ll \nu_{max}$ corresponds to the mean number of produced pairs and it is given by:
\begin{equation}
\overline{n}_{k} = \bigl|v_{p}(\tau) \bigr|^2 = \biggl(\frac{\nu}{\nu_{max}}\biggr)^{m_{T} -4}, \qquad \nu< \nu_{max},
\label{FFF4}
\end{equation}
where $m_{T}$ denotes, in practice, to the spectral slope of $\Omega_{gw}(\nu,\tau_{0})$ in a given frequency interval (see Eq. (\ref{FFF7}) and discussion thereafter). According to the parametrization of Eq. (\ref{FFF4}) in the conventional lore where the consistency relations are enforced $m_{T} \simeq - r_{T}/8 + {\mathcal O}(r_{T}^2)$ \cite{AA7,AA8,AA9} (see also \cite{AA11,AA12,AA13}). There are, however, different physical situations where $m_{T}> 0$ or even $m_{T}>1$ \cite{AA13}; in these cases the spectral energy density increase at high frequencies while the averaged multiplicity for $\nu\ll \nu_{max}$ is comparatively less suppressed than in the standard lore where $m_{T} \to 0$. The analysis of specific cases suggests that in front of the power law of Eq. (\ref{FFF4}) a model-dependent numerical factor (be it $\gamma_{1} = {\mathcal O}(1)$) may appear; however $\gamma_{1}$ can be absorbed into a redefinition of $\nu_{max}$.  The pair production process described by Eqs. (\ref{FFF1})--(\ref{FFF2}) implies that for $\nu > \nu_{max}$ the averaged multiplicity is instead suppressed when the frequencies of the gravitons exceed the maximal frequency \cite{PARKTH,BIRREL}:
\begin{equation}
\frac{\bigl|v_{p}(\tau) \bigr|^2 }{1 + \bigl|v_{p}(\tau) \bigr|^2 } = e^{- \gamma (\nu/\nu_{max})}, \qquad\qquad \nu> \nu_{max}.
\label{FFF5}
\end{equation}
The degree of exponential suppression appearing in Eq. (\ref{FFF5}) depends on $\gamma$ which is a further numerical factor ${\mathcal O}(1)$ that is controlled by the smoothness of the transition between the 
inflationary and the postinflationary phase. Either $\gamma_{1}$ or $\gamma$ can be reabsorbed in the redefinition of $\nu_{max}$, but not both; for this reason we choose to keep $\gamma$ as a free parameter and this choice corresponds to the strategy already employed in explicit numerical calculations of the amplification coefficients \cite{mg0,mg1}. Equation (\ref{FFF5}) applies in all the situations where the evolution of the expansion rate is continuous across the various transitions and it implies that the production of particles falls off faster than any inverse power of $\nu$ (or of $k$) \cite{BIRREL}. If  Eqs. (\ref{FFF3})--(\ref{FFF4}) and (\ref{FFF5}) are considered as complementary limits of a single expression the mean number of pairs produced from the vacuum takes the following suggestive form:
\begin{equation}
\overline{n}(\nu, \tau_{0}) = \gamma \frac{x^{m_{T} - 3}}{e^{\gamma\,x} -1},\qquad\qquad x = (\nu/\nu_{max}).
\label{FFF6}
\end{equation}
The averaged multiplicity introduced in Eq. (\ref{FFF6}) correctly interpolates between the low-frequency regime described by Eq. (\ref{FFF4}) and the high-frequency domain of Eq. (\ref{FFF5}). 

\subsection{The maximal frequency and the single pair production}
From the averaged multiplicity of the pairs of Eq. (\ref{FFF6}) it is possible to deduce a general constraint on the maximal frequency of the produced gravitons. In short the logic is that, in spite of the values of $m_{T}$ and $\gamma$,  the maximal frequency must correspond to the production of a single pair of gravitons with opposite three-momenta since for $\nu \gg \nu_{max}$ the production rate must be exponentially suppressed. The spectral energy density of the gravitons 
can be expressed in terms of the mean number of produced pairs $\overline{n}(\nu,\tau_{0})$:
\begin{equation}
\Omega_{gw}(\nu, \tau) =  \frac{1}{\rho_{crit}} \, \frac{ d \langle \rho_{gw} \rangle }{d \ln{\nu} } = \frac{128 \pi^3}{3} \frac{\nu_{max}^{4}}{H^2 \, M_{P}^2 \, a^{4}}
 \biggl(\frac{\nu}{\nu_{max}}\biggr)^{4}  \, \overline{n}(\nu, \tau).
 \label{FFF7}
 \end{equation} 
 In Eq. (\ref{FFF7}) $H$ denotes the Hubble rate, $a$ is the scale factor\footnote{ We recall that $H = \dot{a}/a$ where the overdot indicates a derivation with respect to the cosmic time coordinate $t$. In what 
 follows we shall also employ different time parametrizations. For instance in the conformal time coordinate we would have 
 that $a\, H = {\mathcal H}$ where ${\mathcal H} = a^{\prime}/a$; the prime denotes, in this case, a derivation 
 with respect to the conformal time coordinate $\tau$ and, by definition, $a(\tau) \, d\tau = d\,t$. Later on a further generalization of $\tau$ is introduced but its use is limited to sections \ref{sec3} and \ref{sec4}.}. In Eq. 
 (\ref{FFF7}) we have that $M_{P} = G^{-1/2} = 1.22 \times 10^{19} \, \, \mathrm{GeV}$ is the Planck mass scale.
 In our discussions we shall be using natural units $\hbar=c =\kappa_{B}=1$ where $\kappa_{B}$ denotes 
the Boltzmann constant.  When the powers of $\hbar$ are correctly restored\footnote{The correct $\hbar$ dependence is obtained by considering that the energy of a single graviton is given by $\hbar\, \omega$ which w simply write as $k$ in units 
 $\hbar = c=1$. Another $\hbar$ is hidden in the definition of Planck mass. Taking into account these effects we have that, overall, $\Omega_{gw}(\nu,\tau_{0}) \propto \hbar^2$ as suggested in \cite{BB10}.}  $\Omega_{gw}(\nu, \tau)$ is ${\mathcal O}(\hbar^2)$ showing that we are dealing here with an effect that vanishes in the limit $\hbar\to 0$ \cite{BB10}. This observation fits perfectly with the logic of the present analysis and correctly suggests that relic gravitons are a unique laboratory for the exploration of quantum gravitational effects \cite{AA13} especially at high-frequencies \cite{REC2}.
 \begin{figure}[!ht]
\centering
\includegraphics[height=8cm]{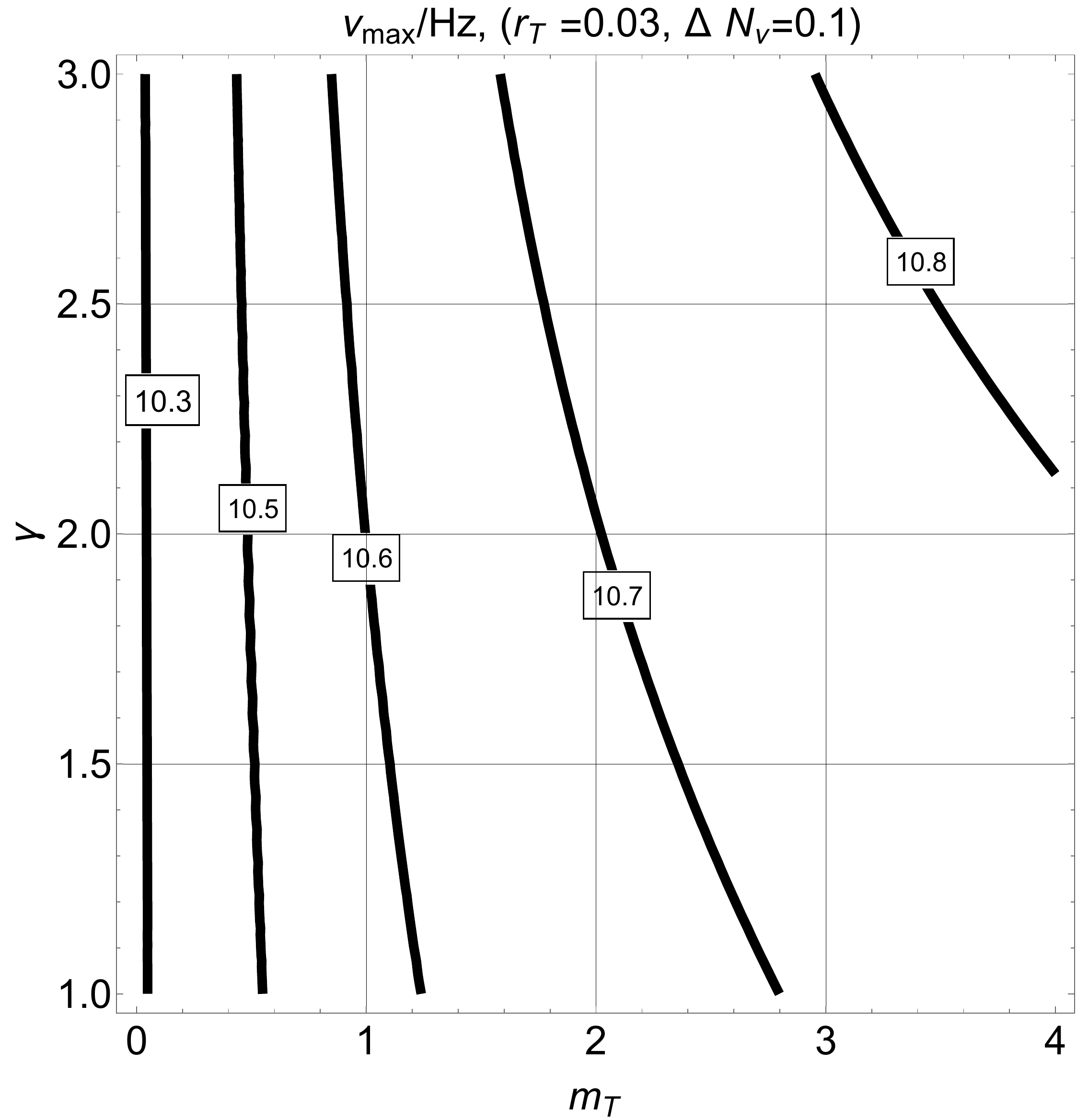}
\includegraphics[height=8cm]{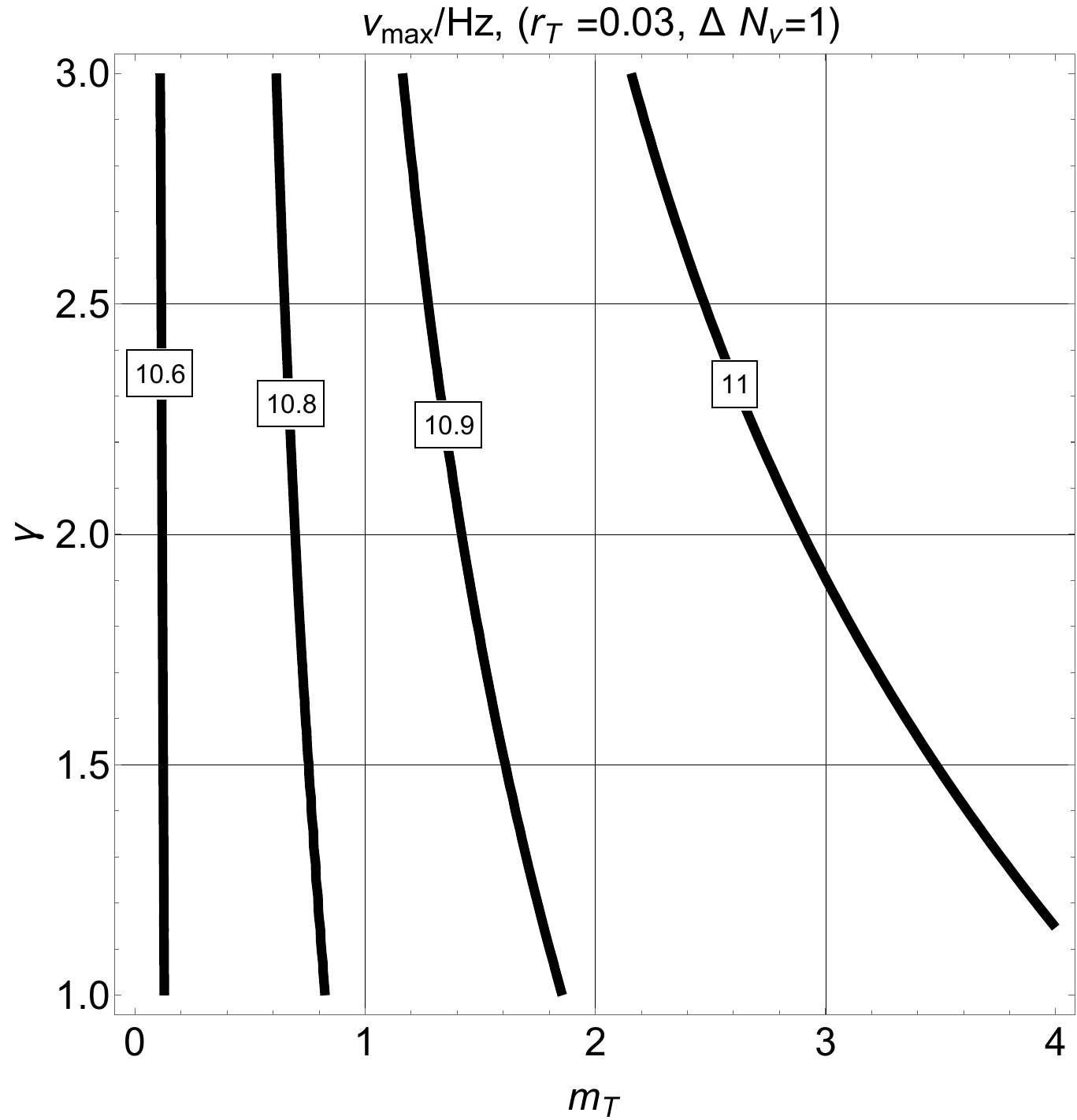}
\caption[a]{The different labels in both plots correspond to the common logarithms of $(\nu_{max}/\mathrm{Hz})$ determined from Eqs. (\ref{FFFB2})--(\ref{FFFB3}) and (\ref{FFFB4}). In both plots the minimal value of $m_{T}$ corresponds to the one of the concordance scenario supplemented by the consistency conditions (i.e. $m_{T} \simeq - r_{T}/8$) whereas the value of $r_{T}$ is consistent with the current determinations (i.e. $r_{T} < 0.035$ \cite{AA11,AA12}). Both plots, obtained 
for different values of $\Delta N_{\nu}$,  support the bound of Eq. (\ref{FFFB4}).}
\label{FIGURE1}      
\end{figure}
Although  Eqs. (\ref{FFF1})--(\ref{FFF2}) separately hold for each tensor polarization, 
the two polarizations have been correctly taken into account in the prefactor of Eq. (\ref{FFF7}).
Inserting now  Eq. (\ref{FFF6}) into Eq. (\ref{FFF7}) and rearranging the various terms, 
the following expression for $h_{0}^2 \, \Omega_{gw}(\nu, \tau_{0})$ is readily obtained:
\begin{equation}
h_{0}^2 \, \Omega_{gw}(\nu, \tau_{0}) = 3.659 \times 10^{-41} \biggl(\frac{\nu_{max}}{100\, \mathrm{Hz}}\biggr)^{4}\,\, F(x, m_{T}, \gamma),
\label{FFF8}
\end{equation}
where, as in Eq. (\ref{FFF6}), $x= (\nu/\nu_{max})$ and $F(x, m_{T}, \gamma)$ denotes the following dimensionless function
\begin{equation}
F(x, m_{T}, \gamma) = \gamma \, \frac{x^{m_{T}+1}}{e^{\gamma x} -1}, \qquad x= (\nu/\nu_{max}).
\label{FFF8a}
\end{equation}
Equations (\ref{FFF8})--(\ref{FFF8a}) can now be inserted into the  big-bang nucleosynthesis 
constraint  \cite{BBN1,BBN2,BBN3}
\begin{equation}
h_{0}^2  \int_{\nu_{bbn}}^{\infty}\, \Omega_{gw}(\nu,\tau_{0}) \,\,d\ln{\nu} = 5.61 \times 10^{-6} \Delta N_{\nu} \biggl(\frac{h_{0}^2 \,\Omega_{\gamma0}}{2.47 \times 10^{-5}}\biggr),
\label{FFFB1}
\end{equation}
where $\Omega_{\gamma\,0}$ is the critical fraction of photons 
in the concordance paradigm and $\Delta N_{\nu}$ accounts for the supplementary neutrino species. The standard BBN  results are in agreement with the observed abundances for $\Delta N_{\nu} \leq 1$; we shall choose $\Delta N_{\nu} = 0.1$. In Eq. (\ref{FFFB1}) $\nu_{bbn}= {\mathcal O}(10^{-2})$ nHz is the big-bang nucleosynthesis frequency \cite{AA13}.
By then combining Eqs. (\ref{FFF6})--(\ref{FFF7}) and  (\ref{FFF8})--(\ref{FFFB1}) we can obtain the following bound
\begin{equation}
3.65 \times 10^{-41} \,\, \biggl(\frac{\nu_{max}}{100 \, \mathrm{Hz}}\biggr)^{4} \,\, {\mathcal I}(m_{T},\gamma) \,\,<\,\, 5.61 \times 10^{-6} \Delta N_{\nu} \biggl(\frac{h_{0}^2 \Omega_{\gamma0}}{2.47 \times 10^{-5}}\biggr),
\label{FFFB2}
\end{equation}
where ${\mathcal I}(m_{T},\gamma)$ now indicates the integral 
\begin{equation}
{\mathcal I}(m_{T},\gamma) = \gamma \int_{\nu_{bbn}/\nu_{max}}^{\infty}\, \frac{x^{m_{T}}}{e^{\gamma x} -1} \, \, d x.
\label{FFFB3}
\end{equation}
Because of the exponential suppression due to Eq. (\ref{FFF5}), 
${\mathcal I}(m_{T},\gamma) $ converges for $\nu \to \infty$ even if, in the lower limit 
of integration, the integral inherits a potential logarithmic divergence when $m_{T} \to 0$.
This possibility is however not a problem since it only happens in the case of the conventional 
lore where we do know (see section \ref{sec3}) that $\nu_{max} = {\mathcal O}(200) \, \mathrm{MHz}$.
In spite of this direct determination, we must bear in mind that $x$ is very small but never goes to zero. 
A similar comment holds for the value of $m_{T}$ which is of the order of $- r_{T}/8$ and only 
goes to zero in the case of an exact de Sitter stage of expansion; this possibility is however 
excluded since in the exact de Sitter case the scalar modes of the geometry are not excited \cite{AA13}.
Finally for  $m_{T} >0$ the lower bound of integration can be ignored for all practical purposes.
All in all from Eq. (\ref{FFFB3}) we obtain the following upper limit on $\nu_{max}$ 
\begin{equation}
\nu_{max} \leq 6.257 \times 10^{10} \,\, \biggl[\frac{\Delta N_{\nu}}{{\mathcal I}(m_{T}, \gamma)}\biggr]^{1/4} \, \mathrm{Hz} < \mathrm{THz},
\label{FFFB4}
\end{equation}
with $[\Delta N_{\nu}/{\mathcal I}(m_{T}, \gamma)]< {\mathcal O}(10)$. 
\begin{figure}[!ht]
\centering
\includegraphics[height=8cm]{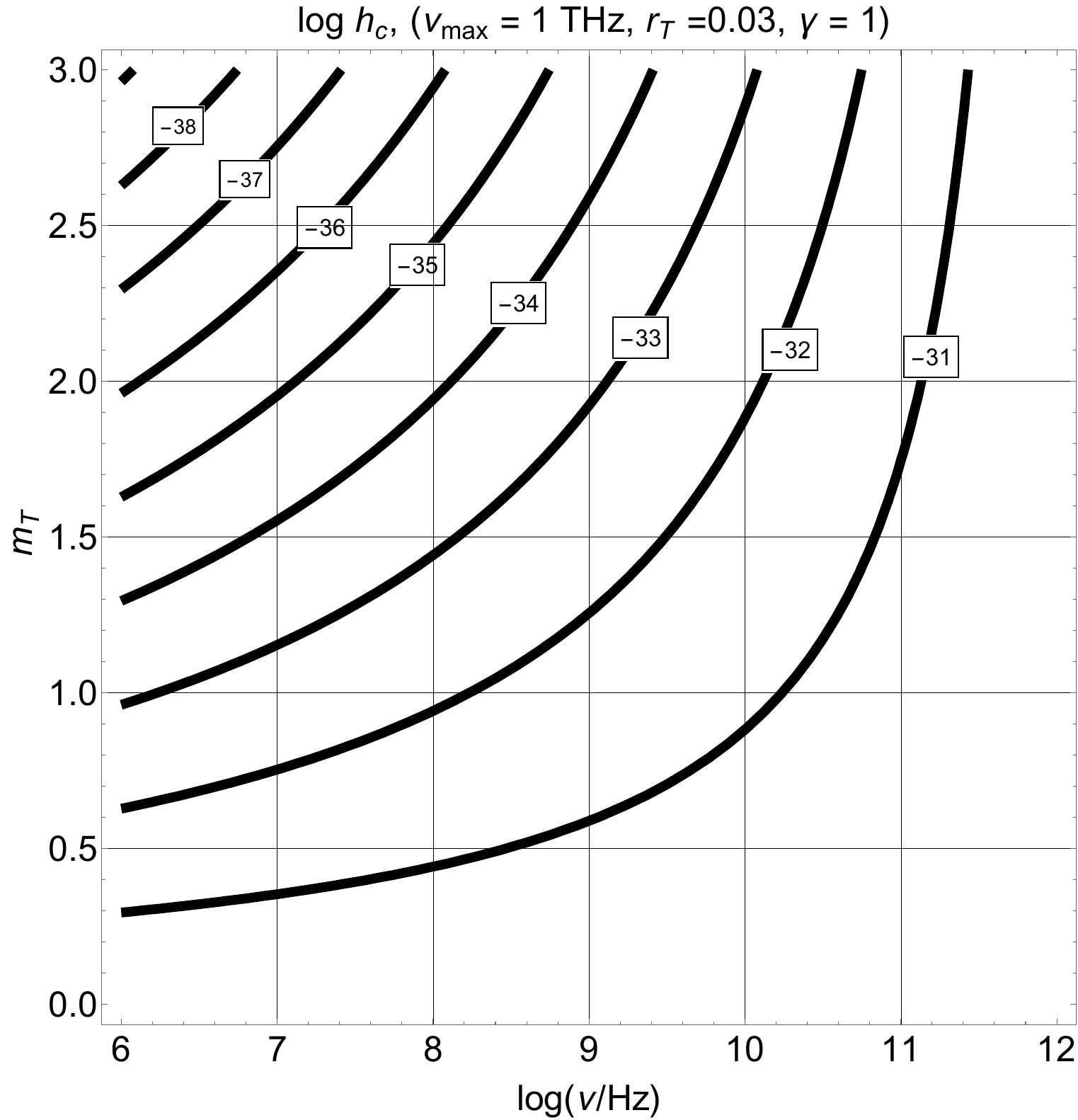}
\includegraphics[height=8cm]{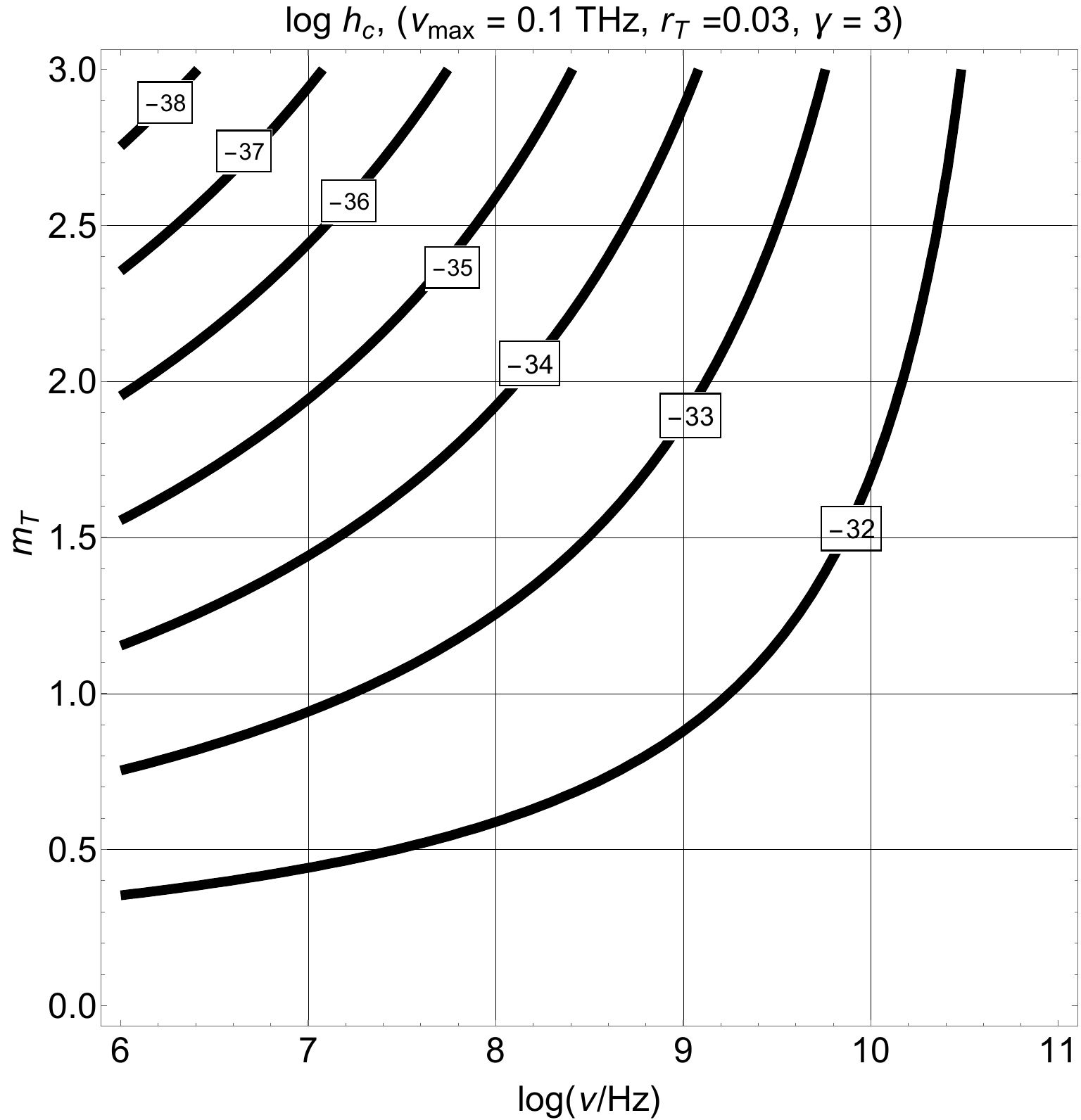}
\caption[a]{In both plots we illustrate the common logarithm of the chirp amplitude as a function of the frequency and of $m_{T}$. In  left plot $\nu_{max} = 1\, \mathrm{THz}$ whereas in the right plot $\nu_{max} = 0.1\, \mathrm{THz}$. From both plots it s clear that the chirp amplitude 
potentially resolved by high-frequency detectors should be at least ${\mathcal O}(10^{-33})$ or smaller.}
\label{FIGURE2}      
\end{figure}
The bound of Eq. (\ref{FFFB4}) is graphically illustrated in Fig. \ref{FIGURE1}  by computing 
numerically the integral that appears in Eq. (\ref{FFFB3}). The different 
contours of (\ref{FIGURE1}) correspond to curves of constant maximal frequency and the labels 
denote the common logarithm of $(\nu_{max}/\mathrm{Hz})$. At most $\nu_{max}$ 
can be ${\mathcal O}(\mathrm{THz})$ and this observation has an immediate practical implication:
if we use the interpolating expression of the averaged multiplicity 
together with the constraint on $\nu_{max}$ the chirp amplitude in high-frequency 
domain (say between the MHz and the GHz) can be determined. Indeed, recalling that the chirp amplitude may be related to the spectral energy density as $\Omega_{gw}(\nu,\tau_{0}) = 2 \, \pi^2 \, \nu^{2}\,  h_{c}^2(\nu,\tau_{0})/( 3 H_{0}^2)$ \cite{AA13}
we have that $h_{c}^2(\nu,\tau_{0}) = 64 \pi\, \biggl(\frac{\nu_{max}}{M_{P}}\biggr)^2 \, F(\nu/\nu_{max}, m_{T}-2, \gamma)$; this expression can also be written as:
\begin{equation}
h_{c}(\nu,\tau_{0}) = 7.64\times 10^{-41} \, \biggl(\frac{\nu}{100\, \mathrm{Hz}}\biggr) \, \sqrt{F(\nu/\nu_{max}, m_{T}-2, \gamma)}.
\label{FFFB6a}
\end{equation}
In Fig. \ref{FIGURE2} the bound on $\nu_{max}$ is illustrated by computing the common logarithm of the chirp amplitude from Eq. (\ref{FFFB6a}) in the high-frequency domain, i.e. between the MHz and the THz. The result of Eq. (\ref{FFFB4}) together with the Fig. \ref{FIGURE2} determine the minimal chirp amplitude required for the direct detection of gravitons in the high-frequency domain 
\begin{equation}
h^{(min)}_{c}(\nu, \tau_{0}) <  {\mathcal O}(10^{-32}) \biggl(\frac{\nu}{0.1\, \mathrm{THz}}\biggr)^{-1 +m_{T}/2}.
\label{FFFB7}
\end{equation}
From Eq. (\ref{FFFB7}) we can see that a sensitivity ${\mathcal O}(10^{-20})$ or even ${\mathcal O}(10^{-24})$ in the chirp amplitude for frequencies in the MHz or GHz regions is irrelevant for a direct or indirect detection of high-frequency gravitons. In this high-frequency domain microwave cavities \cite{FGF0,FGF1,FGF2,FGF3,FGF4,FGF5,FGF6,FGF7} operating in the MHz and GHz regions could be used for this purpose \cite{ST3,ST3a}. Depending on the value of $m_{T}$, Fig. \ref{FIGURE2} shows that $h_{c}^{(min)}$ must be at least ${\mathcal O}(10^{-32})$ (or smaller) for a potential detection of cosmic gravitons in the THz domain. In particular, for $m_{T} > 2$ the allowed  $h^{(min)}_{c}$ gets larger as the frequency increases; conversely for $m_{T} \leq 2$ frequencies smaller than the THz can be experimentally more convenient. When $m_{T} \to 1$ (as it happens in the case of a postinflationary stiff phase)  the chirp amplitude in the MHz range could be ${\mathcal O}(10^{-28})$. When $m_{T} \to 2$ (as it may happen in ekpyrotic scenarios) we would have instead that $h^{(min)}$ is effectively frequency independent  \cite{EK1,EK2}.  Finally when $m_{T} \to 3$ (as it could happen for the pre-big bang scenario \cite{EK3}) the chirp amplitude at lower frequencies gets even smaller. 
The implications of the bound (\ref{FFFB4}) are then essential  for an accurate assessment of the required sensitivities of high-frequency instruments.

\subsection{The multiplicity distribution}
 The statistics of the graviton pairs can also be derived by using simple group theoretical considerations \cite{sch1,sch2,pere} that will be specifically recalled later on in section \ref{sec4}. It turns out that the multiplicity distribution of the produced pairs from the vacuum follows the  Bose-Einstein (i.e. geometric) probability distribution (see also \cite{AA5,AA6}). For the preliminary purposes of the present section we can go back to Eqs. (\ref{FFF1})--(\ref{FFF2}) and observe that 
the state $| z_{\vec{k}} \rangle $ (annihilated by $\widehat{a}_{\vec{k}}$) obeys the condition 
$ u_{k} \, \widehat{b}_{\vec{k}} \, = \, v_{k} \, \widehat{b}^{\,\,\dagger}_{-\vec{k}} $; from this observation, in the case of pair production from the vacuum, we may deduce that 
\begin{equation}
| z_{\vec{k}} \rangle = {\mathcal N}_{\vec{k}} \, \exp{[ (u_{k}/v_{k}) \,\widehat{b}_{\vec{k}}^{\,\,\dagger} \, \widehat{b}_{-\vec{k}}^{\,\,\dagger} ]}
\,\, | 0_{\vec{k}} \,\, 0_{- \vec{k}} \rangle,
\label{STAT1}
\end{equation}
where $ | 0_{\vec{k}} \,\, 0_{- \vec{k}} \rangle$ is the two-mode vacuum state. The normalization of $| z_{\vec{k}} \rangle$ implies that ${\mathcal N}_{\vec{k}} = 1/\sqrt{|u_{k}|^2} = 1/\sqrt{ 1 + \overline{n}_{k}}$ so that 
\begin{equation}
| z_{\vec{k}} \rangle = \frac{1}{\sqrt{ 1 + \overline{n}_{k}}} \,\, \sum_{m=0}^{\infty} \biggl(\frac{v_{k}}{u_{k}}\biggr)^{m} \,\, |\,m_{\vec{k}}\, m_{- \vec{k}} \rangle. 
\label{STAT2}
\end{equation}
 With respect to $| 0_{\vec{k}} \,\, 0_{- \vec{k}} \rangle$ (i.e. the vacuum annihilated by $\widehat{b}_{\vec{k}}$ and $\widehat{b}_{-\vec{k}}$) the state $| \, z_{\vec{k}}\rangle$  is a superposition of gravitons with opposite three-momenta. Thus, the probability amplitude to find $n$ pairs with opposite three-momentum is ultimately given by:
\begin{equation}
\langle \, n_{\vec{k}} \, n_{- \vec{k}} | z_{\vec{k}} \,\rangle = \frac{1}{\sqrt{|u_{k}|^2}} \biggl(\frac{v_{k}}{u_{k}}\biggr)^{n},
\label{STAT3}
\end{equation}
so that the corresponding probability distribution associated with the diagonal elements of the 
corresponding density matrix is of Bose-Einstein type:
\begin{equation}
p_{n} = \bigl| \langle \, n_{\vec{k}} \, n_{- \vec{k}} | z_{\vec{k}} \,\rangle\bigr|^2 = \frac{1}{|u_{k}|^2} \biggl(\frac{|v_{k}|^2}{|u_{k}|^2}\biggr)^{n} = \frac{\overline{n}_{k}^{n}}{(\overline{n}_{k} +1 )^{n +1}}.
\label{STAT4}
\end{equation}
Equation (\ref{STAT4}) is a direct consequence of Eq. (\ref{STAT1}) and it is therefore a consequence 
of the unitary quantum mechanical evolution encoded in Eqs. (\ref{FFF1})--(\ref{FFF2}). The Bose-Einstein distribution, per se,  does not imply that the underlying 
state is given by a chaotic or even thermal mixture. This is the reason why it does not make 
much sense to say that the correlations associated with the relic gravitons can be easily reproduced by of artificially concocted classical sources somehow close to thermal mixtures. The second remark is that, in the derivation of the bound and of the statistics of the relic gravitons, we considered the production from the vacuum and this choice is compatible with a sufficiently long inflationary stage of expansion. If the duration of inflation minimal (or nearly minimal) the averaged multiplicity also depends on the  averaged multiplicities of the initial state; see, for instance, Refs. \cite{instate1}. The argument involving the  maximal frequency remains however the same but it is applied to the total averaged multiplicity. If a number of initial pairs is present the multiplicity distribution is negative binomial; see, in this respect, the discussion contained in appendix \ref{APPA}.

All in all we showed that the maximal frequency of the relic gravitons must necessarily be smaller than the THz and, in a quantum mechanical perspective, the maximal frequency corresponds to the minimal number of produced quanta. Roughly speaking this requirement implies the production of one graviton pair for each mode of the field. The quantum mechanical considerations developed in this section  
are not affected by the details of the expansion rate but only by the properties of the averaged multiplicity and, in the following section, this conclusion will be corroborated by a purely classical discussion. From a purely classical viewpoint the maximal frequency of cosmic gravitons corresponds to the wavelengths that exited the Hubble radius at the end of inflation and reentered at the onset of the postinflationary evolution. While these scales obviously experienced a smaller amplification in comparison with the wavelengths that exited the Hubble radius at the beginning of the inflationary evolution, as we are going to see in section \ref{sec3} the $\nu_{max}$ derived in this manner depends on the details of the postinflationary evolution and generally agrees with the THz 
constraint.

\subsection{The effective description}
As already mentioned at the beginning of this section the relic gravitons are fields 
coming from the quantization of the tensor modes of the geometry on a fixed background geometry \cite{BIRREL,PTOMS}. This is why the results derived here cannot be extended, strictly speaking, to Planckian curvature regimes where the distinction between a classical background 
and the corresponding quantum fluctuations is not meaningful. Instead of general 
discussion it seems more appropriate to analyze the effective description 
of conventional inflationary scenarios. The perspective adopted here follows 
the discussion of Ref. \cite{REC3} (see also Refs. \cite{AA13} and \cite{REC3a,REC3b}).
The first point to appreciate, is that the dynamical evolution of a large class of inflationary models is conventionally described in terms of a scalar-tensor theory of gravity whose effective Lagrangian density contains a single 
inflaton field $\varphi$ 
\begin{equation}
{\mathcal L}_{0} = \sqrt{- G} \biggl[ - \frac{ \overline{M}_{P}^2\,\, R}{ 2} + 
\frac{1}{2} G^{\alpha\beta} \partial_{\alpha} \varphi \partial_{\beta} \varphi - V(\varphi)\biggr],
\label{ONE3}
\end{equation}
where $\overline{M}_{P} = M_{P}/\sqrt{8 \pi}$ is the reduced Planck mass
while $V(\varphi)$ denotes the inflaton potential and $G$ is the determinant of the four-dimensional 
metric $G_{\alpha\beta}$. Equation  (\ref{ONE3}) represents in fact the first term of a 
generic effective field theory where the higher derivatives are suppressed by the negative powers of a large mass $M$ associated with the fundamental theory that underlies the effective description \cite{REC3}. In terms of the appropriate dimensionless scalar $\phi = \varphi/M$ Eq. (\ref{ONE3}) 
can be rephrased as:
\begin{equation}
{\mathcal L}_{0}=  \, \sqrt{-G} \biggl[ - \frac{\overline{M}_{P}^2}{2} R + \frac{M^2}{2} g^{\mu\nu} \partial_{\mu} \phi \partial_{\nu} \phi - M_{P}^2 U(\phi)\biggr], 
\qquad U(\phi) = \frac{V(M \phi)}{M_{P}^2},
\label{ONE3a}
\end{equation}
The leading correction to Eq. (\ref{ONE3a}) consists of all possible terms containing four derivatives and they can be parametrized in the following manner: 
\begin{eqnarray}
&& \Delta\,{\mathcal L} = \sqrt{-G} \biggl[ c_{1}(\phi) \bigl(G^{\alpha\beta} \partial_{\alpha} \phi \partial_{\beta} \phi\bigr)^2 
+ c_{2}(\phi) G^{\mu\nu}\, \partial_{\mu} \phi \, \partial_{\nu} \phi\, \Box \phi + 
c_{3}(\phi) \bigl( \Box \phi \bigr)^2 
\nonumber\\
&& + c_{4}(\phi) \, R^{\mu\nu} \, \partial_{\mu} \phi \,\partial_{\nu} \phi
+ c_{5}(\phi) \, R\, G^{\mu\nu} \,\partial_{\mu} \phi \,\partial_{\nu} \phi
+ c_{6}(\phi) R \, \Box \phi + c_{7}(\phi) R^2 + c_{8}(\phi) \, R_{\mu\nu} \,R^{\mu\nu} 
\nonumber\\
&& + c_{9}(\phi) R_{\mu\alpha\nu\beta} \, R^{\mu\alpha\nu\beta} + c_{10}(\phi) C_{\mu\alpha\nu\beta} \, C^{\mu\alpha\nu\beta}
+ c_{11}(\phi)  R_{\mu\alpha\nu\beta} \, \widetilde{\,R\,}^{\mu\alpha\nu\beta} + c_{12}(\phi) C_{\mu\alpha\nu\beta} \, \widetilde{\,C\,}^{\mu\alpha\nu\beta}\biggr],
\label{THREEa}
\end{eqnarray}
where $R_{\mu\alpha\nu\beta}$ and $C_{\mu\alpha\nu\beta}$ denote the Riemann and  Weyl tensors while $ \widetilde{\,R\,}^{\mu\alpha\nu\beta}$ and $\widetilde{\,C\,}^{\mu\alpha\nu\beta}$ are the corresponding duals; the $c_{i}(\phi)$ (with $i = 1,\,.\,.\,, 12$) are all dimensionless. As long as the effective description is tenable we can separate 
the evolution of the background from the one of the corresponding fluctuations. 
The (cosmic) time derivative of the scalar field $\varphi$ appearing in Eq. (\ref{ONE3}) 
can be notoriously written as $\dot{\varphi} = \sqrt{2\, \epsilon} \, \overline{M}_{P} \, H$ 
where $H= \dot{a}/a$ denotes the Hubble rate and $\epsilon = - \dot{H}/H^2$ is the so-called 
slow-roll parameter that is minute during the inflationary stage; note that the overdot denotes a 
derivation with respect to the cosmic time\footnote{The cosmic time parametrization and the conformal time coordinate (already introduced in Eqs. (\ref{FFF1})--(\ref{FFF2}) and denoted by $\tau$) will be employed when appropriate. The connection between the two is given by $a(\tau) d\tau = dt$. In section \ref{sec4} we shall even introduce a more general parametrization referred to as the $\eta$-time.} coordinate $t$ (see also the discussion after Eq. (\ref{CCC1})). 
While the general situation may be more complicated, in the context of conventional inflationary scenarios there are two complementary situations. The first case corresponds to 
$\epsilon \leq {\mathcal O}(0.01)$: in this situation the mass scale $M = {\mathcal O}(\epsilon) \overline{M}_{P}$. There is also a second possibility and it corresponds to $M= {\mathcal O}(\overline{M}_{P})$ with $\epsilon \ll 1$. As we shall see (see also Eq. (\ref{CRC2d}) and discussion thereafter) according to the consistency relations 
(valid in the case of single-field scenarios), $\epsilon \simeq r_{T}/16$. Since, at the moment,
$r_{T} < 0.06$ (or even $r_{T} <0.03$) we can say, with a fair degree of confidence, that $\epsilon \ll 1$. 

From Eq. (\ref{THREEa}) various other interesting conclusions can be drawn. For instance the leading correction to the two-point function of the scalar mode of the geometry comes from the terms containing four-derivatives of the inflaton field while in the case of the 
tensor modes (which are the ones discussed here) the leading corrections stem from  $C_{\mu\alpha\nu\beta} \,\widetilde{C}^{\mu\alpha\nu\beta}$ and $R_{\mu\alpha\nu\beta} \,\widetilde{R}^{\mu\alpha\nu\beta}$ which are typical of Weyl and Riemann gravity.
Both terms break parity and are therefore capable of polarizing the stochastic backgrounds of the 
relic gravitons \cite{REC3a} by ultimately affecting the dispersion relations of the two circular polarizations. In what follows we shall neglect the terms that break parity and the general form 
of the effective action for the tensor modes of the geometry will be the one 
of Eq. (\ref{DDD1}) (see section \ref{sec4} and discussion therein). We stress though 
that thanks to the two form factors ${\mathcal D}_{1}(\tau)$ and ${\mathcal D}_{2}(\tau)$ Eq. (\ref{DDD1}) may also account for higher-order corrections that do not break parity. The parity breaking terms, however, can be always included and could polarize the relic backgrounds, as suggested in Ref. \cite{REC3a}.

\renewcommand{\theequation}{3.\arabic{equation}}
\setcounter{equation}{0}
\section{Classical production and the maximal frequency}
\label{sec3}
Although the bound deduced in Eq. (\ref{FFFB4}) only relies on quantum mechanical considerations, in a classical perspective the maximal frequency of the spectrum should depend on the timeline of the expansion rate. 
This is why the arguments of section \ref{sec2} are now corroborated and complemented by a classical discussion of different scenarios.
\begin{figure}[!ht]
\centering
\includegraphics[height=8cm]{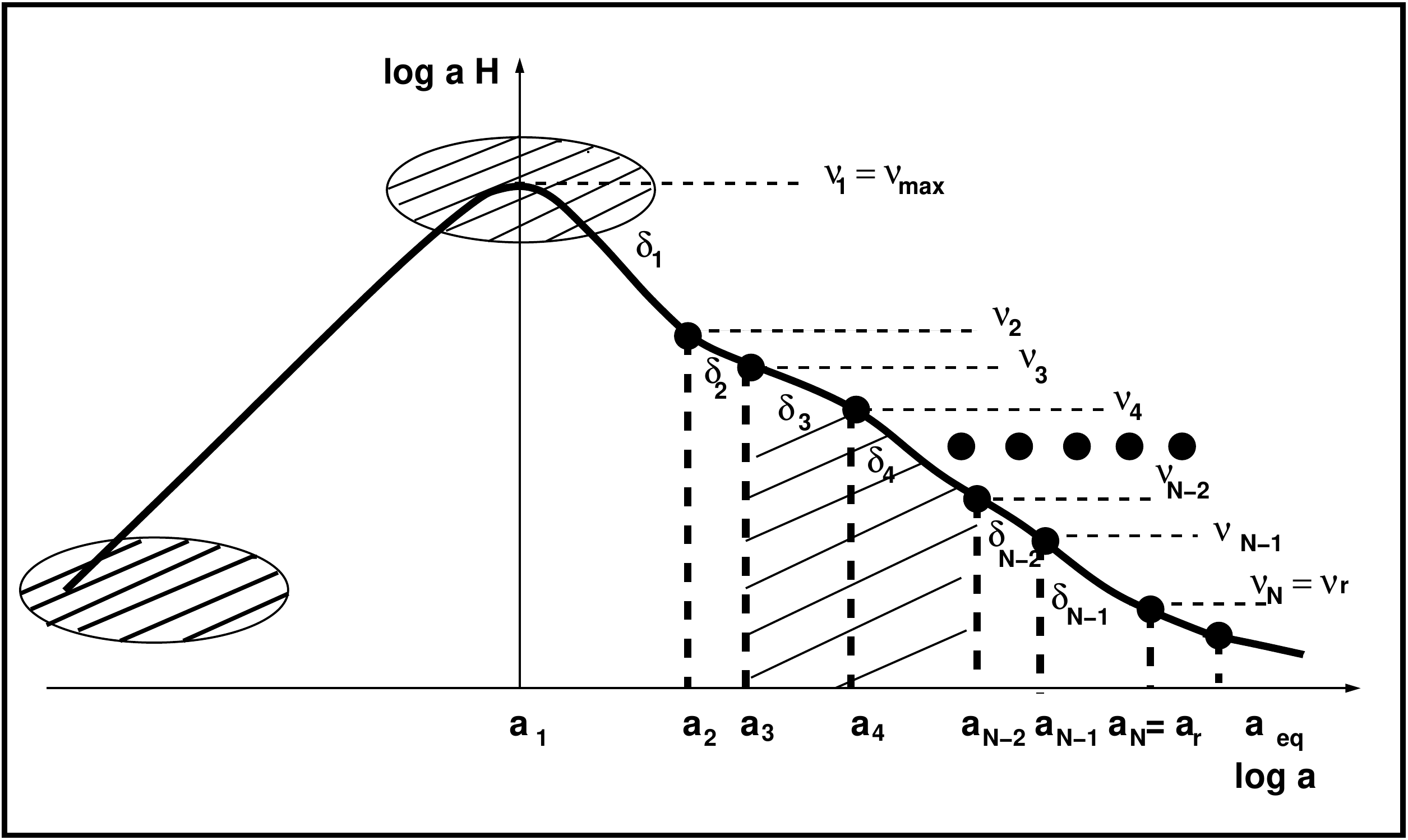}
\caption[a]{The logarithm of the comoving expansion rate is illustrated as a function of the logarithm 
of the scale factor. In this cartoon we assume that the background first inflates (with $a \, H \propto a$) and then always expands in a decelerated manner. The maximum of the thick curve corresponds to the transition between the 
inflationary stage and the decelerated evolution.  The dashed blob close to the maximum is instead of the same order of the maximal frequencies of the spectrum. The dashed area at the extreme left of the plot correspond to the region 
where the smallest frequencies of the spectrum are normalized before exiting the Hubble radius.    The various 
$\delta_{i}$ denote the expansion rates during the postinflationary evolution where various phases may appear. In case
all the $\delta_{i}$  go to $1$, the standard postinflationary evolution dominated by radiation is 
recovered. }
\label{FIGURE3}      
\end{figure}
While the considerations of section \ref{sec2} are model-independent, the developed here are admittedly model-dependent even if the different examples illustrated in this section
hopefully encompass all the relevant physical situations associated with the concordance 
paradigm and with its extensions. In the cartoon of Fig. \ref{FIGURE3} the logarithm of the comoving 
expansion rate $a \, H = {\mathcal H}$ is illustrated\footnote{As usual $H$ denotes the standard 
Hubble rate and ${\mathcal H} = a^{\prime}/a$; the prime 
indicates throughout a derivation with respect to the conformal time coordinate. See also Eq. (\ref{FFF7}) and discussion thereafter.}.
When a conventional inflationary stage of expansion is followed by a decelerated epoch
$a\, H$ has a maximum. In this context the maximal frequency of the spectrum coincides 
with $a_{max} \, H_{max}$. All the frequencies (or wavenumbers) smaller than ${\mathcal O}(a_{max}\, H_{max})$ are amplified depending on the rates of expansion experienced not only during inflation but also in the decelerated stage. For the sake of illustration, in Fig. \ref{FIGURE3} the evolution $a\, H$ has been partitioned in various stages characterized by different values of $\delta$ that represents the rate of expansion in the conformal time coordinate $\tau$. 
If $\delta \to 1$ between $a_{1}$ and $a_{eq}$ the postinflationary evolution 
is dominated by radiation and $a\,H \propto a^{-1}$. This 
possibility is the one commonly endorsed in the context 
of the concordance paradigm. However it has been argued that prior to the 
formation of light nuclei \cite{BBN1,BBN2,BBN3} the 
expansion rate can be very different fro radiation and, in this case, $\nu_{max}$ 
can increase. In what follows we shall first remind of the way the maximal frequency 
of the spectrum is determined. We shall then swiftly consider the conventional 
case where the postinflationary evolution is dominated by radiation and then 
compare the possible alternatives with the THz bound discussed in the 
previous section. 

\subsection{The classical evolution}
For the sake of concreteness we refer to the following classical 
action describing the generalized evolution of the tensor modes of the geometry:
\begin{equation}
S_{g} = \frac{1}{8\, \ell_{P}^2} \int d^{3} x \int d\tau \biggl[ {\mathcal D}_{1}(\tau) \partial_{\tau} h_{i\, j} \partial_{\tau}  h^{i\, j} - 
{\mathcal D}_{2}(\tau) \partial_{k} h_{i\, j} \partial^{k} h^{i\, j} \biggr],
\label{DDD1}
\end{equation}
where $\ell_{P} = \sqrt{8\, \pi G}= \overline{M}_{P}^{-1}$. We note that 
$\overline{M}_{P}$ is the reduced Planck mass that is explicitly related to the Planck mass already introduced in 
Eqs. (\ref{FFF7}) and (\ref{FFFB6a}) as  $\overline{M}_{P} = M_{P}/\sqrt{8 \pi}$. The tensor modes of the geometry appearing in Eq. (\ref{DDD1}) are solenoidal and traceless, i.e.  $\partial_{i} \, h^{i\, j} =0$ and $h_{i}^{\,\,i} =0$. Equation (\ref{DDD1}) implicitly includes all the different models where the propagation 
of the tensor modes only contains two derivatives; the parametrization of the time coordinate can be rescaled so that the general form of the tensor action gets simpler in terms of the $\eta$-time:
 \begin{equation}
S_{g} = 
 \frac{1}{8 \ell_{P}^2} \int d^{3} x\, \int d\eta \, b^2(\eta)\, \biggl[ \partial_{\eta}  h_{i\, j} 
 \partial_{\eta} h^{i\, j} -  \partial_{k} h_{i\, j} \,\partial^{k} h^{i\, j}\biggr].
 \label{DDD2}
 \end{equation}
The $\eta$-time and the generalized scale factor $b(\eta)$ are defined, respectively, as: 
\begin{equation}
d\tau \, \sqrt{{\mathcal D}_{2}(\tau)/{\mathcal D}_{1}(\tau)} =d\eta, \qquad b(\eta) = [{\mathcal D}_{1}(\eta) \, {\mathcal D}_{2}(\eta)]^{1/4}.
\label{DDD3}
\end{equation}
When ${\mathcal D}_{1}(\tau) = {\mathcal D}_{2}(\tau) = a^2(\tau)$ the $\eta$-time coincides with the standard conformal time coordinate. If we now introduce the rescaled tensor amplitude $\mu_{i\,j} = b\, h_{i\, j}$ and express Eq. (\ref{DDD3}) in terms of $\mu_{i\, j}$ we obtain a simpler form of the action \ref{DDD2}
\begin{eqnarray}
S_{g} = \frac{1}{8 \ell_{P}^2} \int d^{3} x \int d\eta \biggl[ \partial_{\eta} \mu_{i\,j} \, \partial_{\tau} \mu^{i\,j} 
+ {\mathcal F}^2 \,\mu_{i\,j} \mu^{i\,j}  - \partial_{k} \mu_{i\, j} \partial^{k} \mu^{i\, j} 
- {\mathcal F}\biggl(\mu_{i\,j} \partial \mu^{i\, j} + \mu^{i\, j} \partial_{\eta} \mu_{i\, j}\biggr)  \biggr],
\label{DDD4}
\end{eqnarray}
where\footnote{We stress that the overdot denotes now a derivation with respect to $\eta$ but in case of potential ambiguities we shall always indicate explicitly the argument of the derivative.} ${\mathcal F}= \dot{b}/b= \partial_{\eta}b/b$.  In the general relativistic situation when ${\mathcal D}_{1}(\tau) = {\mathcal D}_{2}(\tau) = a^2(\tau)$ the derivative with respect to $\eta$ coincides with a derivative with respect to $\tau$ so that ${\mathcal F}= {\mathcal H} = a^{\prime}/a$ and the prime denotes a derivation with respect to the conformal time coordinate $\tau$. The canonical momentum deduced from 
Eq. (\ref{DDD4}) is given by  $\pi_{i\, j} = (\partial_{\eta} \mu_{i\, j} - {\mathcal F} \mu_{i\, j})/(8 \ell_{P}^2)$ and 
 the tensor Hamiltonian is:
\begin{eqnarray}
H_{g}(\eta) = \int d^{3}x\, \biggl[ 8 \ell_{P}^2 \pi_{i\,j} \, \pi^{i\,j} + \frac{1}{8 \ell_{P}^2} \partial_{k} \mu_{i\, j} \partial^{k} \mu^{i\, j} + {\mathcal H} \biggl(\mu_{i\, j} \, \pi^{i\, j} + \pi_{i\, j} \, \mu^{i\, j}\biggr) \biggr].
\label{DDD5}
\end{eqnarray}
In Fourier space the classical fields are represented as:
\begin{eqnarray}
\mu_{i\, j}(\vec{x}, \eta) &=& \frac{\sqrt{2} \ell_{P}}{(2\pi)^{3/2}} \sum_{\alpha} \int d^{3} k \,\,e^{(\alpha)}_{i\,j}(\hat{k})  \,\, \widehat{\mu}_{\vec{k},\,\alpha}(\eta) \, e^{-i \vec{k}\cdot\vec{x}},
\label{DDD6}\\
\widehat{\pi}_{i\, j}(\vec{x}, \eta) &=& \frac{1}{4 \,\sqrt{2}\, \ell_{P}\, (2\pi)^{3/2}} \sum_{\alpha} \int d^{3} k \,\,e^{(\alpha)}_{i\,j}(\hat{k})  \,\, \widehat{\pi}_{\vec{k},\,\alpha}(\eta) \, e^{-i \vec{k}\cdot\vec{x}},
\label{DDD7}
\end{eqnarray}
so that the index $\alpha$ runs over the two tensor polarizations $e^{\oplus}_{i\, j}(\hat{k}) = \hat{m}_{i} \, \hat{m}_{j} - \hat{n}_{i} \, \hat{n}_{j}$ and  $e^{\otimes}_{i\, j}(\hat{k}) = \hat{m}_{i} \, \hat{n}_{j} + \hat{n}_{i} \, \hat{m}_{j}$. As usual $\hat{m}$, $\hat{n}$ and $\hat{k}$ are a triplet of mutually orthogonal unit vectors (obeying $\hat{m}\times \hat{n} = \hat{k}$). Using now Eqs. (\ref{DDD6})--(\ref{DDD7}) into Eq. (\ref{DDD5}) the classical Hamiltonian becomes:
\begin{equation}
H_{g}(\eta) = \frac{1}{2} \int d^{3} p \sum_{\alpha=\oplus,\otimes} \biggl[ \pi_{-\vec{p},\,\alpha}\, \pi_{\vec{p},\,\alpha} + p^2\,\mu_{-\vec{p},\,\alpha}\, \mu_{\vec{p},\,\alpha} + 2 {\mathcal F} \pi_{-\vec{p},\,\alpha}\,\mu_{\vec{p},\,\alpha} \biggr].
\label{DDD8}
\end{equation}
The evolution of the classical fields deduced from Eq. (\ref{DDD8}) is finally given by:
\begin{eqnarray}
\partial_{\eta}\mu_{\vec{k}, \, \alpha} = \pi_{\vec{k},\,\alpha} + {\mathcal F} \,\, \mu_{\vec{k}, \, \alpha}, 
\nonumber\\
\partial_{\eta} \pi_{\vec{k}, \, \alpha}  = - k^2 \widehat{\mu}_{\vec{k},\,\alpha} - {\mathcal F}\,\, \widehat{\pi}_{\vec{k}, \, \alpha},
\label{DDD9}
\end{eqnarray}
If the equations for $\mu_{\vec{k}, \, \alpha}$ and $\pi_{\vec{k},\,\alpha}$ are decoupled, they become 
\begin{equation}
\ddot{\mu}_{\vec{k}, \, \alpha} + \biggl[ k^2 - \frac{\ddot{b}}{b} \biggr] \widehat{\mu}_{\vec{k}, \,\alpha}=0, \qquad \widehat{\pi}_{\vec{k},\,\alpha} = \dot{\mu}_{\vec{k}, \, \alpha} - {\mathcal F} \,\mu_{\vec{k}, \, \alpha},
\label{DDD10}
\end{equation}
in the time coordinate $\eta$. Equation (\ref{DDD10}) implies 
that the maximal amplified frequency (or, in equivalent terms, the minimal amplified wavelength of the spectrum) 
corresponds to those modes for which $k^2$ is of the order of $|\ddot{b}/b|$. Indeed when $k^2 \ll |\ddot{b}/b|$ 
the corresponding modes are said to be superadiabatically amplified while in the opposite 
range (i.e. $k^2 \gg |\ddot{b}/b|$) the classical fields oscillate and no amplification takes place. It is 
therefore clear that the maximal wavenumber of the spectrum must be deduced from the condition 
$k^2 \simeq |\ddot{b}/b|$. This condition will now b specifically discussed when the timeline 
of the expansion rate corresponds to the cartoon of Fig. \ref{FIGURE3}. 

\subsection{The maximal frequency in the conventional lore}
In the case of conventional inflationary scenarios in their extensions 
the $\eta$-time coincides with the conformal time coordinate and $b \to a$. The pump field 
appearing in Eq. (\ref{DDD10}) can then be written as: 
\begin{equation}
\frac{a^{\prime\prime}}{a} =  a^2 H^2 (2 - \epsilon), \qquad \epsilon = - \dot{H}/H^2,
\label{CCC1}
\end{equation}
where $\epsilon$ is the standard slow-roll parameters and the overdot denotes throughout the remaining part of this section a derivation with respect to the cosmic time coordinate $t$ defined as $a(\tau) \, d\tau = d\,t$. During the inflationary stage of expansion $\epsilon \ll 1$ whereas at the end of inflation 
\begin{equation}
\epsilon = - \frac{\dot{H}}{H^2} =\biggl(\frac{3\dot{\varphi}^2}{\dot{\varphi}^2 + 2\,V}\biggr)
\to 1,
\label{CCC2}
\end{equation}
where $\varphi$ is the inflaton and the end of the inflationary stage which takes place when 
$\dot{\varphi}^2$ and $V$ are comparable. Recalling now the cartoon of Fig. \ref{FIGURE3} together with Eq. (\ref{CCC1}) we conclude that the maximal wavenumber of the spectrum is proportional to $a\,H$ evaluated at the end of the inflationary stage and that the field amplitudes of Eq. (\ref{DDD10}) experience their minimal amplification when 
\begin{equation}
 k^2 = k_{max}^2 \simeq a_{max}^2 \, H_{max}^2 = a_{1}^2 \, H_{1}^2.
\label{CRC2}
\end{equation}
It then follows that $\nu_{max} = H_{max} (a_{max}/a_{0})/(2 \pi)$ where $a_{0}$ denotes the scale factor at the present time; we shall be assuming that  $a_{0} =1$ so that, at the present time, physical and comoving frequencies are bound to coincide. Broadly speaking the classical value of $\nu_{max}$ depends on two separate physical quantities: {\it (i)} the maximal expansion rate $H_{max}$ (coinciding, for the 
present purposes, with the rate at the end of inflation) and {\it (ii)} the total amount 
of redshift $(a_{max}/a_{0})$ that ultimately depends on the {\em whole} postinflationary 
expansion history. 

The total amount of redshift determined during a radiation stage (extending from $a_{max}$ to $a_{eq}$) does not maximize the value of $\nu_{max}$ and it does not minimize it either.  To clarify this statement we now estimate the value of $\nu_{max}$ when the postinflationary evolution is dominated by radiation\footnote{In the following two subsections we are going to consider more general situations and eventually compare the obtained values with the quantum bound of section \ref{sec2}.} between $a_{1}$ and $a_{eq}$ (see Fig. \ref{FIGURE3}). 
For this purpose we first remind two simple relations, namely 
\begin{equation}
 \frac{a_{0} \, H_{0}}{a_{eq} \, H_{eq}} = \frac{1}{(2 \, \Omega_{R\, 0})^{1/4}} \sqrt{\frac{H_{0}}{H_{eq}}}, \qquad \frac{a_{eq}\, H_{eq}}{a_{r} \, H_{r}} = \bigl(g_{s,\, r}/g_{s,\, eq}\bigr)^{1/3} \, \bigl(g_{\rho,\, eq}/g_{\rho,\,r}\bigr)^{1/4} \, \sqrt{\frac{H_{eq}}{H_{r}}},
 \label{CRC2a}
\end{equation}
where  $g_{\rho}$ is the number of relativistic degrees of freedom in the plasma while $g_{s}$ denotes the effective number of relativistic degrees of freedom appearing in the entropy density. In the conventional situation of the standard model of particle interactions \cite{CL1} $g_{s,\, r}= g_{\rho,\, r} = 106.75$ and $g_{s,\, eq}= g_{\rho,\, eq} = 3.94$. Therefore from Eq. (\ref{CRC2a}) we may also deduce that the value of the maximal frequency is given by:
\begin{equation}
\overline{\nu}_{max} = \frac{ (2 \Omega_{R0})^{1/4}}{2 \pi} \,\, \biggl(\frac{g_{s,\, eq}}{g_{s, \, r}}\biggr)^{1/3} \,\biggl(\frac{g_{\rho,\, r}}{g_{\rho, \, eq}}\biggr)^{1/4} \,\, \sqrt{H_{max} \, H_{0}},
\label{CRC2b}
\end{equation}
where $\overline{\nu}_{max}$ now indicates {\em the value of the maximal frequency corresponding 
to a conventional postinflationary evolution dominated by radiation down to the equality time}; furthermore, with standard notations $\Omega_{R0}= {\mathcal O}(10^{-5})$ stands for the present critical fraction of relativistic species in the concordance scenario. The value of $H_{max}$ can be immediately estimated from $H_{k}$ which is the value of the expansion rate where a given wavelength crosses the Hubble radius during inflation:
\begin{equation}
H_{k} = \frac{H_{max}}{1 - \epsilon_{k}} \biggl|\frac{k}{a_{max} \, H_{max}} \biggr|^{ - \frac{\epsilon_{k}}{1 - \epsilon_{k}}} \simeq H_{max}  \biggl|\frac{k}{a_{max} \, H_{max}} \biggr|^{ - \epsilon_{k}}\biggl[1 + {\mathcal O}(\epsilon_{k})\biggr],
\label{CRC2c}
\end{equation}
where $\epsilon_{k} = \epsilon(1/k)$ and $H_{k}= H(1/k)$ are both viewed as a function 
of the conformal time coordinate and their value is fixed when the corresponding wavelength crosses the Hubble radius during inflation\footnote{In more technical terms this particular stage will be referred to as the exit (i.e. $\tau_{ex}(k)$) since the corresponding wavelength exits the Hubble radius; the $\tau_{ex}(k)$ is opposed to the reentry time where 
the given wavelength becomes again shorter than the Hubble radius. Both $\tau_{ex}(k)$ and $\tau_{re}(k)$ are $k$-dependent quantities.}. We now remind that the amplitude of the power spectrum of curvature inhomogeneties ${mathcal A}_{{\mathcal R}}$ also estimates $H_{k}$ according to the following relation:
\begin{equation}
{\mathcal A}_{{\mathcal R}} = \frac{H_{k}^2}{\pi\, \epsilon_{k} \, M_{P}^2} \simeq \frac{16\, H_{max}^2}{\pi \, r_{T}\, M_{P}^2},
\label{CRC2d}
\end{equation}
where the second equality follows from the consistency relations stipulating that 
$r_{T} \simeq 16\, \epsilon_{k}$. All in all, thanks to Eqs. (\ref{CRC2c})--(\ref{CRC2d}) the value 
of $\overline{\nu}_{max}$ given in Eq. (\ref{CRC2b}) can be estimated in more explicit terms:
\begin{equation}
\overline{\nu}_{max} =171.46\, \biggl(\frac{{\mathcal A}_{{\mathcal R}}}{2.41\times 10^{-9}}\biggr)^{1/4}\,\,
\biggl(\frac{r_{T}}{0.03}\biggr)^{1/4} \,\, \biggl(\frac{h_{0}^2 \, \Omega_{R\,0}}{4.15\times 10^{-5}}\biggr)^{1/4} \biggl(\frac{g_{s,\, eq}}{g_{s, \, r}}\biggr)^{1/3} \,\biggl(\frac{g_{\rho,\, r}}{g_{\rho, \, eq}}\biggr)^{1/4}\,\,\mathrm{MHz}.
\label{CRC3}
\end{equation}
The estimate of $\overline{\nu}_{max}$ is then of the order of $200$ MHz but 
a more accurate determination of the maximal frequency depends on $r_{T}$. In Eq. (\ref{CRC3}) 
we assumed that $g_{s,\, r}= g_{\rho,\, r} = 106.75$ and $g_{s,\, eq}= g_{\rho,\, eq} = 3.94$; in this 
case the contribution of the relativistic species is ${\mathcal O}(0.75)$. Note also that 
if we double the value of $r_{T}$ (i.e. $r_{T} =0.06$) the $\overline{\nu}_{max}$ gets ${\mathcal O}(200) $ MHz. 
The value of  $\overline{\nu}_{max}$ is always much smaller that the $\mathrm{THz}$ so that the quantum bound of Eq. (\ref{FFF4}) is always satisfied. 

\subsection{The maximal frequency and the postinflationary history}
The maximal frequency of the relic gravitons depends on $H_{max}$ but a modified 
postinflationary evolution may increase the value given in Eq. (\ref{CRC3}) by few orders\footnote{In this sense the dependence of $\overline{\nu}_{max}$ on $r_{T}$, $g_{s}$ and $g_{\rho}$ spelled out in Eq. (\ref{CRC3}) is only meaningful as long as we pretend that the expansion history after inflation is always dominated by radiation.} of magnitude and potentially contradict the quantum bound deduced in Eq. (\ref{FFF4}). To investigate this possibility we
follow the notations established in Fig. \ref{FIGURE3} and assume that between the 
end of inflation and the dominance of radiation there are $N$ different stages of expansion 
that are arbitrarily different from radiation. 
It is not difficult to show that the value of the maximal frequency becomes, in this case:
\begin{equation}
\nu_{max} = \overline{\nu}_{max}\, \prod_{i=1}^{N-1} \,\, \xi_{i}^{\beta_{i}},
\label{CRC4}
\end{equation}
where $\xi_{i}$ and $\beta_{i}$ measure, respecively, the duration of each of the postinflationary 
stages and the corresponding expansion rate; while $\xi_{i}$ is the ratio between two successive 
expansion rates, $\beta_{i}$ depends on the $\delta_{i}$ of Fig. \ref{FIGURE3}:
 \begin{equation}
\xi_{i} = H_{i+1}/H_{i} < 1, \qquad  \beta_{i}\, =\, (\delta_{i} -1)/[2\, (\delta_{i} + 1)].
\label{CRC5}
\end{equation}
In Eq. (\ref{CRC5}) $H_{i}$ denotes the rate during the {\em i-th} stage of expansion and the result of Eq. (\ref{CRC3})
is recovered when all the $\beta_{i} \to 0$ implying that all the $\delta_{i} \to 1$: in this situation 
the evolution is dominated by radiation from $H_{max}$ down to $H_{eq}$. This also means that 
the value of $\nu_{max}$ given in Eq. (\ref{CRC5}) may exceed $200$ MHz provided, at least,
one of the various $\delta_{i}$ gets smaller than $1$.  The results of Eqs. (\ref{CRC4})--(\ref{CRC5}) are obtained 
when between $H_{max}$ and $H_{bbn}$ there are $N$ different expanding stages; we 
conservatively assumed that $H_{r} \to H_{bbn} = 10^{-42} \, M_{P}$ corresponding to a typical nucleosynthesis temperature 
${\mathcal O}(10)$ MeV.  As already stressed in Fig. \ref{FIGURE3} we conventionally posit that $H_{N}$ coincides with the onset of the radiation dominated phase, i.e. $H_{N}= H_{r}$ and $a_{N}= a_{r}$. 

For the sake of concreteness we now consider the simplest situation implicitly contained in 
Eqs. (\ref{CRC4})--(\ref{CRC5}), namely the case $N= 3$; according to Eq. (\ref{CRC4}) the expression of $\nu_{max}$ becomes
\begin{equation}
\nu_{max} =  \,\,\xi_{1}^{\frac{\delta_{1} -1}{2 (\delta_{1}+1})}\,\,\xi_{2}^{\frac{\delta_{2} -1}{2 (\delta_{2}+1})}\,\, \overline{\nu}_{max}.
\label{CRC6}
\end{equation}
By looking again at Fig. \ref{FIGURE3} there are two further typical frequencies in the spectrum namely $\nu_{2}$ and $\nu_{3} = \nu_{r}$:
\begin{equation}
\nu_{2} = \sqrt{\xi_{1}} \,\, \xi_{2}^{(\delta_{2} -1)/[2 (\delta_{2}+1)]} \, \,\overline{\nu}_{max},\qquad \nu_{r}= \sqrt{\xi_{1}}\, \sqrt{\xi_{2}}\, \overline{\nu}_{max} = \sqrt{\xi} \,\overline{\nu}_{max},
\label{CRC7}
\end{equation}
where, by definition, $\xi= \, \xi_{1} \, \xi_{2}$. In the case of $N$ intermediate stages preceding  the dominance of radiation at $a_{r}$, between $\nu_{max}$ and $\nu_{r}$
there will be $N-2$ intermediate frequencies corresponding to specific breaks in the spectral energy density as discussed in Ref. \cite{mg3}.
Thus for $N=3$ the postinflationary contribution to $\nu_{max}$ is maximized 
when the $\delta_{i}$ and $\xi_{i}$ take their minimal values:
\begin{equation}
\delta_{1} = \delta_{2}\to  1/2, \qquad\qquad \xi_{1} \, \xi_{2} = H_{bbn}/H_{max}.
\label{CRC8}
\end{equation}
The common value of $\delta_{1}$ and $\delta_{2}$ corresponds to the slower expansion rate 
allowed in the primeval plasma. For instance in a perfect fluid the maximal value of the barotropic index (be it $w_{max}$) corresponds to the expansion rate, i.e.  $\delta_{min} = 2/(3 w_{max} +1)$ and since, at most, $w_{max} \to 1$ we obtain, as anticipated $\delta_{min} \to 1/2$.  The expansion rate can be slower than radiation also when the energy-momentum tensor is dominated by the oscillations of the inflaton. Assuming the minimum of the potential is located in $\varphi =0$
the inflaton potential  can be parametrized as $V(\varphi)\simeq V_{1} \Phi^{2 q}$ (with $\Phi = \varphi/M_{P}$).  If the postinflationary evolution is dominated by the inflaton oscillations, the averaged evolution of the comoving horizon may mimic the timeline of a stiff epoch and the graviton spectra  \cite{mg0,mg1,mg3}. During the coherent oscillations of $\varphi$ the energy density of the scalar field is roughly constant and, in average, the expansion rate is $\delta=  (q+1)/(2 q-1)$ \cite{osc1,osc2,osc3,osc4,osc5}. Again the minimal value of $\delta$ is ${\mathcal O}(1/2)$ and this happens when $q\gg 1$.

The same argument presented in the case of $N= 3$ is easily repeated for an arbitrary number of intermediate phases; in this case the generalized values of the intermediate frequencies are:
\begin{eqnarray}
\nu_{m} &=& \sqrt{\xi_{1}} \, .\,.\,.\, \sqrt{\xi_{m-1}}\,  \prod_{m\,=\,1}^{N -1} \xi_{i}^{\frac{\delta_{i} -1}{2 (\delta_{i}+1})}\,\, \overline{\nu}_{max},
\label{CRC9}\\
\nu_{r} &=& \nu_{N} = \sqrt{\xi_{1}}\,\sqrt{\xi_{2}} \, .\,.\,.\, \sqrt{\xi_{N-2}}\,\sqrt{\xi_{N-1}} \,\,\overline{\nu}_{max}.
\label{CRC10}
\end{eqnarray}
In Eq. (\ref{CRC9}) each of the different values of $m \leq N -1$ correspond to the frequencies of the 
breaks illustrated in Fig. \ref{FIGURE3}. Finally, the generalization of Eq. (\ref{CRC8}) reads:
 \begin{equation}
\overline{\xi}= \xi_{1} \, \xi_{2} \, .\, .\,.\, \xi_{N-2} \, \xi_{N-1}= H_{r}/H_{max}, \qquad \delta_{1} = \delta_{2}= \,.\,.\,.=\delta_{N-2} = \delta_{N-1} \to 1/2.
\label{CRC11}
\end{equation}
In other words, because all the $\delta_{i}$ are equal we obtain that the product of all the $\xi_{i}$ (denoted by $\overline{\xi}$ in Eq. (\ref{CRC11})) is ultimately raised to the same power implying that the contribution of the whole decelerated stage of expansion of Eq. (\ref{CRC4}) is maximized by a single expanding stage characterized by $\overline{\delta} < 1$:
\begin{equation}
\prod_{i=1}^{N-1} \,\, \xi_{i}^{\beta_{i}} \leq \biggl(\frac{H_{r}}{H_{max}}\biggr)^{ \overline{\beta}} < \biggl(\frac{H_{bbn}}{H_{max}}\biggr)^{\overline{\beta}}, 
\qquad \overline{\beta} = \frac{\overline{\delta}-1}{2 (\overline{\delta} +1)} \to - \frac{1}{6}.
\label{CRC12}
\end{equation}
 Thanks to Eqs. (\ref{CRC9})--(\ref{CRC10}) and (\ref{CRC11})--(\ref{CRC12}) we therefore obtain the following bound on $\nu_{max}$
\begin{equation}
\nu_{max} < 10^{6} \, \biggl(\frac{H_{max}}{M_{P}}\biggr)^{2/3}\,\, \biggl(\frac{h_{0}^2 \Omega_{R0}}{4.15\times 10^{-5}}\biggr)^{1/4} \, \, \mathrm{THz},
\label{CRC13}
\end{equation}
where it has been assumed that $H_{r} \to H_{bbn} = 10^{-42} \, M_{P}$; this choice implies, as already stressed, that the background was dominated by radiation already for temperatures ${\mathcal O}(10)$ MeV. Furthermore since $H_{max} = {\mathcal O}(H_{k}) = {\mathcal O}(10^{-6}) \, M_{P}$ (see Eq. (\ref{CRC2d})) the limit of Eq. (\ref{CRC13}) suggests that $ \nu_{max} < {\mathcal O}(100)\, \mathrm{THz}$. If we now consider together the two limits of Eqs. (\ref{FFFB4}) and (\ref{CRC13}) we conclude that the quantum bound is always more constraining than the one obtained with classical arguments. 
\begin{figure}[!ht]
\centering
\includegraphics[height=8cm]{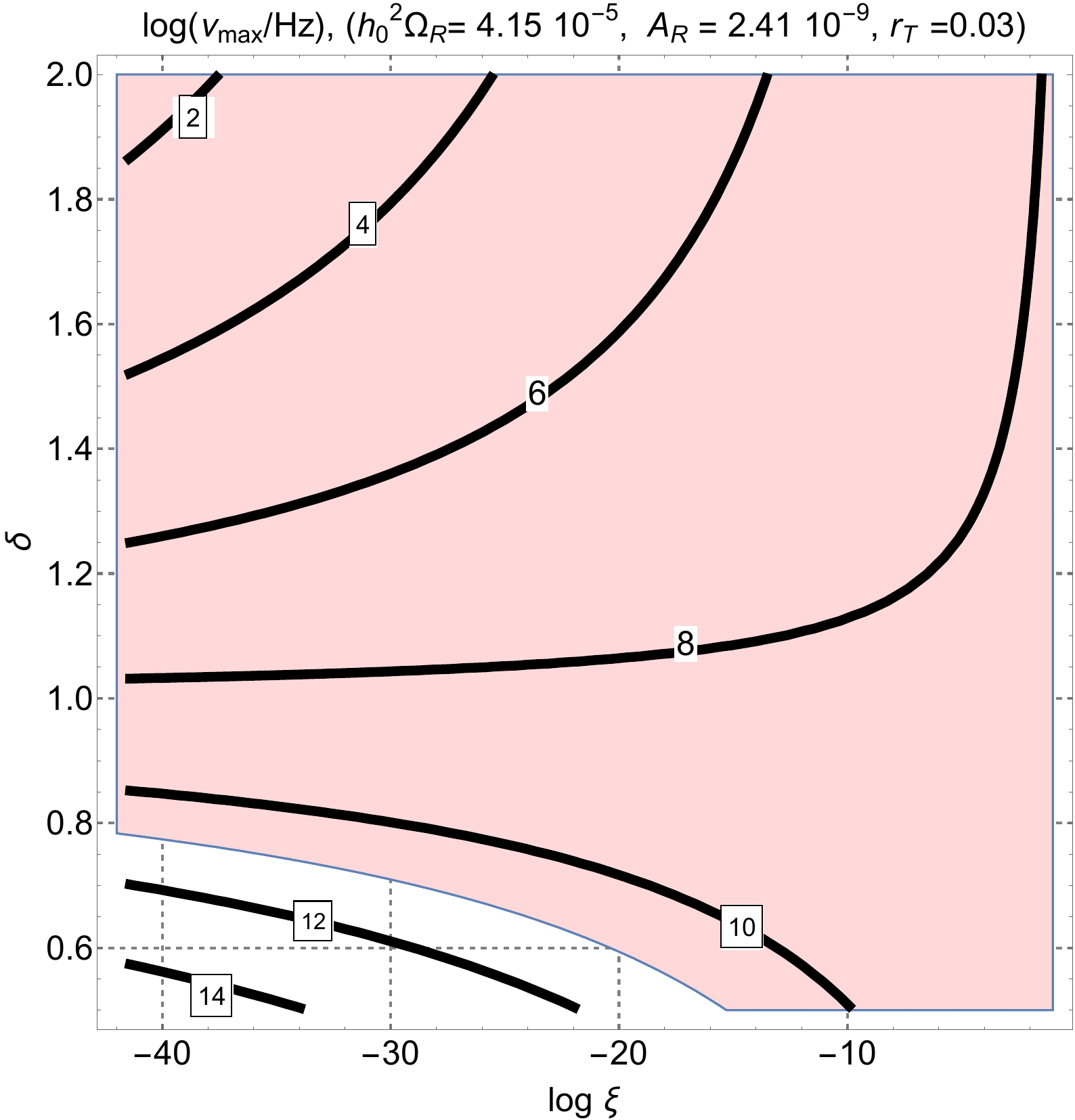}
\includegraphics[height=8cm]{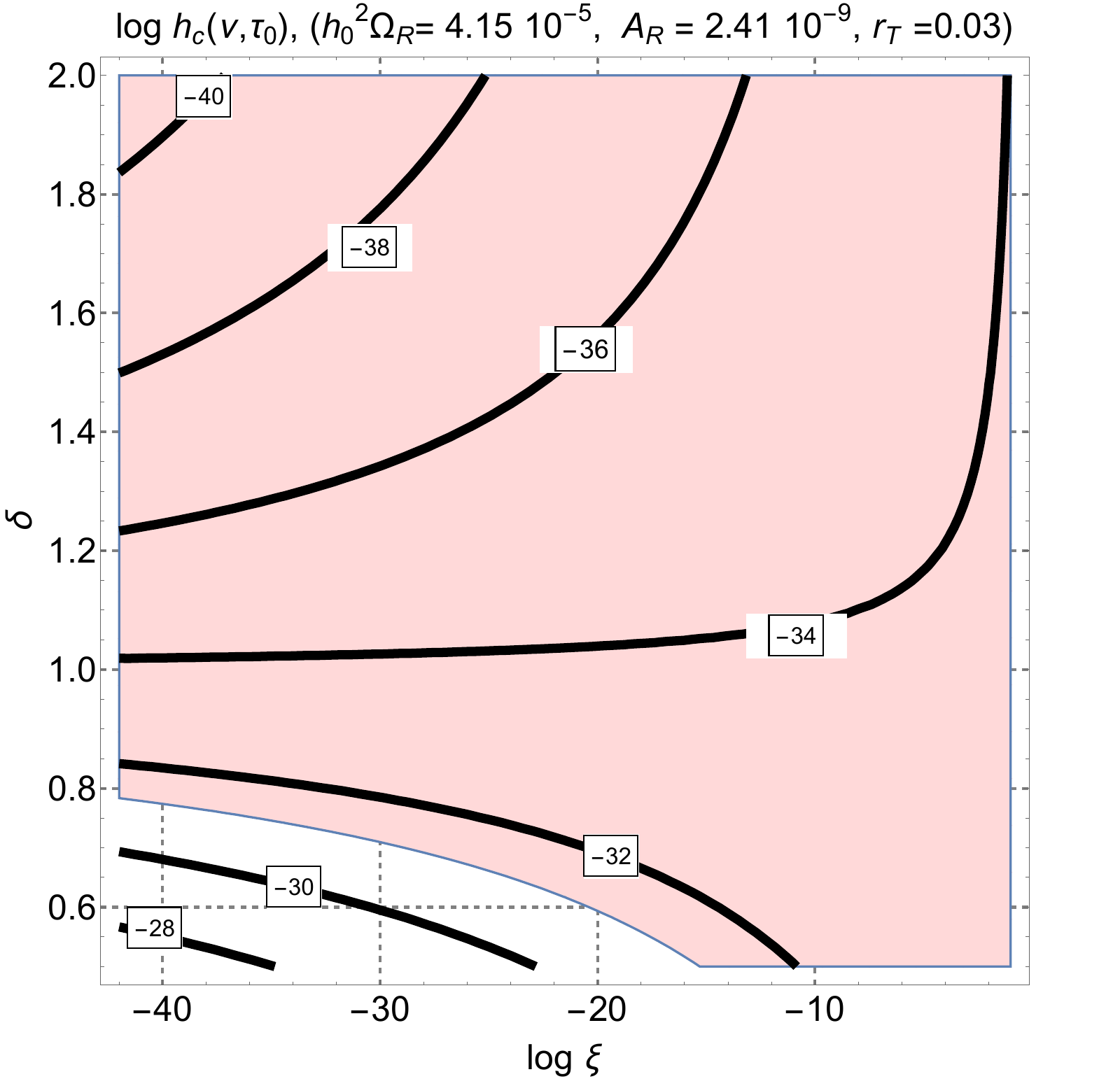}
\caption[a]{In the plot at the left the labels appearing on the various contours illustrate the common logarithm of $\nu_{max}$ expressed in Hz.  In the right plot we report instead, for comparison, the common logarithms of the chirp amplitudes computed for the same range of parameters. In both plots on the horizontal and on the vertical axes refer, respectively, to the values of $\log{\xi}$ and of $\delta$. In both plots the shaded area corresponds to the region where 
all the phenomenological constraints are concurrently satisfied.  The shaded region of the right plot defines, in practice, the minimal sensitivities required for any instrument aiming at detecting high-frequency gravitons in the high-frequency domain. As we can see $h_{c}^{(min)}$ should be ${\mathcal O}(10^{-32})$ or smaller. This analysis corroborates the estimate already obtained in Eq. (\ref{FFFB7}) with the quantum mechanical argument propounded in section \ref{sec2}.}
\label{FIGURE4}      
\end{figure}
To illustrate this issue even further, in the left plot of Fig. \ref{FIGURE4} we report the common logarithm of $\nu_{max}$ (expressed in Hz) as a function of the expansion rate $\delta$ and of the length of the postinflationary phase $\xi$. The lower portion of Fig. \ref{FIGURE4} refers to the choice that maximizes the deviation of $\nu_{max}$ from $\overline{\nu}_{max}$. The shaded region in both plots of Fig. \ref{FIGURE4} defines the portion of the parameter space  where all the phenomenological constraints discussed of section are concurrently satisfied and, in particular, the bounds coming from big-bang nucleosynthesis \cite{BBN1,BBN2,BBN3}. It is clear that, within the shaded region,  the quantum 
bound of Eq. (\ref{FFF4}) is always satisfied. The results of Fig. \ref{FIGURE4} graphically demonstrate that expansion rate slower than radiation may shift from $\nu_{max}$ from the  MHz region to the THz domain, as qualitatively argued from the results of Eqs. (\ref{CRC4})--(\ref{CRC5}) and (\ref{CRC6}). The shift in $\nu_{max}$ towards higher frequencies 
should not excessively increase the spectral energy density of the created gravitons at $\nu_{max}$; in the case of a single $\delta$-phase this quantity is computed to be
\begin{equation}
 \Omega_{gw}(\nu_{max}, \tau_{0}) = \frac{8}{3 \pi} \, \Omega_{R0} \biggl(\frac{H_{max}}{M_{P}}\biggr)^{4/(\delta+1)} \, \biggr(\frac{H_{r}}{M_{P}}\biggr)^{2(\delta -1)/(\delta+1)}.
\label{CRC14}
\end{equation}
 Equation (\ref{CRC14}) demonstrates that when $\delta<1$ (so that the postinflationary the expansion is slower than radiation) the value of $H_{r}$ cannot be too small otherwise $\Omega_{gw}(\nu_{max},\tau_{0})$ would violate various constraints including the one associated with nucleosynthesis bounds \cite{mg3}.  Furthermore the values of $\nu_{max}$ appearing in each contour of Fig. \ref{FIGURE4}  increase when $\delta <1$ but they decrease when $\delta >1$. Between these two cases the former is clearly more constraining than the latter especially in the light of the quantum bound of Eq. (\ref{FFF4}). When $\delta > 1$ the postinflationary expansion rate is faster than radiation and the maximal frequency gets always smaller than ${\mathcal O}(200)$ MHz. This 
is why when $\delta > 1$  the bound of Eq. (\ref{FFF4}) is automatically satisfied. For the sake of illustration in the right plot of Fig. \ref{FIGURE4} the various contours correspond to different values of the 
common logarithms of the chirp amplitudes in the region of the parameter space where all the phenomenological constraints 
are satisfied. As we can see this result is in fact compatible with the one directly 
obtained from the quantum bound (see Eq. (\ref{FFFB7}) and discussion therein) since $h_{c}^{(min)} \leq {\mathcal O}(10^{-32})$. 

\subsection{Less conventional possibilities}
So far  the estimates of $\nu_{max}$ have been considered in the context of conventional inflationary models and we are now going to discuss few complementary cases related, respectively, to the refractive index of the relic gravitons, to the 
bouncing scenarios and to the thermal spectrum of thermal gravitons.  
The tensor modes of the geometry may inherit an effective refractive index when they propagate in curved backgrounds \cite{REFR1,REFR2} and the inclusion of a refractive index generically leads to growing spectra of relic gravitons at intermediate frequencies \cite{REFR3}.
During the last three years the pulsar timing arrays reported a series of repeated evidences of gravitational radiation (with stochastically distributed Fourier amplitudes) at a benchmark frequency of the order of $30$ nHz and characterized by spectral energy densities (in critical units) ranging between $10^{-8}$ and $10^{-9}$ \cite{BB3,BB4}. While it is still unclear if the signal observed by the pulsar timing arrays is really due to a diffuse background of gravitational radiation, the presence of a dynamical refractive index during a conventional inflationary phase modify the exit of the intermediate scales 
and leads to a sharp excess in the nHz range \cite{BB5a}. A sufficient condition for this to happen is that the refractive index is always larger than $1$ during inflation \cite{REFR3,BB5a}. As a consequence during inflation the $\eta$-time and the conformal time coordinates evolve differently. More precisely we have that, in this case, the effective scale factor $b(\eta)$ appearing in Eqs. (\ref{DDD2})--(\ref{DDD3}) is given by $b(\eta) = a(\eta)/\sqrt{n(\eta)}$ where $n(\eta)$ denotes the profile of the refractive index of the relic gravitons. Since the refractive index goes to $1$ in the postinflationary phase the maximal frequency of the graviton spectrum is in the range ${\mathcal O}(0.1)$ GHz (see Ref. \cite{BB5a} and discussion therein). This figure fully agree with the THz bound of Eq. (\ref{FFF4}).

The second possibility examined here involves the bouncing scenarios\footnote{The quantum mechanical bound derived in Eq. (\ref{FFF4}) applies also in this case as long as we can identify an asymptotic vacuum where the classical fluctuations have been suppressed.}. There exist at the moment various 
versions of the bouncing scenarios all based on different and sometimes opposite premises \cite{BBBO4,BBBO5,BBBO6,BBBO7,BBBO8}. The spectra of the relic gravitons in bouncing models 
have been reviewed in Ref. \cite{AA13} within a unified perspective but what matters 
for the present considerations is that the bouncing scenarios, in spite of their diverse premises, 
can always be divided in two broad categories \cite{TTT3}: the bounces of the scale factor (where the 
Hubble rate goes to zero at least once) and the curvature bounces where it is the derivative of the Hubble rate 
that passes through zero (at least once). Both cases may suffer serious drawbacks from the viewpoint 
of the effective four-dimensional description \cite{TTT3} but the relevant point, for the present ends, is that after the bouncing regime the maximal curvature scale at the onset of the decelerated stage of expansion could even be as large as the Planck mass. Depending on the specific scenario we may have that $H_{max} \leq M_{P}$ implying $\nu_{max} < \mathrm{THz}$ 
for all the cases of a consistent postinflationary evolution \cite{AA13}.

Up to now we demonstrated that the cosmic gravitons produced in conventional and unconventional inflationary scenarios must satisfy the quantum bound of Eq. (\ref{FFFB4}) and, a fortiori, the classical bound of Eq. (\ref{CRC13}). It is now useful to  verify this conclusion by considering scenarios based on different physical premises. In the case of a bouncing scenario where the postinflationary stage of expansion is dominated by radiation $\nu_{max}$ can be estimated as:
\begin{equation}
\nu_{max} = 0.1 \, \biggl(\frac{H_{max}}{M_{P}}\biggr)^{1/2} \,\, \mathrm{THz} < \mathrm{THz}, 
\label{HH6}
\end{equation}
where $H_{max} < M_{P}$. If $H_{max}$ coincides with the string scale we would 
have approximately that $H_{max} = M_{s} = {\mathcal O}(10^{-2})\, \mathrm{M}_{P}$ 
so that $\nu_{max}$ may range between $10$ and $100$ GHz \cite{AA13}.

Last but not least we consider the thermal graviton spectrum and show that
also in this case the THz bound is satisfied.The temperature of the gravitons always undershoots the one of the thermal photons; in particular we denoting with $T_{g\, 0} $ and $ T_{\gamma\,0}$ the current values of the graviton and photon 
temperatures they can be related as $T_{g\, 0} = \epsilon_{g} \,\, T_{\gamma\,0}$ where $\epsilon_{g}< 1$ 
depends on the number of relativistic species at the time of graviton decoupling \cite{AA13}. 
The maximal frequency of the thermal black-body spectrum is then estimated ad $\nu_{max} = 73.943 \mathrm{GHz}$ 
where we assumed that the temperature of the gravitons is $T_{g\, 0} = 0.905 \, \mathrm{K}$ as it follows 
by assuming that all the species of the standard model of particle interactions were 
in thermal equilibrium at the scale of graviton decoupling. 

\subsection{The classical and the quantum bounds: final assessment}
We can now summarize the logic pursued so far by recalling, in a physical perspective, 
the results of Eqs. (\ref{FFFB4}), (\ref{CRC13}) and (\ref{HH6}). 
From the requirement that only one pair of gravitons are produced we first 
obtained the quantum bound of Eq. (\ref{FFFB4}). We already stressed 
that this bound does depend on the production mechanism but it is 
insensitive to the cosmological evolution. The quantum bound generally 
implies that $\nu_{max} < {\mathcal O}(\mathrm{THz})$. The quantum discussion has been corroborated by the classical derivation of this section where the maximal frequency 
is obtained by scaling the maximal curvature scale at the present time. From this 
requirement we obtained that, within a classical approach, $\nu_{max}$ should 
be  smaller than ${\mathcal O}(10^{6})\mathrm{THz}$. This bound 
obtained in Eq. (\ref{CRC13}) is less constraining than the quantum limit. The result 
of Eq. (\ref{HH6}) (valid in the case of bouncing scenarios) is again of the order of 
Eq. (\ref{FFFB4}). This means that the result of Eq. (\ref{FFFB4}) is the relevant one from the physical viewpoint. The upper limit on the maximal frequency of the spectrum is then obtained 
by requiring that only a single pair of gravitons is produced. The interesting aspect of this 
result, as stressed in section \ref{sec2}, is that the multiplicity distribution of the gravitons pairs 
is geometric (i.e. Bose-Einstein), as already argued long ago with a slightly different logic \cite{PARKTH}.

\renewcommand{\theequation}{4.\arabic{equation}}
\setcounter{equation}{0}
\section{Unitarity and the averaged multiplicity}
\label{sec4}
The conclusions based on Eqs. (\ref{FFF1})--(\ref{FFF2}) are rooted in the unitary evolution of the field operators describing the gravitons and the same remark holds 
for the quantum theory of parametric amplification as whole \cite{louisell,mollow1,mollow2}. 
For this reason the bounds of \ref{sec2} as well as the multiplicity distributions of the gravitons in the different regimes cannot be reliably deduced if unitarity is either partially or totally lost. In this sense the largeness of the averaged multiplicities of the gravitons does not imply per se a classical evolution or a partial loss of quantum coherence as we could superficially argue. The unitarity of the evolution ultimately implies that the commutation relations are preserved as the background passes through different evolutionary stages. The quantumness of the created gravitons is encoded in the degrees of quantum coherence especially in the high-frequency regime \cite{REC2}. In this section the unitarity 
will then be used as a constraint on the approximation schemes employed to evaluate the averaged multiplicity and the other correlation properties of the created gravitons. We shall then demonstrate that the gravitons truly appear in a macroscopic 
quantum state characterized by a large averaged multiplicity. A number of complementary results related to the discussions
of the present section can be found in the appendices \ref{APPB} and \ref{APPC}.

\subsection{Quantum parametric amplification}
The pair production from the vacuum  described by Eqs. (\ref{FFF1})--(\ref{FFF2}) and the quantum theory of parametric amplification have actually the same physical content. Even if the general argument
implying the bound on the maximal frequency does not depend on the details 
of the parametric amplification, the averaged multiplicities and the correlation properties are quantitatively 
different. The standard tenets of the quantum theory of parametric amplification are analyzed in this section with the purpose of deducing the averaged multiplicities in a framework where unitarity is explicitly enforced. We therefore start from the Hamiltonian of Eq. (\ref{DDD8}) and promote the classical fields to the status of operators.  In Fourier space the operators 
$\widehat{\mu}_{\vec{k},\,\alpha}$ and $\widehat{\pi}_{\vec{k},\,\alpha}$ 
can be expressed in terms of the corresponding creation and annihilation operators: 
\begin{equation}
\widehat{\mu}_{\vec{p} \, \alpha} = \frac{1}{\sqrt{2 p}} \biggl[ \widehat{a}_{\vec{p},\,\alpha} + \widehat{a}^{\dagger}_{-\vec{p},\,\alpha} \biggr], \qquad \widehat{\pi}_{\vec{p} \,\alpha} = - i\,\sqrt{\frac{p}{2}} 
\biggl[ \widehat{a}_{\vec{p},\,\alpha} - \widehat{a}^{\dagger}_{-\vec{p},\,\alpha} \biggr].
\label{DNT1}
\end{equation}
Since $[\widehat{a}_{\vec{k},\, \alpha}, \, \widehat{a}_{\vec{p},\, \beta}^{\dagger}] = \delta_{\alpha\beta}\, \delta^{(3)}(\vec{k}- \vec{p})$ we also have that $\widehat{\mu}_{\vec{k},\,\alpha}$ and $\widehat{\pi}_{\vec{k},\,\alpha}$ obey 
the standard commutation relations $[\hat{\mu}_{\vec{k},\, \alpha}(\eta), \, \hat{\pi}_{\vec{p},\, \beta}(\eta)] = i \, \delta_{\alpha\beta}\,\delta^{(3)}(\vec{k} + \vec{p})$.  After the classical fields are replaced by the corresponding operators  in Eq. (\ref{DDD8}), the Hamiltonian operator becomes: 
\begin{eqnarray}
\widehat{H}_{g}(\eta) &=& \frac{1}{2} \int d^{3} p \sum_{\alpha=\oplus,\otimes} \biggl\{ \widehat{\pi}_{-\vec{p},\,\alpha}\, \widehat{\pi}_{\vec{p},\,\alpha} + p^2 \widehat{\mu}_{-\vec{p},\,\alpha}\, \widehat{\mu}_{\vec{p},\,\alpha} 
+ {\mathcal F} \biggl[ \widehat{\pi}_{-\vec{p},\,\alpha}\, \widehat{\mu}_{\vec{p},\,\alpha} + 
\widehat{\mu}_{-\vec{p},\,\alpha}\, \widehat{\pi}_{\vec{p},\,\alpha} \biggr] \biggr\},
\label{DNT2}
\end{eqnarray}
where the sum runs, as usual, over the two polarizations of the gravitons.
Equation (\ref{DNT2}) is written directly in the $\eta$-time parametrization but the results reported 
in this section obviously hold also for the more specific situation where the $\eta$-time coincides with the conformal time coordinate $\tau$. For instance the Hamiltonian operator in the conformal time parametrization can be formally obtained 
from Eq. (\ref{DNT2}) by replacing ${\mathcal F}$ with ${\mathcal H}$ and by changing 
the derivations with respect to the $\eta$-time into derivatives with respect to $\tau$ in the 
definitions of the corresponding canonical momenta.  The presence of both  $\widehat{a}_{\vec{p},\,\alpha}$ and $\widehat{a}_{-\vec{p},\,\alpha}^{\dagger}$ in  Eq. (\ref{DNT1}) implies that the gravitons are produced in pairs of opposite three-momenta from a state where the total three-momentum vanishes; indeed inserting 
Eq. (\ref{DNT1}) into Eq. (\ref{DNT2}) we have
\begin{eqnarray}
\widehat{H}_{g}(\eta) &=& \frac{1}{2} \int d^{3} p \sum_{\alpha=\oplus,\, \otimes} \biggl\{ p \,\,\biggl[ \widehat{a}^{\dagger}_{\vec{p},\,\alpha} \widehat{a}_{\vec{p},\,\alpha} 
+ \widehat{a}_{- \vec{p},\,\alpha} \widehat{a}^{\dagger}_{-\vec{p},\,\alpha} \biggr] 
+ \lambda \,\,\widehat{a}^{\dagger}_{-\vec{p},\,\alpha} \widehat{a}_{\vec{p},\,\alpha}^{\dagger} 
+ \lambda^{\ast} \,\,\widehat{a}_{\vec{p},\,\alpha} \widehat{a}_{-\vec{p},\,\alpha} \biggr\},
\label{DNT3}
\end{eqnarray}
where we introduced the convenient notation $\lambda = i {\mathcal F}$.  In Eq. (\ref{DNT3}) we could insert a coherent component that would be linear in the creation and annihilation operators. This further component (neglected in the forthcoming discussion and eventually corresponding to an external force) could follow from a coupling proportional to $\mu_{i\, j} \, \Pi^{i\,j}$ (where $\Pi^{i\,j}$ denotes the anisotropic stress).  Equation (\ref{DNT3}) describes the production of pairs
of gravitons with opposite three-momenta and the  three terms quadratic in the creation and annihilation operators correspond in fact, up to a multiplicative constant, to the generators of the $SU(1,1)$ group defined as 
It can be easily checked that the same commutation relations appearing in Eq. (\ref{APPB10})
are verified in the continuous mode case:
\begin{eqnarray}
L_{+} &=& \frac{1}{2}\sum_{\beta} \int d^{3} k \,\, \widehat{a}^{\dagger}_{- \vec{k},\, \beta}\,\,  \widehat{a}^{\dagger}_{ \vec{k},\, \beta}, \qquad\qquad L_{-} = \frac{1}{2}\sum_{\beta} \int d^{3} k \,\, \widehat{a}_{\vec{k},\, \beta}\,\,  \widehat{a}_{- \vec{k},\, \beta},
\nonumber\\
L_{0} &=& \frac{1}{4}\sum_{\beta} \int d^{3} k \,\,\biggl[  \widehat{a}^{\dagger}_{\vec{k},\, \beta}\,\,  \widehat{a}_{ \vec{k},\, \beta} +  \widehat{a}_{- \vec{k},\, \beta}\,\,  \widehat{a}^{\dagger}_{- \vec{k},\, \beta}\biggr].
\label{DNT4}
\end{eqnarray}
It can be readily verified that $[L_{+},\, L_{-}] = - 2 L_{0}$ and that 
$[ L_{0},\, L_{\pm}] = \, \pm\, L_{\pm}$. The operators $L_{0}$ and $L_{\pm}$ are all 
dimensionless since the creation and annihilation operators defined for a continuum of modes are all 
dimensional as it can be argued from their commutator $[\widehat{a}_{\vec{k},\, \alpha}, \, \widehat{a}_{\vec{p},\, \beta}^{\dagger}] = \delta_{\alpha\beta}\, \delta^{(3)}(\vec{k}- \vec{p})$. We should also observe that the presence of the prefactors 
in Eq. (\ref{DNT4}) is required to avoid the double counting that might arise in the continuum case.
Bearing in mind this remark we shall often switch from the continuous mode description 
to the single-mode approximation where few modes of the field will be separately discussed. This approach is often employed in quantum optics and, in doing so, we shall implicitly assume a 
finite volume where the creation and annihilation operators are dimensionless (see, in this respect, appendix \ref{APPB}).
The evolution equations for $\widehat{a}_{\vec{p}}$ and $\widehat{a}_{-\vec{p},\,\alpha}^{\dagger}$ in the Heisenberg description follow from the Hamiltonian 
(\ref{DNT3}) and they are: 
\begin{equation}
\partial_{\eta} \widehat{a}_{\vec{p},\,\alpha}=  - i\, p \, \widehat{a}_{\vec{p},\,\alpha} -  i \, \lambda \widehat{a}_{- \vec{p},\,\alpha}^{\dagger},\qquad\qquad
\partial_{\eta}\widehat{a}_{-\vec{p},\,\alpha}^{\dagger} =  i\, p \, \widehat{a}_{-\vec{p},\,\alpha}^{\dagger} +  i \, \lambda^{\ast} \widehat{a}_{\vec{p},\,\alpha},
\label{DNT5}
\end{equation}
where $\partial_{\eta}$ indicates a derivation with respect to the $\eta$-time coordinate.
The solution of Eq. (\ref{DNT5}) is usually written \cite{mollow1,mollow2} in terms of two (complex) functions $u_{p,\,\alpha}(\eta)$ and $v_{p,\,\alpha}(\eta)$:
 \begin{eqnarray}
\widehat{a}_{\vec{p},\,\alpha}(\eta) &=& u_{p,\,\alpha}(\eta)\,\, \widehat{b}_{\vec{p},\,\alpha}-  
v_{p,\,\alpha}(\eta)\,\, \widehat{b}_{-\vec{p},\,\alpha}^{\dagger},  
\label{DNT6}\\
\widehat{a}_{-\vec{p},\,\alpha}^{\dagger}(\eta) &=& u_{p,\,\alpha}^{\ast}(\eta) \,\,\widehat{b}_{-\vec{p},\,\alpha}^{\dagger}  -  v_{p,\,\alpha}^{\ast}(\eta)\,\, \widehat{b}_{\vec{p},\,\alpha}.
\label{DNT7}
\end{eqnarray}
Barring for the polarization dependence, Eqs. (\ref{DNT6})--(\ref{DNT7}) effectively 
coincide with Eqs. (\ref{FFF1})--(\ref{FFF2}) and if the parametrization of Eqs. (\ref{DNT6})--(\ref{DNT7}) is inserted into Eq. (\ref{DNT5}) the evolution equations for $u_{p,\,\alpha}(\eta)$ and $v_{p,\,\alpha}(\eta)$ are obtained:
\begin{equation}
\partial_{\eta} \,u_{p,\,\alpha} = - i p\, u_{p,\,\alpha}  + i \lambda\,\,  v_{p,\,\alpha}^{\ast}, 
\qquad \partial_{\eta} v_{p,\,\alpha}= - i p\, v_{p,\,\alpha} + i \lambda \,\,u_{p,\,\alpha}^{\ast}.
\label{DNT8}
\end{equation}
As repeatedly mentioned in section \ref{sec2}, the functions $u_{p,\,\alpha}(\eta)$ and $v_{p,\,\,\alpha}(\eta)$ 
are subjected to the unitarity condition $|u_{p,\,\alpha}(\eta)|^2 - |v_{p\,\,\alpha}(\eta)|^2 =1$ 
which is fully analog to Eq.  (\ref{FFF3}). Since  $u_{p,\,\alpha}(\eta)$ and $v_{p,\,\alpha}(\eta)$ are two complex functions subjected to the condition $|u_{p,\,\alpha}(\eta)|^2 - |v_{p\,\,\alpha}(\eta)|^2 =1$ they can be parametrized in terms of one amplitude and 
two phases as $u_{k,\,\alpha}(\eta) = e^{- i\, \delta_{k,\,\alpha}} \cosh{r_{k,\,\alpha}}$ 
and  as $v_{k,\,\alpha}(\eta) =  e^{i (\theta_{k,\,\alpha} + \delta_{k,\,\alpha})} \sinh{r_{k,\,\alpha}}$ (see, in this respect, Eq. (\ref{APPC1}) and discussion thereafter).  If we now  insert 
Eqs. (\ref{DNT6})--(\ref{DNT7}) into  Eq. (\ref{DNT1}) we obtain the full form of the quantum fields 
and of the conjugate momenta: 
\begin{eqnarray}
\widehat{\mu}_{i\,j}(\vec{x},\eta) &=& \frac{\sqrt{2} \ell_{P}}{(2\pi)^{3/2}} \, \sum_{\alpha} \int d^{3} \, k \, e^{(\alpha)}_{i\,j}(\hat{k}) \biggl[ f_{k,\, \alpha}(\eta) \, \widehat{b}_{\vec{k}, \, \alpha} \, e^{- i \vec{k} \cdot\vec{x}} + f_{k,\, \alpha}^{\ast}(\eta) \, \widehat{b}_{\vec{k}, \, \alpha}^{\,\dagger} \, e^{ i \vec{k} \cdot\vec{x}} \biggr],
\label{DNT9}\\
\widehat{\pi}_{m\,n}(\vec{x},\eta) &=& \frac{1}{4 \, \sqrt{2} \ell_{P}\, (2\pi)^{3/2}} \, \sum_{\beta} \int d^{3} \, k \, e^{(\beta)}_{m\,n}(\hat{k}) \biggl[ g_{k,\, \beta}(\eta) \, \widehat{b}_{\vec{k}, \, \beta} \, e^{- i \vec{k} \cdot\vec{x}} + g_{k,\, \beta}^{\ast}(\eta) \, \widehat{b}_{\vec{k}, \, \beta}^{\,\dagger} \, e^{ i \vec{k} \cdot\vec{x}} \biggr],
\label{DNT10}
\end{eqnarray}
where $f_{k,\,\alpha}(\eta)$ and $g_{k\, \beta}(\eta)$ are the two (complex) mode functions. From 
Eqs. (\ref{DNT9})--(\ref{DNT10}) we can define $\widehat{\mu}_{i\, j}(\vec{q}, \eta)$ and $\widehat{\pi}_{m\, n}(\vec{q}, \eta)$\begin{equation}
\widehat{\mu}_{i\, j}(\vec{q}, \eta) = \int \frac{d^{3} x}{(2\pi)^{3/2}} \, \widehat{\mu}_{i\, j}(\vec{x}, \eta) e^{i \vec{q}\cdot\vec{x}}, \qquad \widehat{\pi}_{m\, n}(\vec{p}, \eta) = \int \frac{d^{3} x}{(2\pi)^{3/2}} \, \widehat{\pi}_{i\, j}(\vec{x}, \eta) e^{i \vec{p}\cdot\vec{x}}.
\label{DNT11}
\end{equation} 
The explicit forms of $\widehat{\mu}_{i\, j}(\vec{q}, \eta)$ and $\widehat{\pi}_{m\, n}(\vec{p}, \eta)$ are:
\begin{eqnarray}
&&\widehat{\mu}_{i\, j}(\vec{q}, \eta) = \sqrt{2}\, \ell_{P} \sum_{\alpha} \biggl[ f_{q,\, \alpha}(\eta) \, \widehat{b}_{\vec{q},\, \alpha} e^{(\alpha)}_{i\,j}(\hat{q}) + f_{q,\, \alpha}^{\ast}(\eta) \, \widehat{b}_{-\vec{q},\, \alpha}^{\,\dagger} e^{(\alpha)}_{i\,j}(-\hat{q}) \biggr],
\label{DNT12}\\
&&\widehat{\pi}_{m\, n}(\vec{p}, \eta) = \frac{1}{4\, \sqrt{2}\, \ell_{P} }\sum_{\beta} \biggl[ g_{p,\, \beta}(\eta) \, \widehat{b}_{\vec{p},\, \beta} e^{(\beta)}_{m\,n}(\hat{p}) + g_{p,\, \beta}^{\ast}(\eta) \, \widehat{b}_{-\vec{p},\, \beta}^{\,\dagger} e^{(\beta)}_{i\,j}(-\hat{p}) \biggr].
\label{DNT13}
\end{eqnarray}
The commutator of the field operators and of the conjugate momenta is then given by:
\begin{equation}
[\widehat{\mu}_{i\,j}(\vec{q},\eta), \, \widehat{\pi}_{m\,n}(\vec{p},\eta)] = i \, {\mathcal S}_{i\, j\,m\, n}(\widehat{q}) \, \delta^{(3)}(\vec{q} + \vec{p}),
\label{DNT14}
\end{equation}
where ${\mathcal S}_{i\, j\,m\, n}(\widehat{q})$ accounts for the sum over the polarizations 
\begin{equation}
{\mathcal S}_{i\, j\,m\, n}(\widehat{q}) = \frac{1}{4} \bigg[p_{i\,m}(\hat{q})\, p_{j\,n}(\hat{p}) + 
p_{i\,n}(\hat{q})\, p_{j\,m}(\hat{p}) - p_{i\,j}(\hat{q}) p_{m\,n}(\hat{q})\biggr], 
\label{DNT15}
\end{equation}
and $p_{i\,j}(\hat{q})= (\delta_{i\,j} - \widehat{q}_{i} \, \widehat{q}_{j})$ is the standard solenoidal 
projector. It is important to remark that the canonical form of the commutation relations of Eq. (\ref{DNT14}) 
holds since the mode functions $f_{q,\,\alpha}(\eta)$ and $g_{q,\,\alpha}(\eta)$
obey the standard Wronskian normalization condition
\begin{equation}
f_{q,\,\alpha}(\eta) g_{q,\,\alpha}^{\ast}(\eta) - f_{q,\,\alpha}^{\ast}(\eta) \, g_{q,\,\alpha}(\eta) = i,
\label{DNT16}
\end{equation}
which is, in turn,  a direct consequence of the requirement that $|u_{q,\,\alpha}(\eta)|^2 - |v_{q\,\,\alpha}(\eta)|^2 =1$; 
indeed $u_{q,\,\alpha}(\eta)$ and $v_{q\,\,\alpha}(\eta)$ are directly related to $f_{q,\,\alpha}(\eta)$ and $g_{q,\,\alpha}(\eta)$ as:
\begin{equation}
u_{q,\,\alpha}(\eta) = \frac{q \, f_{q,\, \alpha}(\eta)\, + \, i g_{q,\,\alpha}(\eta)}{\sqrt{2 q}}, \qquad\qquad
v_{q,\,\alpha}^{\,\ast}(\eta) = - \frac{q \, f_{q,\, \alpha}(\eta)\, - \, i g_{q,\,\alpha}(\eta)}{\sqrt{2 q}}.
\label{DNT17}
\end{equation}
Equations (\ref{DNT15})--(\ref{DNT16}) and (\ref{DNT17}) demonstrate that the unitarity condition $|u_{p,\,\alpha}(\eta)|^2 - |v_{p\,\,\alpha}(\eta)|^2 =1$ guarantees that the commutation relation between the field operators and the conjugate moments are preserved throughout the dynamical evolution. 

Let us finally consider the conventional situation of the Ford-Parker action \cite{AA3} where the effective theory
does not contain parity-violating terms which could be however present in more general situations \cite{REC3,REC3a}; in this case we may replace the $\eta$-time with the conformal time and consider the two polarizations as independent. The averaged multiplicity $\overline{n}(k,\tau)$ accounting for the pairs of gravitons with opposite three-momenta for each tensor polarization follows then from Eqs. (\ref{DNT6})--(\ref{DNT7}) 
\begin{equation}
\langle \hat{N}_{k} \rangle = \langle \widehat{a}_{\vec{k}}^{\dagger} \widehat{a}_{\vec{k}} + \widehat{a}_{-\vec{k}}^{\dagger} \widehat{a}_{-\vec{k}} \rangle = 2 \, \overline{n}(k,\tau), \qquad\qquad  \overline{n}(k,\tau) =  |v_{k}(\tau)|^2.
\label{DNT18}
\end{equation}
The largeness of the averaged multiplicity is sometimes used to argue that the final state of the evolution 
of the relic gravitons is classical and the argument is, in short, the following. Since $|v_{k}(\tau)|^2\gg 1$ 
we also have that $|u_{k}(\tau)|^2 \gg 1$ and this means that the field operators effectively commute:
\begin{equation}
|u_{k}(\tau)|^2 \simeq |v_{k}(\tau)|^2, \qquad\qquad f_{k}(\tau) \, g_{k}^{\ast}(\tau) \simeq g_{k}(\tau) f_{k}^{\ast}(\tau) ,\qquad\qquad [\widehat{\mu}_{i\,j}(\vec{k},\tau), \, \widehat{\pi}_{m\,n}(\vec{p},\tau)] \simeq 0.
\label{DNT19}
\end{equation}
The heuristic argument of Eq. (\ref{DNT19}) is however self-contradictory since it suggests 
that unitarity is approximately lost every time a large number of pairs is produced. This is not the way 
we should look at the quantum theory of parametric amplification. On the contrary, as we are going to see in the 
next subsection, when the approximation scheme is accurate the violations of unitarity are neither 
explicit nor implicit even if the averaged multiplicity of the produced pairs is very large. 

\subsection{Approximate forms of the averaged multiplicities}
In the standard case where $\eta \to \tau$ and ${\mathcal F} \to {\mathcal H}$ the estimate of the averaged multiplicity is actually based on the approximate solutions of Eq. (\ref{DNT8}) . The strategy is to  avoid the real parts 
of  $u_{p,\,\alpha}$ and $v_{p,\,\alpha}$ up to the very end of the calculation since the evolution is simpler in terms of the (complex) linear combinations $(u_{p,\,\alpha} -v^{\ast}_{p,\,\alpha})$ and $(u_{p,\,\alpha} + v^{\ast}_{p,\,\alpha})$ obeying, respectively, the following pair of equations: 
\begin{eqnarray}
&& (u_{p,\,\alpha} -v^{\ast}_{p,\,\alpha})^{\prime\prime} + \biggl[ k^2 - \frac{a^{\prime\prime}}{a}\biggr] (u_{p,\,\alpha} -v^{\ast}_{p,\,\alpha}) =0, 
\label{DNT20}\\
&&  (u_{p,\,\alpha} +v^{\ast}_{p,\,\alpha})^{\prime\prime} + \biggl[ k^2 - a \biggl(\frac{1}{a}\biggr)^{\prime\prime}\biggr] (u_{p,\,\alpha} +v^{\ast}_{p,\,\alpha}) =0.
\label{DNT21}
\end{eqnarray}
Equations (\ref{DNT20})--(\ref{DNT21}) correspond to the evolution of the mode 
functions of the field and can be solved with the WKB approximation \cite{REC4} (see also Ref. \cite{AA13}).
Since both polarizations obey the same equation the index $\alpha$ can be dropped
and the solutions of Eqs. (\ref{DNT20})--(\ref{DNT21}) become:
\begin{eqnarray}
u_{k}(\tau) =  \frac{1}{2} \bigl[A_{k}(\tau_{ex},\tau_{re}) + i\, B_{k}(\tau_{ex},\tau_{re})\bigr] e^{- i \, k (\Delta \tau + \tau_{ex})},
\label{DNT22}\\
v_{k}^{\ast}(\tau) = - \frac{1}{2}  \bigl[A_{k}(\tau_{ex},\tau_{re}) - i\, B_{k}(\tau_{ex},\tau_{re})\bigr] e^{i \, k (\Delta \tau + \tau_{ex})},
\label{DNT23}
\end{eqnarray}
where $\Delta \tau = (\tau- \tau_{re})$. In Eqs. (\ref{DNT22})--(\ref{DNT23})  $\tau_{ex}(k)$ and $\tau_{re}(k)$ 
coincide, respectively,  with the times where a given wavelength exited and reentered the Hubble radius.
The turning points are implicitly determined  from $ k^2 = a^2\, H^2 [ 2 - \epsilon(\tau)]$
where, as usual, $\epsilon = - \dot{H}/H^2$ denotes the slow-roll parameter. If $\epsilon \neq 2$ we have that $k \tau_{re} ={\mathcal O}(1)$ and $k \tau_{ex} = {\mathcal O}(1)$. If a given scale reenters during radiation we will have that $\epsilon\to 2$in the vicinity of the turning point and $k\tau_{re} \ll 1$. The explicit 
expressions of  $A_{k}(\tau_{ex},\tau_{re})$ and $B_{k}(\tau_{ex},\tau_{re})$ are:
\begin{eqnarray}
A_{k}(\tau_{ex},\tau_{re}) &=& \biggl(\frac{a_{re}}{a_{ex}}\biggr) \, Q_{k}(\tau_{ex}, \tau_{re}),
\label{DNT24}\\
B_{k}(\tau_{ex}, \tau_{re}) &=& \biggl(\frac{{\mathcal H}_{re}}{k} \biggr) \,  \biggl(\frac{a_{re}}{a_{ex}}\biggr) \, Q_{k}(\tau_{ex}, \tau_{re})
- \biggl(\frac{a_{ex}}{a_{re}}\biggr) \biggl(\frac{{\mathcal H}_{ex} + i k}{k}\biggr),
\label{DNT25}\\
Q_{k}(\tau_{ex}, \tau_{re}) &=& 1 - ({\mathcal H}_{ex} + i k) \int_{\tau_{ex}}^{\tau_{re}} \frac{a_{ex}^2}{a^2(\tau^{\prime})} \,\, d \tau^{\prime}.
\label{DNT26}
\end{eqnarray}
In Eqs. (\ref{DNT24})--(\ref{DNT26}), with obvious notations, ${\mathcal H}_{ex} = {\mathcal H}(\tau_{ex})$, 
and ${\mathcal H}_{re} = {\mathcal H}(\tau_{re})$. From the explicit expression of Eqs. (\ref{DNT22})--(\ref{DNT23}) and (\ref{DNT25})--(\ref{DNT26}) it necessarily follows that $|u_{k} (\tau) |^2 - |v_{k}(\tau)|^2 =1$. This also means 
that the explicit form of the mode functions:
\begin{eqnarray}
f_{k}(\tau) &=& \frac{e^{- i k \tau_{ex}}}{\sqrt{ 2 k}} \biggl[ A_{k}(\tau_{ex},\tau_{re}) \, \cos{k \,\Delta\tau} + B_{k}(\tau_{ex},\tau_{re}) \, \sin{k \,\Delta\tau} \biggr],
\label{DNT27}\\
g_{k}(\tau) &=& e^{- i k \tau_{ex}}\, \sqrt{\frac{k}{2}}\, \biggl[ - A_{k}(\tau_{ex},\tau_{re}) \, \sin{k \,\Delta\tau} + B_{k}(\tau_{ex},\tau_{re}) \, \cos{k \,\Delta\tau} \biggr],
\label{DNT28}
\end{eqnarray}
implies that $f_{k}(\tau) g_{k}^{\ast}(\tau) - f_{k}^{\ast}(\tau) \, g_{k}(\tau) = i$, as it can be easily verified with 
some lengthy but straightforward calculation. If the background expands between $a_{ex}$ and $a_{re}$ we have that all 
the terms containing the ratio $(a_{re}/a_{ex}) \gg 1$ superficially dominate 
against those proportional to $(a_{ex}/a_{re}) \ll 1$. If we use this logic in a simplified (and incorrect)
manner we would simply keep the dominant terms and completely neglect the subdominant ones.
This sketchy approach violates unitarity and a the correct strategy is instead to expand systematically $u_{k}(\tau)$ 
and $v_{k}(\tau)$ with the constraint that $|u_{k} (\tau) |^2 - |v_{k}(\tau)|^2 =1$. This analysis can be found in appendix 
\ref{APPC}. The idea is that, after some preliminary rearrangement of Eqs. (\ref{DNT24})--(\ref{DNT25}), the obtained expressions are expanded in terms of the small parameter of the problem, namely $a_{ex}/a_{re}\ll 1$ by requiring that, order by order in the expansion, the unitarity is always enforced. The result of this approach is 
\begin{eqnarray}
u_{k}(\tau) = \frac{e^{- i k\tau}}{2} \biggl[ i +  q_{ex}(1 - i) \,  {\mathcal I}(\tau_{ex}, \tau_{re}) \, \biggr] (q_{re} - \, i) \biggl(\frac{a_{re}}{a_{ex}}\biggr) + ( 1 - i \, q_{ex}) \biggl(\frac{a_{ex}}{a_{re}}\biggr)+ {\mathcal O}\biggl[ \biggl(\frac{a_{ex}}{a_{re}}\biggr)^{5}\biggr],
\nonumber\\
v_{k}(\tau) = \frac{e^{- k \tau}}{2} \biggl[ - 1 +  (q_{ex} - i){\mathcal I}(\tau_{ex}, \tau_{re}) \biggr] (q_{re} - \, i) \biggl(\frac{a_{re}}{a_{ex}}\biggr) + ( 1 + i \, q_{ex}) \biggl(\frac{a_{ex}}{a_{re}}\biggr)+ {\mathcal O}\biggl[ \biggl(\frac{a_{ex}}{a_{re}}\biggr)^{5}\biggr],
\label{APPC11}
\end{eqnarray}
where $q_{ex} = a_{ex} \, H_{ex}/k$ and $q_{re} = a_{re} \, H_{re}/k$. We can immediately verify that, to the given order in the perturbative expansion (i.e. $(a_{re}/a_{ex})^{5}$), Eq. (\ref{APPC11}) implies that $|u_{k}(\tau)|^2 - |v_{k}(\tau)|^2=1$ so that unitarity is not lost while keeping the leading terms of the expansion. From Eq. (\ref{APPC11}) it also follows that 
\begin{eqnarray}
|u_{k}(\tau)|^2 + |v_{k}(\tau)|^2 &=& (1 + q_{ex}^2) \, q_{re}\, {\mathcal I}(\tau_{ex}, \tau_{re}) - q_{ex} q_{re} 
\nonumber\\
&+& \frac{(1+ q_{re}^2)}{2}
\biggl\{1 + {\mathcal I}(\tau_{ex}, \tau_{re})\biggl[{\mathcal I}(\tau_{ex}, \tau_{re}) ( 1 + q_{ex}^2) - 2 q_{ex}\biggr] \biggl(\frac{a_{re}}{a_{ex}}\biggr)^2  \biggr\}
\nonumber\\
&+& \frac{( 1 + q_{ex}^2)}{2} \biggl(\frac{a_{ex}}{a_{re}}\biggr)^2 .
\label{APPC11aa}
\end{eqnarray}
In Eq. (\ref{APPC11aa}) the first contribution is, in practice, ${\mathcal O}(1)$ unless the reentry takes place during 
radiation in which case $q_{re}\gg 1$; the second contribution in the same equation is the dominant one 
since $a_{re} \gg a_{ex}$. Finally the third term in Eq. (\ref{APPC11aa}) is always subleading. All the three 
contributions are however essential to the given order in the perturbative expansion  
since they all contribute to the unitarity of the evolution. Thanks to Eq. (\ref{APPC11aa}) we may 
further simplify the obtained expressions and estimate Eq. (\ref{APPC11}) s:
\begin{equation}
u_{k}(\tau_{ex},\tau_{re}) \,\simeq\,  \frac{1}{2}\biggl[ \biggl(\frac{a_{re}}{a_{ex}}\biggr) + \biggl(\frac{a_{ex}}{a_{re}}\biggr)\biggl], \qquad 
v_{k}(\tau_{ex},\tau_{re}) \,\simeq\, \frac{1}{2}\biggl[ \biggl(\frac{a_{ex}}{a_{re}}\biggr) - \biggl(\frac{a_{re}}{a_{ex}}\biggr)\biggl],
\label{APPC11a}
\end{equation}
where, again, $|u_{k}(\tau_{ex},\tau_{re})|^2 - |v_{k}(\tau_{ex},\tau_{re})|^2 =1$.
It is true that, according to Eqs. (\ref{APPC11})--(\ref{APPC11a}), $u_{k}$ and $v_{k}$ are of the same 
order in the limit $a_{re}/a_{ex} \gg 1$. It is however incorrect to conclude that unitarity is lost when the average multiplicity is large. All in all, if unitarity  turns out to be violated at the end of a calculation that assumes unitarity from the beginning 
this simply means that the approximation scheme is inaccurate. The averaged multiplicities may certainly be estimated by means of approximations that are only partially accurate but with the proviso that the inaccuracy of the approximation scheme should not be superficially interpreted as a loss of unitarity. Alternatively it is always possible to use directly Eqs. (\ref{DNT22})--(\ref{DNT23}) even though, for all practical purpose, their accuracy is comparable to the one of Eq. (\ref{APPC11}). 
 
As repeatedly mentioned, in the continuous mode case the commutation relations of the creation and annihilation 
operators are obviously given by $[ \widehat{a}_{\vec{k},\,\alpha}, \, \widehat{a}_{\vec{p},\, \beta}] = \delta_{\alpha\beta}\,\,\delta^{(3)}(\vec{k} - \vec{p})$.  The unitary transformation 
corresponding to Eq. (\ref{APPB13}) in the continuous mode case is given by
\begin{equation}
\widehat{a}_{\vec{k},\, \alpha} = {\mathcal R}^{\dagger}(\delta) \, \Sigma^{\dagger}(z) \widehat{b}_{\vec{k}, \alpha} \Sigma(z) \, {\mathcal R}(\delta),
\label{APPB15}
\end{equation}
where ${\mathcal R}(\delta)$ and $ \Sigma(z)$ are the continuous mode generalization 
of the rotation and of the squeezing operators:
\begin{equation}
{\mathcal R}(\delta) = e^{- i \, n(\delta)/2}, \qquad \Sigma(z) = \exp{[\sigma(z)/2]},
\label{APPB16}
\end{equation}
where $\sigma(z)$ and $n(\delta)$ involve both an integral over the modes and a sum over the two tensor 
polarizations:
\begin{eqnarray}
\sigma(z) &=& \sum_{\lambda} \int d^{3} k \,\, \biggl[ z_{k,\,\lambda}^{\ast} \, \widehat{b}_{\vec{k},\,\lambda} 
\widehat{b}_{-\vec{k},\,\lambda} - z_{k,\,\lambda} \, \widehat{b}^{\dagger}_{-\vec{k},\,\lambda} 
\widehat{b}^{\dagger}_{\vec{k},\,\lambda} \biggr],
\nonumber\\
n(\delta) &=& \sum_{\lambda}  \int d^{3}k\, \delta_{k,\,\lambda}\,\biggl[ \widehat{b}_{\vec{k},\,\lambda}^{\dagger} 
 \widehat{b}_{\vec{k},\,\lambda} + \widehat{b}_{-\vec{k},\,\lambda}  \widehat{b}_{-\vec{k},\,\lambda}^{\dagger} \biggr].
 \label{APPB17}
 \end{eqnarray}
Consequently the explicit form of the unitary transformation of Eq. (\ref{APPB15}) is given by
\begin{equation}
\widehat{a}_{\vec{k},\, \alpha} = e^{- i\, \delta_{k,\,\alpha}} \cosh{r_{k,\,\alpha}} \, \widehat{b}_{\vec{k},\,\alpha} -  e^{i (\theta_{k,\,\alpha} + \delta_{k,\,\alpha})} \sinh{r_{k,\,\alpha}} \, \widehat{b}^{\dagger}_{-\vec{k},\,\alpha}.
\label{APPB18}
\end{equation}
If we want to avoid the complication of an integral over the different modes it is often practical to focus the attention 
on one or two modes of the field, as discussed in appendix \ref{APPB}. In the two-mode approximation 
the final state of the particle production process is therefore 
given by 
\begin{equation}
| z \rangle = \Sigma(z) | 0_{+} \, 0_{-} \rangle, \qquad \Sigma(z) = e^{z^{\ast} \, \widehat{b}_{+} \, \widehat{b}_{-} - 
z\widehat{b}_{+}^{\dagger} \, \widehat{b}_{-}^{\dagger}},
\label{SQ1}
\end{equation}
where $[\widehat{b}_{I}, \widehat{b}_{J}^{\dagger}] = \delta_{I,\,J}$ where $I, \, J = +,\,-$; this notation 
should not be confused  with the tensor polarizations since, in this context, the $\pm$ are related to the 
sign of a (single) comoving three-momentum. Equation (\ref{SQ1}) may also contain a further 
contribution from the rotation operator (i.e. the two-mode analog of ${\mathcal R}(\delta)$ appearing in Eq. (\ref{APPB16})). We finally note that in Eq. (\ref{SQ1}) $z = r\, e^{i\theta}$. If we know 
match the approximations leading to Eqs. (\ref{APPC11})--(\ref{APPC11a}) with 
the explicit form of the squeezing operator of Eq. (\ref{SQ1}) we would have, using the results of 
appendix \ref{APPC} that 
\begin{eqnarray}
\widehat{a}_{+} &=& \Sigma^{\dagger}(z) \, \widehat{b}_{+} \Sigma(z) = \cosh{r} \,\, \widehat{b}_{+} - e^{i \, \theta} \sinh{r} \,\, \widehat{b}_{-}^{\dagger},
\label{SQ2}\\
\widehat{a}_{-} &=& \Sigma^{\dagger}(z) \, \widehat{b}_{-} \Sigma(z) = \cosh{r} \,\, \widehat{b}_{-} - e^{i \, \theta} \sinh{r} \,\, \widehat{b}_{+}^{\dagger},
\label{SQ3}
\end{eqnarray}
with $ r = \ln{(a_{re}/a_{ex})}$. From Eqs. (\ref{SQ2})--(\ref{SQ3}) we clearly have 
that the averaged multiplicity of the produced pairs is given by $ \overline{n} = \sinh^2{r}$ 
that ultimately coincides with the result of Eqs. (\ref{APPC11})--(\ref{APPC11a}). 

\renewcommand{\theequation}{5.\arabic{equation}}
\setcounter{equation}{0}
\section{The entropy of the relic gravitons and the optical equivalence}
\label{sec5}
 When the high-frequency gravitons are produced either 
from the vacuum or from some other initial state unitarity is never lost even if the averaged multiplicity of the pairs may get very large. Consequently, if the gravitons are produced from a pure initial state (like the vacuum) the corresponding density matrix remains idempotent and then no entropy is associated to the multiparticle final state.  Formally this observation 
follows from the unitarity of the operators given in Eqs. (\ref{SQ1})--(\ref{SQ2}) as well as from the Hermiticity of the 
corresponding Hamiltonian. The quantum theory of parametric amplification however provides a natural cosmological arrow associated with an entanglement entropy \cite{ST8,ST9,ST9a,ST9b,ST10,ST11}. In our context the growth of the averaged multiplicity of the produced gravitons is naturally associated with the increase of the entanglement entropy of the gravitational field. The entropy will now be estimated when the density matrix is reduced because of the incomplete 
measurement of the final state. In an idealized experiment based, for instance, on the 
analysis of the gravitational analog of the Hanbury Brown-Twiss correlations \cite{HBT1}
only one of the two gravitons of the pair is typically detected. In this situation the resulting 
entanglement entropy is proportional to the logarithm of the averaged multiplicity.
Since the final multiplicity is always large,  the entropy is effectively proportional to the squeezing parameter $r$ \cite{ST10,ST11}. The reduction of the density matrix in different bases leads in fact to the same von Neumann entropy whose integral over all the modes of the spectrum is dominated by the maximal frequency. It is interesting that from the THz bound deduced in section \ref{sec2} the total integrated entropy of the relic gravitons always undershoots the cosmic microwave background entropy.

\subsection{Pure final state}
When a portion of the system is unobservable (or when observations are confined to a subset of the 
physical degrees of freedom) information is lost and the total density matrix can then be 
reduced. The total Hilbert space of our problem contains different subspaces and the operators 
associated with the opposite momenta of a graviton pair effectively act on separated subspaces. 
For this reason we can focus on a single pair of gravitons so that the associated operators will be $\widehat{b}_{+}$ and $\widehat{b}_{-}$. In a quantum optical context $\widehat{b}_{+}$ and $\widehat{b}_{-}$ may represent the signal and the idler mode \cite{MANDL,HBT4}. Let us then focus on the two-mode operator of Eq. (\ref{SQ1}) and recall, for the sake of concreteness, that $r = \ln{(a_{ex}/a_{re})}$. Let us note, preliminarily, that $\Sigma(z)$ can be factorized as the product of the exponentials of the group generators as:
 \begin{equation}
\Sigma(z) = \exp{\biggl[ -\frac{z}{|z|} \tanh{|z|} \,\,\, L_{+}\biggr]}\times \exp{[- 2 \ln{\cosh|z|}\,\,\, L_{0}]}\times \exp{\biggl[ \frac{z^{*}}{|z|} \, \tanh{|z|} \,\,\,L_{-}\biggr]},
\label{ENTR1}
\end{equation}
see also Eq. (\ref{APPB12}) and discussion therein. In Eq. (\ref{ENTR1}) the operators $L_{0}$ and $L_{\pm}$ have been introduced:
\begin{equation}
L_{+} = \widehat{b}_{+}^{\,\dagger}\,\widehat{b}_{-}^{\,\dagger}, \qquad\qquad L_{-} = \widehat{b}_{+}\,\widehat{b}_{-},
\qquad\qquad L_{0} = \frac{1}{2} ( \widehat{b}_{+}^{\,\dagger} \,\widehat{b}_{+} + \widehat{b}_{-}\,\widehat{b}_{-}^{\,\dagger}),
\label{ENTR2}
\end{equation}
and they obey the commutation relations of the $SU(1,1)$ Lie algebra discussed in appendix \ref{APPB}, namely  $[L_{+}, \, L_{-}] = - 2 \, L_{0}$ and $[L_{0}, \, L_{\pm}] = \pm \, L_{\pm}$. In the two-mode approximation of  Eqs. (\ref{ENTR1})--(\ref{ENTR2}) the operator $\widehat{b}_{+}^{\, \dagger}$ creates a graviton of momentum $+\vec{k}$ while $\widehat{b}_{-}^{\,\dagger}$ creates a graviton with momentum $- \vec{k}$; the two-mode Fock state is, by definition,
\begin{equation}
| n_{+}\,\, n_{-} \rangle = \frac{(\widehat{b}_{+}^{\,\dagger})^{n_{+}}}{\sqrt{n_{+}\, !}}\,\, \frac{(\widehat{b}_{-}^{\,\dagger})^{n_{-}}}{\sqrt{n_{+}\, !}}\,\, |0_{+}\,\, 0_{-}\rangle. 
\label{ENTR3}
\end{equation}
When the relic gravitons are produced in pairs of opposite three-momenta we have that $n_{+} = n_{-}$
and from the  explicit expressions of the group generators (see Eq. (\ref{ENTR2})) we have that their action 
on the two-mode vacuum is given by $L_{-} \, | 0_{+} \, 0_{-}\rangle =0$  while for $L_{0}$ we have instead $L_{0} \, | 0_{+} \, 0_{-}\rangle = | 0_{+} \, 0_{-}\rangle/2$. The entanglement entropy of the final state does not change if further phases appear in the final state. This can happen if a rotation operator acts on the two-mode vacuum. In what follows this possibility will be neglected 
but can be treated within the same approach. In the two-mode approximation adopted here, the final state of particle production 
is characterized by the total density matrix 
\begin{equation}
\widehat{\rho} = | \,z \,\rangle \langle \,z\,|= \Sigma(z) \,\,| 0_{+} \, 0_{-}\rangle \,\,\langle 0_{-}\, 
 0_{+}| \,\,\Sigma^{\dagger}(z).
 \label{ENTR3a}
 \end{equation}
 It follows from Eq. (\ref{ENTR3a}) that $\widehat{\rho}^2 = \widehat{\rho}$. Furthermore, since 
 the states are correctly normalized we also have that $\mathrm{Tr} \widehat{\rho}^2 = \mathrm{Tr} \widehat{\rho} =1$.
 
 \subsection{Operators acting on a portion of the total Hilbert space}
 Since $\widehat{b}_{+}$ and $\widehat{b}_{-}$ may represent the signal and the idler mode, we 
 shall keep this terminology just for the sake of conciseness and bear in mind that, ultimately, these 
 two modes do not simply arise as ingredients of a quantum measurement but represent in fact gravitons with opposite three-momenta. Let us then consider some 
 Hermitian observable acting on one of the two Hilbert subspaces; an interesting 
 observable in $\widehat{N}_{+} = \widehat{b}_{+}^{\dagger} \widehat{b}_{+}$. Another 
 Hermitian observable is $\widehat{I}_{+} =  \widehat{b}_{+}^{\dagger}\, \widehat{b}_{+}^{\dagger} \widehat{b}_{+} \widehat{b}_{+}$. If we eventually average these operators and take their 
 ratio we obtain $g_{+}^{(2)}= \langle \widehat{I}_{+} \rangle/ \langle \widehat{N}_{+} \rangle^2$;
 $g_{+}^{(2)}$ which is the degree of second-order coherence appearing in the analysis 
 of the Hanbury-Brown Twiss correlations. 

If we now focus, for simplicity, on $\widehat{N}_{+}$ we can first note that it basically 
operates on the Hilbert subspace of the signal by leaving untouched the idler subspace; in terms 
of Eq. (\ref{ENTR3}) $\widehat{N}_{+}| n_{+}\,\, n_{-} \rangle= n_{+} \, | n_{+}\,\, n_{-} \rangle$. Thanks to the Baker-Campbell-Hausdorff relation of Eq. (\ref{ENTR1}) we can therefore 
write the state $| z\rangle$ as 
\begin{equation}
| z \rangle = \frac{1}{\cosh{r}}  \sum_{n_{\pm}=0}^{\infty}\,\, \bigl(- e^{i \theta } \,\,\tanh{r}\bigr)^{(n_{+}+ n_{-})/2}\, \, \delta_{n_{+} \, n_{-}} |n_{-}\, n_{+} \rangle.
\label{ENTR4}
\end{equation}
This way of writing may seem a bit baroque since, on the one hand, we summed over $n_{+}$ and $n_{-}$ while, on the other hand, we included the $\delta_{n_{+}\, n_{-}}$ that effectively cancels one 
of the two sums and enforces the conservation of the three-momentum. The form of Eq. (\ref{ENTR4}) is however convenient from the algebraic viewpoint because, in this way, 
the action of the operators acting over the subspaces of the total Hilbert space is immediately clear.
Indeed, from Eq. (\ref{ENTR4}) we may compute $\langle z| \widehat{N}_{+}| z \rangle$; the result is
\begin{eqnarray}
\langle z| \widehat{N}_{+}| z \rangle &=& \sum_{n_{\pm}=0}^{\infty}\,\sum_{m_{\pm}=0}^{\infty}
n_{+} \,  \delta_{n_{+} \, n_{-}}\, \delta_{m_{+} \, m_{-}}\,  \delta_{n_{+} \, m_{+}}\, \delta_{n_{-} \, m_{-}} \frac{ \bigl(\tanh{r}\bigr)^{m_{+} + n_{+}}}{\cosh^2{r}} 
\nonumber\\
&=& \sum_{n=0}^{\infty} \, n\, p_{n} = \overline{n}, \qquad p_{n} = \frac{\overline{n}^{n}}{(\overline{n}+1)^{n +1}},
\label{ENTR4a}
\end{eqnarray}
where $\overline{n} = \sinh^2{r}$. The second equality follows directly from the first one 
and the probability distribution $p_{n}$ is of Bose-Einstein type. 
The same discussion of Eq. (\ref{ENTR4a}) can be generalized to any other class of Hermitian observable  $\widehat{I}_{+}$  acting on 
the signal space rather than on the idler space.  In the logic of the Hanbury Brown-Twiss (HBT) measurement this operator may coincide 
with the intensity and its expectation value can then be written as
\begin{equation}
\langle z\, | \widehat{I}_{+}| z \rangle = \sum_{n=0}^{\infty} p_{n} \,\langle \, n\, | \widehat{I}_{+} | n \rangle, \qquad\qquad p_{n}=  \frac{\tanh^{2\,n}{r}}{\cosh^2{r}} = \frac{\overline{n}^{n}}{(\overline{n} + 1)^{n +1}},
\label{ENTR5}
\end{equation}
where, as before, $\overline{n} = \sinh^2{r}$.
\subsection{Reduced density matrix and von Neumann entropy}
The results 
of Eqs. (\ref{ENTR4a})--(\ref{ENTR5}) ultimately imply that from the total density matrix $\widehat{\rho}$ 
we can derive a reduced density matrix by tracing over the idler mode. More specifically 
the explicit form of the total density matrix of Eq. (\ref{ENTR3a}) is 
\begin{equation}
\widehat{\rho} =  \sum_{n_{\pm}=0}^{\infty}\,\sum_{m_{\pm}=0}^{\infty} e^{i q \theta} \,\frac{ \bigl(\tanh{r}\bigr)^{(m_{+} + m_{-})/2 + (n_{+} + n_{-})/2 }}{\cosh^2{r}} \delta_{m_{+}\,m_{-}}\, \delta_{n_{+}\,n_{-}}
|m_{-}\, m_{+} \rangle \langle n_{+}\, n_{-}|,
\label{ENTR4b}
\end{equation}
where $q = (n_{+} - m_{+})/2 + (n_{-} - m_{-})/2$. By definition 
the density matrix of Eq. (\ref{ENTR4b}) describes a pure quantum state. From Eq. (\ref{ENTR4b}) we can however obtain the reduced density operator:
\begin{eqnarray}
\widehat{\rho}_{\mathrm{red}} &=& \mathrm{Tr}_{-} \bigl[ \widehat{\rho}\bigr] 
\nonumber\\
&=& \sum_{k_{-} =0}^{\infty}\sum_{n_{\pm}=0}^{\infty}\,\sum_{m_{\pm}=0}^{\infty} e^{i q \theta} \,\frac{ \bigl(\tanh{r}\bigr)^{(m_{+} + m_{-})/2 + (n_{+} + n_{-})/2 }}{\cosh^2{r}} \delta_{m_{+}\,m_{-}}\, \delta_{n_{+}\,n_{-}}
\langle k_{-} |m_{-}\, m_{+} \rangle \langle n_{+}\, n_{-}|  k_{-} \rangle,
\label{ENTR4c}
\end{eqnarray}
where by definition $\mathrm{Tr}_{-}[.\,.\,.]$ denotes the trace over the 
idler mode. We can now use all the Kroneker deltas of Eq. 
(\ref{ENTR4c}) by recalling that $\langle k_{-} |m_{-}\, m_{+}\rangle = 
\delta_{k_{-}\, m_{-}} |m_{+}\rangle $ and that, similarly, 
$ \langle n_{+}\, n_{-}|  k_{-} \rangle = \delta_{n_{-}\, k_{-}} \langle n_{+} |$. The final result for the reduced density operator becomes therefore 
\begin{equation}
\widehat{\rho}_{\mathrm{red}} = \sum_{n=0}^{\infty} \, p_{n} \, |\,n\, \rangle \langle\, n\,|, \qquad\qquad 
\mathrm{Tr} \widehat{\rho}^2_{\mathrm{red}} = \frac{1}{2 \overline{n} +1} < \mathrm{Tr} \widehat{\rho}_{\mathrm{red}}.
\label{ENTR6}
\end{equation}
where, as in Eq. (\ref{ENTR4a}), the statistical weights $p_{n} = \overline{n}^{n}/(\overline{n} +1)^{n+1}$ correspond to the Bose-Einstein probability distribution even if the averaged multiplicity $\overline{n} = \sinh^2{r}$ is sharply non-thermal. From Eq. (\ref{ENTR4c}) to Eq. (\ref{ENTR6}) we renamed the summation index (i.e. $n_{+}$ became $n$). While the von Neumann entropy associated with Eq. (\ref{ENTR4}) vanishes, Eq. (\ref{ENTR6}) implies that 
\begin{eqnarray}
s &=& - \mathrm{Tr} \bigl[ \widehat{\rho}_{\mathrm{red}} \, \ln{\widehat{\rho}_{\mathrm{red}} }\bigr]
\nonumber\\
&=& - \sum_{n=0}^{\infty} p_{n} \, \ln{p_{n}} = \ln{(\overline{n}+1)} - \overline{n} \ln{\biggl(\frac{\overline{n}}{\overline{n} +1}\biggr)}.
\label{ENTR7}
\end{eqnarray}
From the result of Eq. (\ref{ENTR7}) we see that in the limit $\overline{n}\gg 1$ the entropy gets 
proportional to $\ln{\overline{n}}$, in other words 
\begin{equation}
\lim_{\overline{n} \gg1} \, s(\overline{n}) = \ln{\overline{n}} = 2 \, r = 2 \ln{\biggl(\frac{a_{re}}{a_{ex}}\biggr)}.
\label{ENTR8}
\end{equation}
When observations are performed we shall always observe a graviton with a given three-momentum and we will not be able to distinguish the sign of its three-momentum. This type of measurement can be performed by means of Hanbury-Brown Twiss interferometry \cite{HBT1,HBT2,HBT3} (see also \cite{HBT4}) appropriately extended to the case of the relic gravitons \cite{HBTG1,REC4,HBTG2,HBTG3}. If gravitons are systematically observed in explicit measurements the operators describing the observables will act only on a portion of the Hilbert space, e.g. only gravitons with momentum $\vec{k}$ but not the ones with momentum $- \, \vec{k}$. This observation also implies  that an effective description of the quantum state of the  gravitons can be achieved not only in terms of two-mode squeezing operators but also in terms of the single-mode squeezing operators that are separately acting on the two portions of the Hilbert space (see appendix \ref{APPC} and discussion therein).  The reduction of the density matrix can be performed in different bases \cite{ST9a,ST9b,ST10,ST11};
the results depend on the basis but the asymptotic limit when the averaged multiplicity is large  
is generally proportional to $2\,r$ and hence to the logarithm of $\overline{n}$. The 
coarse grained entropy employed here is quite close to the notion of information theoretic 
entropy introduced many years ago \cite{DDP1,DDP2} with the purpose of reformulating the 
indetermination relations. This idea actually inspired the reduction scheme discussed in Refs. \cite{ST9a,ST9b}. 

Before discussing the integrated entropy, we recall the heuristic argument of Eq. (\ref{DNT19}), already 
criticised in different contexts \cite{ST9a,ST9b,ST10,ST11}. This argument basically suggests 
that unitarity is lost when many particles are produces. As shown before the apparent lack of unitarity 
may just be the result of a failure of the approximation scheme.  When the approximation scheme is accurate the violations of unitarity do not arise even if the averaged multiplicity of the produced pairs is very large; see also, in this respect, appendix \ref{APPC}. As suggested by other authors \cite{ST11} for similar reasons,  the argument stipulating that unitarity is simply lost when large amounts of particles are produced seems both unphysical and technically incorrect.

\subsection{Integrated entropy}
The result of Eq. (\ref{ENTR8}) holds, strictly speaking, in the two-mode case but they are 
easily extended to the field theoretical situation where the density matrix of Eq. (\ref{ENTR4})
can also be written as
\begin{equation}
\hat{\rho}_{\vec{k}} = \frac{1}{\cosh^2{r_{k}}} \sum_{n_{\vec{k}}=0}^{\infty} \sum_{m_{\vec{k}}=0}^{\infty} e^{- i \alpha_{k}(n_{\vec{k}} - m_{\vec{k}})} 
(\tanh{r_{k}})^{n_{\vec{k}} + m_{\vec{k}}} |n_{\vec{k}}\,\, n_{- \vec{k}}\rangle \langle m_{-\vec{k}}\,\,m_{\vec{k}}|,
\label{ENTR9}
\end{equation}
where we have also considered, for completeness, the contribution of the rotation 
operator of Eqs. (\ref{APPB16})--(\ref{APPB17}) bringing a further phase $\delta_{k}$ into the final result. This is why, incidentally, $\alpha_{k} = (2 \delta_{k} - \theta_{k})$. Since the two polarizations evolve independently we just considered a single polarization. In this case the density matrix can be written, in the Fock basis, as: 
\begin{equation}
\hat{\rho} = \sum_{\{n\}} \, P_{\{n\}} \, | \{n\} \rangle \langle \{n\}|,\qquad \sum_{\{n\}} \, P_{\{n\}} =1.
\label{ENTR10} 
\end{equation}
The multimode probability distribution appearing in Eq. (\ref{ENTR10}) is given by:
\begin{equation}
P_{\{n_{\vec{k}}\}} = \prod_{\vec{k}} P_{n_{\vec{k}}} , \qquad P_{n_{\vec{k}}}(\overline{n}_{k})= 
\frac{\overline{n}_{k}^{n_{\vec{k}}}}{( 1 + \overline{n}_{k})^{n_{\vec{k}} + 1 }},
\label{ENTR11}
\end{equation}
where $\overline{n}_{k}$ is the average multiplicity of each Fourier mode. Furthermore, following the standard notation, $ |\{n \}\rangle = |n_{\vec{k}_{1}} \rangle \, | n_{\vec{k}_{2}} \rangle \, | n_{\vec{k}_{3}} \rangle...$ where the ellipses stand for all the occupied modes of the field.  Even though the density matrix describes a mixed state, the average multiplicity does not need to coincide with the conventional Bose-Einstein occupation number determined, for instance, in the framework of the 
grand-canonical ensemble. This situation often occurs in quantum optics for chaotic (i.e. white) light where photons are distributed as in Eq. (\ref{ENTR11}) for each mode of the radiation field but they are produced by atomic sources  kept at an excitation level higher than that in thermal equilibrium.  In connection with Eq. (\ref{ENTR9}) we also note that the density 
matrix can be reduced by considering the phases of the final multiparticle state to be 
unobservable. In this case the right hand side of Eq. (\ref{ENTR9}) can be averaged averaging over $\alpha_{k}$. The reduced density matrix would be given, in this case, by $\hat{\rho}^{\mathrm{red}}_{\vec{k}} = \frac{1}{2\pi} \int_{0}^{2\pi} d \alpha_{k} \hat{\rho}_{\vec{k}}$. The  result obtained 
with this procedure will then lead to 
\begin{equation}
\hat{\rho}^{\mathrm{red}}_{\vec{k}} = \frac{1}{\cosh^2{r_{k}}} \sum_{n_{\vec{k}}=0}^{\infty} 
(\tanh{r_{k}})^{2 n_{\vec{k}}} |n_{\vec{k}}\,\, n_{- \vec{k}}\rangle \langle n_{-\vec{k}}\,\,n_{\vec{k}}|.
\label{ENTR12}
\end{equation}
This observation represents a further reduction scheme of the density matrix that ultimately 
leads to the same entropy of Eq. (\ref{ENTR7}) in the limit of the large averaged multiplicities.

Since we are now dealing with all the modes of the field the entropy associated with a 
single mode of the field must be integrated over the whole spectrum in a given 
fiducial volume. If we consider a fiducial Hubble volume at the present time 
 the total entanglement entropy of the gravitons becomes:
\begin{equation}
S_{g \, 0} = \frac{8}{3} \pi H_{0}^{-3} \int_{k_{min}}^{k_{max}} \frac{d^{3} k}{(2 \pi)^3} \ln{\overline{n}_{k}}.
\label{ENTR13}
\end{equation}
To estimate Eq. (\ref{ENTR13}) we can simply employ the model-independent interpolation 
already suggested in Eq. (\ref{FFF6}). For the illustrative purposes of the present 
discussion it is useful to show that the power-law parametrization of Eq. (\ref{FFF6}) 
is actually compatible with Eqs. (\ref{APPC11})--(\ref{APPC11aa}) and (\ref{APPC11a}) where the averaged multiplcity for $k \ll k_{max}$ 
depends on $(a_{re}/a_{ex}) \gg 1 $. Let us therefore assume that between $a_{ex}$ and $a_{re}$ the background expands
in this case the averaged multiplicity is given by:
\begin{equation} 
\overline{n}_{k} = \frac{1}{4}\biggl(\frac{a_{re}}{a_{ex}}\biggr)^2 \simeq \biggl(\frac{k}{k_{max}}\biggr)^{ - 2 \beta_{re} - 2\beta_{ex}}.
\label{ENTR14}
\end{equation}
The estimate of Eq. (\ref{ENTR14}) follows by assuming that the relevant wavelengths 
cross the Hubble radius for the first time during inflation  (i.e. $ k \tau_{ex} = k/{\mathcal H}_{ex}= {\mathcal O}(1)$) when the scale factor is given, approximately, by $a_{ex} = (- \tau_{1}/\tau_{ex})^{\beta_{ex}} \simeq |k \tau_{1}|^{\beta_{ex}}$; the reentry takes place instead when $a_{re} = (\tau_{re}/\tau_{1})^{\beta_{re}} \simeq |k \tau_{1}|^{-\beta_{re}}$.
In the conventional situation $\beta_{ex} = 1/(1- \epsilon)$ (where $\epsilon$ is the slow-roll parameter) and $\beta_{ex} = {\mathcal O}(1)$: this means that $\overline{n}_{k}\simeq (k/k_{max})^{-4}$.  If  the wavelength reenters the Hubble radius during a stiff phase we have instead 
that $\overline{n}_{k}\simeq (k/k_{max})^{-3}$ since $\beta_{re} = {\mathcal O}(1/2)$. Therefore the 
spectral energy density of Eq. (\ref{FFF7})  is comparatively larger at high frequencies when $\overline{n}_{k}\simeq (k/k_{max})^{-3}$ than in the 
conventional case the averaged multiplicity scales as $(k/k_{max})^{-4}$. While more accurate determinations of the averaged multiplicity can be presented in slightly different 
frameworks \cite{REC2}, the considerations based on Eq. (\ref{ENTR14}) are consistent with Eq. (\ref{FFF6}).
Thanks to the explicit form of Eq. (\ref{ENTR4}) the integration variable appearing in Eq. (\ref{ENTR13}) can be rescaled and the total entropy of the gravitons is 
\begin{equation}
S_{g\, 0}= \frac{32}{3} \,\pi^2\, \biggl(\frac{\nu_{max}}{H_{0}}\biggr)^3 {\mathcal K}(m_{T}), \qquad {\mathcal K}(m_{T}) = (4 - m_{T}) \int_{\nu_{min}/\nu_{max}}^{1} x^2 \, \ln{x} \, d \, x.
\label{ENTR15}
\end{equation}
Since in the integral of Eq. (\ref{ENTR15}) the value of $m_{T}$ is not essential and the 
lower limit of integration goes to zero (at least in practice since $\nu_{p} \simeq \nu_{min} = 
3\, 10^{-18} \mathrm{Hz}$) we have that $S_{g\, 0} ={\mathcal O}(10) (\nu_{max}/H_{0})^{3}$. 
Since the total entropy is always dimensionless the result for $S_{g\, 0}$ can now be compared with the well known result of the thermal entropy of the cosmic microwave background computed within the same fiducial Hubble volume 
\begin{equation}
S_{\gamma\, 0} = \frac{4}{3} \pi \, H_{0}^{-3}\, s_{\gamma}, \qquad \qquad s_{\gamma}= 
\frac{4}{45} \pi^2 T_{\gamma\, 0}^3.
\label{ENTR16}
\end{equation}
Since $T_{\gamma\, 0} = T_{\gamma 0} = (2.72548 \pm 0.00057) \, \mathrm{K}$ \cite{CMBT} we also have that 
\begin{equation}
T_{\gamma 0} = 356.802 \biggl(\frac{T_{\gamma 0}}{2.72548\,\, \mathrm{K}}\biggr) \, \, \mathrm{GHz}.
\label{ENTR17}
\end{equation}
If we now require that $S_{g\, 0} < S_{\gamma\, 0}$ we have from Eqs. (\ref{ENTR15})--(\ref{ENTR16}) and (\ref{ENTR17}) that 
\begin{equation}
S_{g\, 0} < S_{\gamma\, 0}\qquad \Rightarrow \qquad \nu_{max} < \mathrm{THz}.
\label{ENTR18}
\end{equation}
It is interesting that the maximal frequency of the spectrum deduced in section \ref{sec2}
also implies that the entanglement entropy of the gravitons undershoots the total entropy of the 
cosmic microwave background.

\subsection{Superfluctuant and subfluctuant operators}
The considerations developed in the previous sections suggest that the multiparticle 
state of the gravitons is macroscopic but still quantum mechanical and it seems 
difficult to reproduce the correlations in a completely classical picture. This is especially
true for the second-order degrees of quantum coherence that have been 
thoroughly discussed in the past \cite{HBTG1,REC4,HBTG2,HBTG3}. To make this statement more accurate 
we observe here that the final state of the relic gravitons cannot be mimicked 
in terms of a semiclassical quasiprobability distribution. This means that the 
final state of the relic gravitons is inherently quantum mechanical. Indeed 
the optical equivalence theorem in quantum optics \cite{MANDL} would suggest an equivalence between the expectation value of an operator in Hilbert space and the expectation value of its associated function in the phase space formulation with respect to a positive semi-definite quasiprobability distribution \cite{PDIST1,PDIST2,PDIST3}.  In the present context the quasi-probability distribution is not necessarily positive semidefinite, as we are now going to show.
Let us consider, for this purpose, the operator $\Sigma(r)$ corresponding to the one 
of Eq. (\ref{ENTR1}) in the case $\theta\to 0$. We may argue that such a choice can always
be implemented either as a result of the observational set-up or because of other dynamical considerations. 
In this case the two-mode operator can be always written
\begin{equation}
\Sigma(r) = e^{r ( b_{+} \, b_{-} - b_{+}^{\dagger} \, b_{-}^{\dagger})} = e^{r (b^2 - b^{\dagger\, 2})/2} \, 
e^{r (\widetilde{b}^2 - \widetilde{b}^{\dagger\, 2})/2}.
\label{PDS1}
\end{equation}
For notational convenience, we shall suppress the caret appearing on top of each 
operator; this is necessary since it is practical to decompose $b_{+}$ and $b_{-}$ as\footnote{It would be 
excessive to add a caret on top of the tilde in the definition of $\widetilde{b}$; this is why we shall suppress 
the caret.}
\begin{equation}
b_{+} = \frac{b +\, i \, \widetilde{b}}{\sqrt{2}}, \qquad b_{+} = \frac{b -\, i \, \widetilde{b}}{\sqrt{2}},
\label{PDS2}
\end{equation}
where $[b,\, b^{\dagger}] =1$, $[\widetilde{b},\, \widetilde{b}^{\dagger}] =1$ and $[ b, \, \widetilde{b} ] =0$.
If the original operator in Eq. (\ref{PDS1}) would contain a $\theta$ dependence 
we could always add an arbitrary phase in the definitions of Eq. (\ref{PDS2}); this phase could be 
fixed to eliminate the original $\theta$-dependence possibly appearing in Eq. (\ref{PDS1}). We can finally 
relate $b$ and $\widetilde{b}$ with the quadrature operators that we simply denote as
\begin{eqnarray}
x &=& \frac{b + b^{\dagger}}{\sqrt{2}}, \qquad p = - i \, \frac{(b - b^{\dagger})}{\sqrt{2}}
\nonumber\\
\widetilde{x} &=& \frac{\widetilde{b} + \widetilde{b}^{\dagger}}{\sqrt{2}}, \qquad \widetilde{p} = - i \, \frac{(\widetilde{b} - \widetilde{b}^{\dagger})}{\sqrt{2}}.
\label{PDS3}
\end{eqnarray}
The quadrature operators are customarily defined with a further $\sqrt{2}$ in the denominator. 
Since we find it more inspiring, we prefer here the comparatively older notation of Ref. \cite{STO} implying $[x, p] = \, i$ and $[\widetilde{x}, \widetilde{p}] =i$. In this language the operator $\Sigma(r)$ can then be written as the product of two contraction operators $\Sigma(r) = {\mathcal D}(r)\, \widetilde{{\mathcal D}}(r)$ where 
\begin{equation}
 {\mathcal D}(r) = e^{ i\,r (x \, p+ p\, x)/2}, \qquad \qquad \widetilde{{\mathcal D}}(r) = e^{ i\,r (\widetilde{x} \, \widetilde{p}+ \widetilde{p}\, \widetilde{x})/2}.
\label{PDS4}
\end{equation}
From Eq. (\ref{PDS4}) it is clear that the action of $ {\mathcal D}(r)$ and $\widetilde{{\mathcal D}}(r)$
on the quadrature operators implies a contraction of $x$ and $\widetilde{x}$ and a corresponding 
dilatation of $p_{r}$ and $\widetilde{p}$
\begin{eqnarray}
x \to X_{r} &=& {\mathcal D}^{\dagger}(r)\, x\, {\mathcal D}(r) = e^{-r} \, x, \qquad p\to P_{r} = {\mathcal D}^{\dagger}(r)\, p\, {\mathcal D}(r) = e^{r} \, p,
\label{PDS5}\\
\widetilde{x} \to \widetilde{X}_{r} &=&\widetilde{ {\mathcal D}}^{\dagger}(r)\, \widetilde{ x}\, \widetilde{ {\mathcal D}}(r) = e^{-r} \, \widetilde{ x}, \qquad \widetilde{ p}\to \widetilde{ P}_{r} = \widetilde{ {\mathcal D}}^{\dagger}(r)\, \widetilde{ p}\, \widetilde{ {\mathcal D}}(r) = e^{r} \, \widetilde{ p}.
\label{PDS6}
\end{eqnarray}
The operators $X_{r}$ and $\widetilde{X}_{r}$ fluctuate below the quantum noise whereas 
$P_{r}$ and $\widetilde{P}_{r}$ fluctuate above it. If we now represent the density operator in the 
two-mode diagonal $P$-representation \cite{PDIST1,PDIST2,PDIST3} 
we have that 
\begin{equation}
\widehat{\rho} = \int \, P(\beta,\, \widetilde{\beta}) \,\,| \beta \, \widetilde{\beta} \rangle \,\langle \widetilde{\beta}\,\beta|\,\, \, d^{2} \beta \, d^2 \widetilde{\beta},
\label{PDS7}
\end{equation}
where $P(\beta,\, \widetilde{\beta})$ is the (two-mode) quasi-probability distribution and $| \beta \, \widetilde{\beta} \rangle $ denotes a two-mode coherent state. In terms of $P(\beta,\, \widetilde{\beta})$ the variances of the operators $X_{r}$ and $\widetilde{X}_{r}$ are given by:
\begin{eqnarray}
(\Delta X_{r})^2 &=&  \frac{e^{-2 r} }{2} = \frac{1}{2} \biggl[ 1 + \int P(\beta, \widetilde{\beta}) (\beta + \beta^{\ast})^2 \, d^{2} \beta \, d^2 \widetilde{\beta}\biggr], 
\label{PDS8}\\ 
(\Delta \widetilde{X}_{r})^2 &=&  \frac{e^{-2 r} }{2} = \frac{1}{2} \biggl[ 1 + \int P(\beta, \widetilde{\beta}) (\widetilde{\beta} + \widetilde{\beta}^{\ast})^2 \, d^{2} \beta \, d^2 \widetilde{\beta}\biggr],
\label{PDS9}
\end{eqnarray}
where, as usual $(\Delta X_{r})^2 = (\langle X_{r}^2 \rangle - \langle X_{r} \rangle^2)$ and similarly for 
$(\Delta \widetilde{X}_{r})^2 $. We can now appreciate from Eqs. (\ref{PDS8})--(\ref{PDS9}) that a positive 
semi-definite $P(\beta, \widetilde{\beta})$ is not compatible with the conditions  $(\Delta X_{r})^2 \ll 1/2$ and 
$(\Delta \widetilde{X}_{r})^2 \ll 1/2$ potentially achieved  when $r \gg 1$; this happens, incidentally, when the 
average multiplicity is large. A similar argument can be repeated with the canonically 
conjugate operators $P_{r}$ and $\widetilde{P}_{r}$. In this case the corresponding variances become:
\begin{eqnarray}
(\Delta P_{r})^2 &=&  \frac{e^{2 r} }{2} = \frac{1}{2} \biggl[ 1 - \int P(\beta, \widetilde{\beta}) (\beta - \beta^{\ast})^2 \,\,\, d^{2} \beta \, d^2 \widetilde{\beta}\biggr], 
\label{PDS10}\\ 
(\Delta \widetilde{P}_{r})^2 &=&  \frac{e^{2 r} }{2} = \frac{1}{2} \biggl[ 1 - \int P(\beta, \widetilde{\beta}) (\widetilde{\beta} - \widetilde{\beta}^{\ast})^2 \,\,\, d^{2} \beta \, d^2 \widetilde{\beta}\biggr].
\label{PDS11}
\end{eqnarray}
We now see from Eqs. (\ref{PDS10})--(\ref{PDS11}) that a positive 
semi-definite $P(\beta, \widetilde{\beta})$ is not compatible with the conditions  $(\Delta P_{r})^2 \gg1/2$ and 
$(\Delta \widetilde{P}_{r})^2 \gg 1/2$, as it happens when $r \gg 1$ and the averaged multiplicity is large.

The simultaneous analysis of Eqs. (\ref{PDS8})--(\ref{PDS9}) and (\ref{PDS10})--(\ref{PDS11}) implies therefore that 
 the diagonal $P$-representation becomes negative over some range of its arguments 
$\beta$ and $\widetilde{\beta}$. Other quantum distribution functions such as the Wigner functions \cite{wig1,wig2},
the complex $P$-functions \cite{COMPP} or the $Q$-functions \cite{Qrep} lead to similar paradoxes either in 
the case of the variances or for the intensity correlations and for initial states containing a final number of gravitons. We therefore acknowledge the viewpoint that the quasi-probability distribution get negative whenever a given quantum state cannot be mimicked by a classical source \cite{COMPP,Qrep}. In spite 
of claims going into the opposite direction, our considerations  show that the relic gravitons appear as macroscopic 
quantum states of many particles that cannot be reproduced with a semiclassical quasi-probability 
distribution as suggested, in quantum optics, by the optical equivalence theorem \cite{MANDL,PDIST1,PDIST2,PDIST3}. 

\newpage
\renewcommand{\theequation}{6.\arabic{equation}}
\setcounter{equation}{0}
\section{Concluding remarks}
\label{sec6}
The evolution of the space-time curvature during the early stages of the cosmological 
evolution excites the tensor modes of the geometry and when the classical fluctuations 
are suppressed (as it happens either during a stage of accelerated expansion or 
in a phase of accelerated contraction) gravitons are produced  from an initial quantum state that likely coincides with the vacuum provided the inflationary epoch is sufficiently long. While the physical premises characterizing the evolution of the space-time curvature might be different, the unitarity of the process and the Hermiticity of the Hamiltonian imply that the gravitons are always produced in pairs of opposite (comoving) three-momentum. Moreover their multiplicity distributions and their associated averaged multiplicities concurrently determine the maximal frequency range that turns out to be ${\mathcal O}(\mathrm{THz})$.  For the derivation of this bound we first considered the properties of the averaged multiplicity and then deduced an interpolating expression that is exponentially suppressed at high-frequencies while it scales as power-law below $\nu_{max}$. Since above the maximal frequency the averaged multiplicity is exponentially suppressed, $\nu_{max}$ roughly corresponds to the production of a single graviton pair: this observation has been used here not as a derived property but rather as an operational definition of the maximal frequency of the spectrum. This is ultimately the logic leading to the THz bound.

In a purely classical perspective $\nu_{max}$ is instead associated with the bunch of (minimal) wavelengths that exited the Hubble radius at the end of inflation and reentered immediately after, i.e. at the onset of the decelerated stage of evolution. In this respect the present analysis shows that the quantum approach to the high-frequency bound is model-independent since it does not require or assumes a specific knowledge of the expansion rates around the onset of the decelerated stage of expansion. For this reason the classical and the quantum strategies are mutually consistent not only in the conventional case but also when the postinflationary expansion rate is slower than radiation as well as in less conventional situations when the curvature scale takes Plankian values. Although in all these cases we would expect $\nu_{max} \gg \mathrm{MHz}$,  the maximal frequency always undershoots the quantum limit in the THz band. The obtained constraint on $\nu_{max}$ translates into an absolute bound of the order of $10^{-33}$ on the minimal chirp amplitudes that should be mandatorily achieved by all classes of  hypothetical detectors aiming at the direct scrutiny of a signal in the region ranging between the MHz and the GHz bands. Minimal chirp amplitudes ${\mathcal O}(10^{-24})$ (possibly obtained by microwave prototypes in the last score year) have therefore  marginal significance. We find this aspect particularly useful for a realistic assessment of the capabilities and of the physical motivations of high-frequency detectors in the forthcoming two decade.

The THz bound and the whole approach of the present investigation are deeply rooted in the quantum mechanical nature of the produced gravitons whose multiparticle final state turns out to be macroscopic but always non-classical as it  follows from the unitary evolution that preserves the quantum coherence. The  apparent losses of quantum coherence  (possibly due to the largeness of the averaged multiplicity of the produced pairs) are then caused by the inaccuracy of the approximation schemes that do not enforce the validity of the commutation relations between canonically conjugated fields throughout all the stages of the dynamical evolution. The multiparticle final state has been also scrutinized in terms of the quasi-probability distributions that may get negative when the statistical properties of a given quantum system are not reproducible as weighted superposition of coherent states. Our analysis shows that the relic gravitons are inherently quantum mechanical and that their quantumness can be measured in terms of an entanglement entropy. Indeed, an entanglement entropy can be introduced in the problem and it is caused  by the loss of the complete information on the underlying quantum field. The reduction of the density matrix in different bases leads to the same von Neumann entropy whose integral over all the modes of the spectrum is dominated again by the maximal frequency. Whenever the THz bound is applied, it turns out that the total integrated entropy of the relic gravitons is comparable with the entropy of the cosmic microwave background but not larger.  As argued in the recent past in different contexts, our considerations suggest that a potential detection of relic gravitons both at low and high frequencies may therefore represent a direct evidence of macroscopic quantum states associated with the gravitational field.

\section*{Aknowledgements}
I wish to thank A. Gentil-Beccot,  F. Gili,  A. Kohls,  L. Pieper, S. Rohr and J. Vigen of the CERN Scientific Information Service for their kind help along the different stages of this investigation. 

\newpage 

\begin{appendix}

\renewcommand{\theequation}{A.\arabic{equation}}
\setcounter{equation}{0}
\section{Multiplicity distributions and their generalizations}
\label{APPA}
The multiplicity distribution of the relic gravitons has been 
a recurrent theme of the present analysis starting with the considerations developed in section \ref{sec2}. The complementary derivation of the multiplicity distribution presented hereunder can be swiftly applied to more general situations like the ones where the gravitons are not necessarily produced from the vacuum. The discussion of this appendix does not apply to a specific cosmological evolution but it depends solely on the pair production mechanism encoded in Eqs. (\ref{FFF1})--(\ref{FFF2}) and in the Hamiltonian of relic gravitons introduced in Eq. (\ref{DNT2}). The first observation of this appendix is that the multiplicity distribution follows directly from a recurrence relation relating  the diagonal elements of the density matrix.  For this purpose we start from Eqs. (\ref{FFF1})--(\ref{FFF3})--(\ref{FFF4}) and recall that, by definition, the diagonal elements of the density matrix are defined as $p_{n} = \langle n| z_{k} \rangle \langle z_{k} | n \rangle$ where, as in section \ref{sec2}, the state $| z_{\vec{k}} \rangle$ is annihilated by $\widehat{a}_{\vec{k}}$ and by 
$\widehat{a}_{-\vec{k}}$; in other words we have that $\widehat{a}_{\vec{k}} \, | z_{\vec{k}} \rangle =0$ and $\widehat{a}_{-\vec{k}} \, | z_{\vec{k}} \rangle =0$. Thanks to Eqs. (\ref{FFF1})--(\ref{FFF2}) the two previous conditions imply:
\begin{equation}
u_{k} \, \widehat{b}_{+} \, | z_{\vec{k}} \rangle= v_{k} \, \widehat{b}^{\dagger}_{-} | z_{\vec{k}} \rangle, \qquad \qquad u_{k} \, \widehat{b}_{-} \, | z_{\vec{k}} \rangle= v_{k} \, \widehat{b}^{\dagger}_{+} | z_{\vec{k}} \rangle.
\label{APPA1}
\end{equation}
If the attention is focused, for instance,  on the first equality appearing in Eq. (\ref{APPA1}), 
the following condition can be readily deduced:
\begin{equation}
u_{k} \,\,\langle n_{-} \, n_{+} | \widehat{b}_{+} | z_{\vec{k}} \rangle = v_{k} \,\,
\langle n_{-} \, n_{+} | \widehat{b}_{-}^{\dagger} | z_{\vec{k}} \rangle.
\label{APPA2}
\end{equation}
By now using the properties of the creation and annihilation operators on the state $ | n_{+}\, n_{-} \rangle$
 Eq. (\ref{APPA2}) can be made more explicit;  if we then focus the corresponding diagonal elements
(i.e. $n_{-} = m+ 1$ and $n_{+} = m$) the recurrence relation for the probability 
of obtaining $m$ pairs from the vacuum becomes:
\begin{equation}
p_{m +1} \,\, |u_{k}|^2 = p_{m}\,\, |v_{k}|^2.
\label{APPA3}
\end{equation}
The solution of Eq. (\ref{APPA3}) is clearly given by the Bose-Einstein distribution 
with average multiplicity $\overline{n}_{k} = |v_{k}|^2$. However, in view of further generalizations 
of Eq. (\ref{APPA3}), it is useful to solve Eq. (\ref{APPA3}) by first deducing the probability 
generating function $P(s)$ associated with the (unknown) probability distribution $p_{m}$. If both sides of Eq. (\ref{APPA3}) are multiplied by $s^{m}$ (with $0\leq s \leq 1$) the obtained result can be summed over $m$
\begin{equation}
|v_{k}|^2 \, \sum_{m=0}^{\infty} s^{m} \, p_{m} = |u_{k}|^2 \, \sum_{m=0}^{\infty} s^{m} \, p_{m +1}, \qquad \qquad P(s) = \sum_{m=0}^{\infty} \, s^{m} p_{m}.
\label{APPA4}
\end{equation}
The second relation appearing in Eq. (\ref{APPA4})  is just the definition of the probability 
generating function and since, by definition, $\sum_{m}\, p_{m} =1$ we also have that $P(1)=1$. We can finally solve Eq. (\ref{APPA4}) and deduce $P(s)$:
\begin{equation}
P(s) = \frac{p_{0} \, (\overline{n}+1)}{ 1 + \overline{n}(1 -s)},
\label{APPA5}
\end{equation}
where $p_{0}$ is the probability of producing zero pairs. Since 
the probability distribution must be normalized we get that 
$P(1) =1$ and also $p_{0} = 1/(\overline{n} +1)$. This means, ultimately, that 
\begin{equation}
P(s) = \frac{1}{ 1 + \overline{n}(1 - s)}.
\label{APPA6}
\end{equation}
Once the analytic form of $P(s)$ is known the probability distribution follows immediately 
by evaluating the $m$-th order derivative of $P(s)$ for $s\to 0$; in our case, as anticipated,
\begin{equation}
p_{m} = \frac{1}{m!} \biggl(\frac{d P}{d s}\biggr)_{s=0} = \frac{\overline{n}_{k}^{m}}{(\overline{n}_{k} +1)^{m +1}}.
\label{APPA7}
\end{equation}
In case the initial state does not coincide with the vacuum the recurrence relation of Eq. (\ref{APPA3}) 
is generalized in the following manner:
\begin{equation}
(m + 1) \,\,p_{m+1} = (a_{k} + b_{k} m) \,\,p_{m}.
\label{APPA8}
\end{equation}
In the limit $a_{k} = b_{k} = |v_{k}|^2/|u_{k}|^2$ Eq. (\ref{APPA8}) coincides with Eq. (\ref{APPA3}) and after multiplying both sides of Eq. (\ref{APPA8}) by $s^{m}$ we can sum 
over $m$; this time the analog of Eq. (\ref{APPA5}) is not simply an algebraic relation but rather a differential equation for $P(s)$
\begin{equation}
( 1 - b_{k} s) \frac{d P}{d s} = a_{k} P(s)\qquad \Rightarrow \qquad P(s) = \frac{c_{1}}{(1 - b_{k} s)^{q}},
\label{APPA9}
\end{equation}
where $q= a_{k}/b_{k}$ and $c_{1}$ is an integration constant that can be fixed from the condition 
$P(1) =1$. The full form of $P(s)$ and of its derivative with respect to $s$ become
\begin{equation}
P(s) = \frac{(1 - b_{k})^{q}}{(1 - b_{k} \, s)^{q}}, \qquad \frac{d P}{d s}\biggl|_{s=1} = \overline{n}_{k} \qquad \Rightarrow\qquad 
b_{k} = \frac{\overline{n}_{k}}{\overline{n}_{k} + q}.
\label{APPA9a}
\end{equation}
The second relation of Eq. (\ref{APPA9a}) is simply the definition of the averaged 
multiplicity given in terms of the derivative of $P(s)$; this observation fixes then $b_{k}$ in terms 
of $\overline{n}_{k}$ and $q$ with the result that, after some algebra, we obtain:
\begin{equation}
P(s) = \frac{q^{q}}{[q + \overline{n}_{k} (1 -s)]^{q}}, \qquad q = a_{k}/b_{k}, \qquad b_{k} = \frac{\overline{n}_{k}}{\overline{n}_{k} + q}.
\label{APPA10}
\end{equation}
Equation (\ref{APPA10})  is the canonical from of the probability generating function of the negative binomial (Pascal)
probability distribution; the probability distribution itself can be obtained by deriving $P(s)$ as already discussed in Eq. (\ref{APPA7})
\begin{eqnarray}
p_{m}(\overline{n},q) &=& \frac{1}{m!} \biggl(\frac{d P}{d s}\biggr)_{s=0} = \left(\matrix{ q + m -1 \cr m }\right)  \biggl(\frac{\overline{n}_{k}}{\overline{n}_{k} + q}\biggr)^{m} \biggl(\frac{q}{\overline{n}_{k}+ q}\biggr)^{q} 
\nonumber\\
&=& \frac{q ( q+ 1)... ( q + m -1)}{m!}  \biggl(\frac{\overline{n}_{k}}{\overline{n}_{k} + q}\biggr)^{m} \biggl(\frac{q}{\overline{n}_{k}+ q}\biggr)^{q}.
\label{APPA11}
\end{eqnarray}
Once more, when $q=1$ Eqs. (\ref{APPA10}) and (\ref{APPA11}) coincide, respectively, with Eqs. (\ref{APPA6}) and (\ref{APPA7}).  The probability distribution of Eq. (\ref{APPA11}) can be expressed as the probability distribution of the sum of an arbitrary number of independent and identically distributed random variables. In this case the probability distribution (and the related characteristic function) are said to be infinitely divisible. Both properties easily follow by appreciating that the characteristic function (\ref{APPA10}) naturally appears as compound Poisson distribution \cite{karlin} and any compound Poisson distribution is infinitely divisible. 

\renewcommand{\theequation}{B.\arabic{equation}}
\setcounter{equation}{0}
\section{Two-mode and single-mode squeezing operators}
\label{APPB}
A two-mode squeezing operator can be described as the product of two degenerate 
 squeezed states; in this way  the properties  of two-mode 
 states can be deduced from the single-mode operators  \cite{sch1,sch2,pere}. On a technical 
ground, under some conditions, the two-mode squeezed squeezing operator can be factorized into the 
tensor product of two degenerate squeezing operators. Historically the degenerate 
squeezing operator has been introduced first \cite{STO,STO2,STO3} even though 
the two-mode squeezing operators appear naturally in the quantum theory of parametric 
amplification \cite{mollow1,mollow2} and in the mode expansion of the tensor modes. 
In short for  single-mode (or degenerate)
states the generators of the  $SU(1,1)$ algebra are given by
\begin{equation}
L^{(1)}_{+} = \frac{1}{2} \,\,\widehat{a}_{1}^{\dagger\, 2}, \qquad L^{(1)}_{-}= 
\frac{1}{2} \,\,\widehat{a}_{1}^{2},\qquad L^{(1)}_{0} = \frac{1}{4}\,\,\biggl(\widehat{a}_{1}^{\dagger} \, \widehat{a}_{1} +  \widehat{a}_{1} \widehat{a}_{1}^{\dagger}\biggr),
\label{APPB1}
\end{equation}
where $[\widehat{a}_{1}, \widehat{a}_{1}^{\dagger}] = 1$; as a consequence the commutation 
relations between the operators of Eq. (\ref{APPB1}) are given by:
\begin{equation}
[L^{(1)}_{+},\, L^{(1)}_{-}] = - \,2 \, L^{(1)}_{0}, \qquad [L^{(1)}_{0}, L^{(1)}_{\pm}] = \pm L^{(1)}_{\pm},
\label{APPB2}
\end{equation}
where $L_{0}^{(1)\,\dagger} = L_{0}$ while $L_{-}^{(1)\,\dagger} = L_{+}^{(1)}$. With the notations 
of Eqs. (\ref{APPB1})--(\ref{APPB2}) we can also write a single-mode Hamiltonian as 
\begin{equation}
\widehat{H}^{(1)} = 2 k L_{0}^{(1)} + \lambda L_{+}^{(1)} + \lambda^{\ast} L_{-}^{(1)}.
\label{APPB3}
\end{equation}
The single-mode (or degenerate) squeezing operator that would follow from the unitary time-evolution  
 generated by the Hamiltonian (\ref{APPB3}) can always
be decomposed in terms of the Baker-Campbell-Hausdorff relation. The standard derivation (obtained in the 
$SU(2)$ case by using the properties of Pauli matrices) can be generalized to the $SU(1,1)$ and this is basically the strategy followed in Refs. \cite{GILMORE,FISHER}. It is however possible to derive the Baker-Campbell-Hausdorff relation in a different manner \cite{truax1} (see also \cite{truax2}), i.e. by reducing  the problem to a system of first-order (nonlinear) differential equations. In both cases the final result can be written as:
\begin{equation}
S^{(1)}(z) =  e^{z^{\ast} L_{-}^{(1)} - z\,L_{+}^{(1)}} = e^{-\Gamma L^{(1)}_{+}}\,\,
 e^{- 2 \ln{\cosh{r}} L^{(1)}_{0}}  \,\, e^{\Gamma^{\ast} L^{(1)}_{-}},
\label{APPB4}
\end{equation}
where $z = e^{i \theta} \,\, r$ and $\Gamma= e^{i \theta} \,\,\tanh{r}$.
It is well known that the single-mode operator $S^{(1)}(z)$ can also be viewed as the generator of a unitary transformation of the creation and annihilation operators:
\begin{equation}
\widehat{c}_{1} = S^{(1)\,\, \dagger}(z) \,\, \widehat{a}_{1} \,\, S^{(1)}(z) = \cosh{r}\,\, \widehat{a}_{1} - e^{ i \theta} \,\,\sinh{r} \,\,\widehat{a}_{1}^{\dagger}.
\label{APPB5}
\end{equation}
In Eq. (\ref{APPB1}) the superscript  distinguishes the generators of the first oscillator from the ones of a putative second oscillator:
\begin{equation}
L^{(2)}_{+} = \frac{1}{2} \widehat{a}_{2}^{\dagger\, 2}, \qquad L^{(2)}_{-}= 
\frac{1}{2} \widehat{a}_{2}^{2},\qquad L^{(2)}_{0} = \frac{1}{4}\biggl(\widehat{a}_{2}^{\dagger} \, \widehat{a}_{2} +  \widehat{a}_{2} \widehat{a}_{2}^{\dagger}\biggr).
\label{APPB6}
\end{equation}
All the relations derived in the case of Eqs. (\ref{APPB3})--(\ref{APPB4}) 
obviously hold in the case the operators defined in Eq. (\ref{APPB5}). If the corresponding 
Hamiltonians are now combined as $\widehat{H} = \widehat{H}^{(1)} + \widehat{H}^{(2)}$
we have that two new sets of two-mode operators can be defined in terms of the following 
 transformation:
\begin{equation}
\widehat{a} = \frac{\widehat{a}_{1} - i \, \widehat{a}_{2}}{\sqrt{2}}, \qquad \widehat{b} = \frac{\widehat{a}_{1} + i \, \widehat{a}_{2}}{\sqrt{2}}.
\label{APPB7}
\end{equation}
We could insert a further global phase at the right-hand side of both 
transformations appearing in Eq. (\ref{APPB7}) but what matters  is that 
the two-mode squeezing operator can therefore be expressed as\footnote{If a further global phase 
is inserted at the right-hand side of Eq. (\ref{APPB7}) the modulus of $z$ remains the same 
but its final phase gets modified by the phase possibly introduced in Eq. (\ref{APPB7}).}
\begin{equation}
\Sigma(z) = e^{ z^{\ast}  \widehat{a} \, \widehat{b}\,\,-z\,\, \widehat{a}^{\dagger} \, \widehat{b}^{\dagger}} = S^{(1)}(z) \, S^{(2)}(z),
\label{APPB8}
\end{equation}
where the second equality follows by inserting into $\Sigma(z)$ the explicit 
form of $\widehat{a}$ and $\widehat{b}$ in terms of $\widehat{a}_{1}$ and $\widehat{a}_{2}$
as dictated by Eq. (\ref{APPB7}). It is also easy to verify that the combinations 
of the group generators still obey the $SU(1,1)$ algebra; more precisely 
we have that  
 \begin{eqnarray}
&& L_{+} = L_{+}^{(1)} +  L_{+}^{(2)} = \widehat{a}^{\dagger} \, \widehat{b}^{\dagger}, \qquad  L_{-} = L_{-}^{(1)} +  L_{-}^{(2)} = \widehat{a} \, \widehat{b},
\nonumber\\
&& L_{0} = L_{0}^{(1)} +  L_{0}^{(2)} = \frac{1}{2}[ \widehat{a}^{\dagger} \widehat{a} + \widehat{b}\, \widehat{b}^{\dagger}].
\label{APPB9}
\end{eqnarray}
The combinations of the single-mode generators lead to the 
results of Eq. (\ref{APPB9}) via the transformation (\ref{APPB7}); moreover, as 
expected, $L_{\pm}$ and $L_{0}$ satisfy  the commutation relations of the $SU(1,1)$ Lie algebra
\begin{equation}
[L_{+},\, L_{-}] = - 2 L_{0},\qquad [L_{0}, L_{\pm}] = \pm L_{\pm}.
\label{APPB10}
\end{equation}
Thanks to Eq. (\ref{APPB7}) we can also verify that the two-mode Hamiltonian is the sum of the two single-mode Hamiltonians, namely:
\begin{equation}
\widehat{H}= \widehat{H}^{(1)} + \widehat{H}^{(2)} = 2 k \, L_{0} + \lambda\, L_{+} + \lambda^{\ast} L_{-}. 
\label{APPB11}
\end{equation}
The factorization of the two-mode squeezing operator in terms of the Baker-Campbell-Hausdorff decomposition follows from the techniques of Refs. \cite{GILMORE,FISHER} and \cite{truax1,truax2}:
\begin{equation}
\Sigma(z) =  e^{z^{\ast}\,\, L_{-} - z\,\,L_{+}} = e^{-\Gamma L_{+}}
 e^{- 2 \ln{\cosh{|z|}} L_{0}}  e^{\Gamma^{\ast} L_{-}},\qquad \Gamma = \frac{z}{|z|} \tanh{|z|}.
\label{APPB12}
\end{equation}
As in the case of Eq. (\ref{APPB5}) the two-mode squeezing operator can be viewed 
as the generator of a unitary transformation involving the two-mode 
creation and annihilation operators:
\begin{equation}
\widehat{c} = \Sigma^{\dagger}(z) \, \widehat{a} \,  \Sigma(z) = \cosh{r} \widehat{a}\, - \, e^{i\theta}\, \sinh{r} \, \widehat{b}^{\dagger}.
\label{APPB13}
\end{equation}
As stressed in Ref. \cite{sch1,sch2} the relation between two-mode operators and (degenerate)
single-mode operators is technically useful but it can be physically misleading. In our problem 
the two modes of $\Sigma(z)$ correspond to gravitons with opposite three-momenta.

\renewcommand{\theequation}{C.\arabic{equation}}
\setcounter{equation}{0}
\section{Explicit form of squeezing parameters in notable limits}
\label{APPC}
The unitary transformation of Eq. (\ref{APPB18}) coincides with the one already introduced 
in Eqs. (\ref{DNT6})--(\ref{DNT7}) and the complex functions 
 $u_{k,\,\alpha}(\tau)$ and $v_{k,\,\alpha}(\tau)$ can then be directly expressed in terms of the two phases  $\delta_{k,\,\alpha}$, $\theta_{k,\,\alpha}$ supplemented by the amplitude $r_{k,\,\alpha}$:
\begin{equation}
u_{k,\,\alpha}(\eta) = e^{- i\, \delta_{k,\,\alpha}} \cosh{r_{k,\,\alpha}}, \qquad\qquad  
v_{k,\,\alpha}(\eta) =  e^{i (\theta_{k,\,\alpha} + \delta_{k,\,\alpha})} \sinh{r_{k,\,\alpha}}.
\label{APPC1}
\end{equation}
If the explicit evolution of $u_{k,\,\alpha}(\tau)$ and $v_{k,\,\alpha}(\tau)$ given in Eqs. (\ref{DNT8}) is 
combined with Eq. (\ref{APPC1}), the evolution of the squeezing parameters in the different physical regimes can be determined. Let us first consider, for the sake of illustration, an expanding de Sitter 
phase where $\eta\to \tau$, $b(\eta) \to a(\tau)$ and ${\mathcal F} \to {\mathcal H}$. 
Solving then Eq. (\ref{DNT2}) with the appropriate initial conditions we obtain for $u_{k}(\tau)$ and $v_{k}(\tau)$:
\begin{equation}
u_{k}(\tau) = \biggl( 1 - \frac{i}{2 k\tau}\biggr) \, e^{- i k \tau}, \qquad v_{k}(\tau) = - \frac{i}{2 k\tau}\, e^{- i k \tau}.
\label{APPC2}
\end{equation}
According to the dictionary implicitly established by Eq. (\ref{APPC1}) we can easily obtain the accurate 
expressions for $r_{k}$, $\theta_{k}$ and $\delta_{k}$:
\begin{eqnarray}
r_{k} &=& - \ln{( - k \tau)} + k^2 \tau^2 + {\mathcal O}(k^4\tau^4),
\label{APPC3}\\
 \delta_{k} &=& - k\tau - \frac{\pi}{2} + \frac{8}{3}k^3 \tau^3 + 
 {\mathcal O}(k^5\tau^5),
\label{APPC4}\\
\theta_{k} &=&  2 k \tau + \pi - \frac{8}{3}k^3 \tau^3 + 
 {\mathcal O}(k^5\tau^5),
\label{APPC5}
\end{eqnarray}
where the polarization index has been suppressed since the result holds 
for both polarization states. The limit $|k\tau| \ll 1$ appearing in Eqs. (\ref{APPC3})--(\ref{APPC4}) and (\ref{APPC5}) 
gives the evolution of the squeezing parameters when the relevant wavelengths 
are larger than the Hubble radius. The expression for the squeezing parameters 
after the various wavelengths have reentered the Hubble radius follows from Eqs. (\ref{DNT24})--(\ref{DNT25}); in those formulas $A_{k}(\tau_{ex},\tau_{re})$ and $B_{k}(\tau_{ex}, \tau_{re})$ are both complex and if we now 
introduce their real and imaginary parts as
\begin{equation}
A_{k}(\tau_{ex},\tau_{re}) = Q_{A}(\tau_{ex},\tau_{re}) - i P_{A}(\tau_{ex},\tau_{re}), \qquad B_{k} = Q_{B}(\tau_{ex},\tau_{re}) - i P_{B}(\tau_{ex},\tau_{re}),
\label{APPC6}
\end{equation}
we can obtain a simplified form of $u_{k}(\tau)$ and $v_{k}(\tau)$.
We remind, in this respect, that $\tau_{ex}$ and $\tau_{re}$ denote the times at which each of the different wavelengths exited and reentered the Hubble radius; for this reason both quantities ultimately depend 
on the modulus of the comoving three-momentum, i.e. $\tau_{ex} = \tau_{ex}(k)$ and $\tau_{re} = \tau_{re}(k)$.
Recalling then Eq. (\ref{APPC6}) the explicit expressions of $u_{k}(\tau)$ and $v_{k}(\tau)$ given in Eqs. (\ref{DNT24})--(\ref{DNT25}) become:
\begin{eqnarray}
&&u_{k}(\tau)=\frac{1}{2} \biggl\{\biggl[(Q_{A}+P_{B})\cos{\beta} + (Q_{B}-P_{A}) \sin{\beta}\biggr] + 
i \biggl[ (Q_{B}- P_{A}) \cos{\beta}- (Q_{A} +P_{B}) \sin{\beta}\biggr]\biggr\}, 
\label{APPC7}\\
&&v_{k}(\tau)=\frac{1}{2}\biggl\{\biggl[(P_{B} -Q_{A})\cos{\beta} -(Q_{B} + P_{A}) \sin{\beta}\biggr] - 
i \biggl[ (Q_{B} +P_{A}) \cos{\beta} + (P_{B} - Q_{A}) \sin{\beta}\biggr]\biggr\},
\label{APPC8}
\end{eqnarray}
where, for the sake of conciseness, the arguments of the functions $(Q_{A},\, P_{A})$ and $(Q_{B},\, P_{B})$ have been
momentarily dropped. In Eqs. (\ref{APPC7})--(\ref{APPC8}) $\beta= k\tau$ while the explicit expressions of $(Q_{A}, \, P_{A})$ and $(Q_{B},\, P_{B})$ are:
\begin{eqnarray}
Q_{A}(\tau_{ex}, \,\tau_{re}) &=& \biggl(\frac{a_{re}}{a_{ex}}\biggr) \biggl[ 1 - q_{ex} \, {\mathcal I}(\tau_{ex}, \tau_{re})\biggr],\qquad
P_{A}(\tau_{ex}, \,\tau_{re}) = \biggl(\frac{a_{re}}{a_{ex}}\biggr) \, {\mathcal I}(\tau_{ex}, \tau_{re}),
\nonumber\\
Q_{B}(\tau_{ex},\, \tau_{re}) &=& q_{re} \biggl(\frac{a_{re}}{a_{ex}}\biggr)  \biggl[ 1 - q_{ex} \, {\mathcal I}(\tau_{ex}, \tau_{re})\biggr] -  q_{ex}\,\biggl(\frac{a_{ex}}{a_{re}}\biggr),
\nonumber\\
P_{B}(\tau_{ex},\, \tau_{re}) &=& q_{re} \biggl(\frac{a_{re}}{a_{ex}}\biggr)  + \biggl(\frac{a_{ex}}{a_{re}}\biggr).
\label{APPC9}
\end{eqnarray}
Three auxiliary quantities have been introduced in Eq. (\ref{APPC9}), namely
\begin{equation}
q_{ex} = a_{ex} \, H_{ex}/k, \qquad q_{re} = a_{re} \, H_{re}/k, \qquad  {\mathcal I}(\tau_{ex}, \tau_{re}) = k \, 
\int_{\tau_{ex}}^{\tau_{re}} \biggl[\frac{a_{ex}}{a(\tau)}\biggr]^2 \, d\tau.
\label{APPC10}
\end{equation}
It is a bit lengthy (but conceptually straightforward) to verify from Eqs. (\ref{APPC7})--(\ref{APPC8}) and (\ref{APPC9})--(\ref{APPC10}) that, indeed, $|u_{k}(\tau)|^2 - |v_{k}(\tau)|^2 =1$. As stressed in section \ref{sec4}, the problem of approximating the expressions of $u_{k}(\tau)$ and $v_{k}(\tau)$ ultimately rests on the enforcement of unitarity. For instance in Eq. (\ref{APPC9}) $(a_{ex}/a_{re}) \ll 1$  in the case of expanding backgrounds. Should we systematically neglect all the terms proportional to $(a_{ex}/a_{re}) \ll 1$? If we would do that 
the results of Eqs. (\ref{APPC7})--(\ref{APPC8}) would simply imply that $u_{k}^2(\tau) \simeq v_{k}^2(\tau)$ 
with an explicit violation of unitarity. The correct approach is instead to use an approximation 
scheme that is fully consistent with unitarity when we evaluate the expansion to a given order in the small parameter 
of the problem; this is the rationale for the result of Eq. (\ref{APPC11}) which is incidentally consistent 
with the strategies that are customarily employed in quantum optics and quantum communication.
Recalling that the turning points $\tau_{re}(k)$ and $\tau_{ex}(k)$ are determined as a solution of the 
equation $k^2 \simeq a^2 \, H^2 [ 2 - \epsilon(a)]$,  when a given wavelength exits the Hubble 
radius during inflation we have that $q_{ex} = {\mathcal O}(1)$ since $\epsilon \ll 1$ and 
the turning points of the WKB approximation are fixed, in practice, by the condition $k^2 = a^2\, H^2$.
Conversely when the reentry takes place far from the radiation stage the same observation can be made since $q_{re} = a_{re} \, H_{re}/k = {\mathcal O}(1)$. However if the reentry occurs when $\epsilon_{re} = {\mathcal O}(2)$
the turning point is not regular and $q_{re} \gg 1$ in all the expressions that appear in Eqs. (\ref{APPC7})--(\ref{APPC8}) and (\ref{APPC9}). 

Taking into account the observations of the previous paragraph, $u_{k}(\tau)$ and $v_{k}(\tau)$ can be expanded 
to a given order in $(a_{re}/a_{ex})$ which is always small if the Universe expands; in doing so we shall keep all the other quantities generic. As already anticipated, the result of this strategy gives directly Eq. (\ref{APPC11}). 
We can immediately verify that, to the given order in the perturbative expansion, Eq. (\ref{APPC11}) 
implies that $|u_{k}(\tau)|^2 - |v_{k}(\tau)|^2=1$ so that unitarity is not lost while correctly accounting for all the leading terms of the expansion. From Eq. (\ref{APPC11}) we can also obtain the expression of $\theta_{k}$:
\begin{equation}
e^{i\theta_{k}} = \frac{u_{k}(\tau) \,v_{k}(\tau)}{|u_{k}(\tau) \,v_{k}(\tau)|} = e^{- 2 i \beta_{k}(\tau)}\bigg\{ \frac{- i + q_{re}}{i + q_{re}} + {\mathcal O}\biggl[\biggl(\frac{a_{re}}{a_{ex}}\biggr)^{2}\biggr]\biggr\}.
\label{APPC12}
\end{equation}
We now have two separate possibilities. If the given wavelength reenters in a decelerated stage different 
from radiation then $q_{re} = {\mathcal O}(1)$  and Eq. (\ref{APPC12}) implies 
that $\theta_{k} = - 2 k \tau - \pi/2$. Conversely if the wavelength reenters precisely during 
radiation, then $q_{re}\gg1$ and  $\theta_{k} = - 2 k \tau$. With similar strategy we can also deduce 
that, approximately, $ e^{2 \,r} \simeq (a_{re}/a_{ex})$.  From the 
evolution equations of $u_{k,\alpha}$ and $v_{k\,\alpha}$ 
we can also deduce the equations governing the squeezing parameters. 
\begin{eqnarray}
r_{p,\alpha}^{\prime} &=& - {\mathcal H} \cos{\theta_{p,\alpha}}, \qquad \delta_{p,\alpha}^{\prime} = p - {\mathcal H} \sin{\theta_{p,\alpha}} \tanh{r_{p,\alpha}}, 
\label{APPC13}\\
\theta_{p,\alpha}^{\prime} &=& - 2 p  + 2 \frac{{\mathcal H} \sin{\theta_{p,\alpha}}}{\tanh{2 r_{p,\alpha}}}.
\label{APPC14}
\end{eqnarray}
Equations (\ref{APPC13})--(\ref{APPC14}) provide useful relations among the squeezing parameters especially 
in the limit where $r_{k} \gg1$. Indeed, by combining Eqs. (\ref{APPC13})--(\ref{APPC14})
we get the following relation 
\begin{equation}
(\theta_{k} + 2 \delta_{k})^{\prime} = \frac{{\mathcal H} \sin{\theta_{k}}}{\sinh{r_{k}} \, \cosh{r_{k}}}.
\label{APPC15}
\end{equation}
When $r_{k} \gg 1$ we have that, in practice, $(\theta_{k} + 2 \delta_{k})^{\prime} =0$. 
But this means that 
\begin{eqnarray}
&& \delta_{k}(\tau) = k \tau + \pi/4 +\alpha, \qquad \theta = - 2 k \tau - \pi/2, \qquad q_{re} = {\mathcal O}(1).
\nonumber\\
&& \delta_{k}(\tau) = k \tau + \alpha, \qquad \theta = - 2 k \tau, \qquad q_{re} \gg 1,
\end{eqnarray}
where $\alpha$ is an integration constant. It is therefore an oversimplification to regard the large squeezing limit as a classical limit. This potentially misleading conclusion follows from the adoption of an  approximation scheme where the inaccuracies of the expansion seem to destroy the unitarity of the evolution which is however always preserved. The same argument holds for the quantum coherence of the system. The graviton pairs created as a result of the quantum theory of parametric amplification are then entangled and this essential quantum mechanical property is preserved by the unitary evolution.
\end{appendix}


\newpage

\begin{thebibliography}{99}

\itemsep -6pt

\bibitem{AA1} L.~P.~Grishchuk,  Sov.\ Phys.\ JETP {\bf 40}, 409 (1975)   [Zh.\ Eksp.\ Teor.\ Fiz.\  {\bf 67}, 825 (1974)].

\bibitem{AA2} L.~P.~Grishchuk,  Annals N.\ Y.\ Acad.\ Sci.\  {\bf 302}, 439 (1977).

\bibitem{AA3}  L. H. Ford and L. Parker, Phys.\ Rev.\ D {\bf 16}, 245 (1977).

\bibitem{AA4} B. L. Hu and L. Parker, Phys. Lett. A {\bf 63}, 217 (1977).

\bibitem{AA5}  L.~Parker,  Phys.\ Rev.\ Lett.\  {\bf 21}, 562 (1968).

\bibitem{AA6}  L.~Parker,  Phys.\ Rev.\  {\bf 183}, 1057 (1969).

\bibitem{AA7}  A. A. Starobinsky, JETP Lett. {\bf 30}, 682 (1979) [Pis'ma Zh. Eksp. Teor. Fiz. {\bf 30}, 719 (1979)].

\bibitem{AA8} L.~F.~Abbott and M.~B.~Wise, Nucl.\ Phys.\ B {\bf 244}, 541 (1984).
   
\bibitem{AA9}  S.~W.~Hawking,  Phys.\ Lett.\  {\bf 150B}, 339 (1985).

\bibitem{AA10}  V. A. Rubakov, M. V. Sazhin, and A. V. Veryaskin, Phys. Lett. B {\bf 115}, 189 (1982).

\bibitem{AA11} N.~Aghanim {\it et al.} [Planck Collaboration], Astron. Astrophys. {\bf 641}, A6 (2020).

\bibitem{AA12}  P.~A.~R.~Ade {\it et al.} [BICEP and Keck Collaborations], Phys. Rev. Lett. {\bf 127}, 151301 (2021).

\bibitem{AA13} M.~Giovannini, Prog. Part. Nucl. Phys. \textbf{112}, 103774 (2020).

\bibitem{BB0a} B.~Abbott {\it et al.} [LIGO Collaboration], Phys.\ Rev.\ D {\bf 69}, 122004 (2004).
  
\bibitem{BB0b} B.~Abbott {\it et al.} [LIGO Collaboration], Phys.\ Rev.\ Lett.\  {\bf 95}, 221101 (2005).  

\bibitem{BB1} B.~P.~Abbott {\it et al.} [LIGO and Virgo Collaborations], Phys. Rev. D {\bf 100}, 061101 (2019).

\bibitem{BB2} R.~Abbott {\it et al.} [KAGRA, Virgo and LIGO Collaborations], Phys. Rev. D {\bf 104}, 022004 (2021).

\bibitem{BB3} D.~J.~Reardon, {\it et al.}, Astrophys. J. Lett. {\bf 951},  L6 (2023)

\bibitem{BB4} G.~Agazie \textit{et al.}, Astrophys. J. Lett. {\bf 951},  L8 (2023).

\bibitem{BB5} R. W. Hellings and G. S. Downs, Astrophys. J. Lett. {\bf  265} L39 (1983).

\bibitem{BB5a} M.~Giovannini, Eur. Phys. J. C \textbf{84}, 67 (2024).

\bibitem{BB6a}   S.~Weinberg,  Phys.\ Rev.\  D {\bf 69}, 023503 (2004).

\bibitem{BB6} D.~A.~Dicus and W.~W.~Repko,  Phys.\ Rev.\  D {\bf 72}, 088302 (2005).
 
\bibitem{BB7} H.~X.~Miao and Y.~Zhang, Phys.\ Rev.\ D {\bf 75}, 104009 (2007).
 
\bibitem{BB8}  K.~W.~Ng, Phys.\ Rev.\ D {\bf 86}, 103510 (2012).
 
 \bibitem{BB9}  B.~A.~Stefanek and W.~W.~Repko, Phys.\ Rev.\ D {\bf 88},  083536 (2013).
 
\bibitem{BB10} L.~M.~Krauss and F.~Wilczek, Phys. Rev. D \textbf{89},  047501 (2014).

\bibitem{REC2} M.~Giovannini, JCAP \textbf{05}, 056 (2023).

\bibitem{ST1} M.~Giovannini,  Phys.\ Rev.\ D {\bf 58}, 083504 (1998).

\bibitem{ST2}   P.~J.~E.~Peebles and A.~Vilenkin,  Phys.\ Rev.\ D {\bf 59}, 063505 (1999).

\bibitem{ST3} M.~Giovannini,  Phys.\ Rev.\ D {\bf 60}, 123511 (1999).

\bibitem{ST3a} M.~Giovannini, Class. Quant. Grav. {\bf 16}, 2905 (1999).

\bibitem{ST4} J.~Haro, W.~Yang and S.~Pan, JCAP {\bf 01}, 023 (2019).

\bibitem{ST5} M.~Gorghetto, E.~Hardy and H.~Nicolaescu, JCAP {\bf 06}, 034 (2021).

\bibitem{ST6} B.~Li and P.~R.~Shapiro, JCAP {\bf 10}, 024 (2021).

\bibitem{ST7}  M.~Giovannini, Phys. Rev. D \textbf{108},  123508 (2023).

\bibitem{ST7a} M. Giovannini, Phys. Lett. B \textbf{854}, 138769 (2024).

\bibitem{TTT1} C. Tolman, Phys. Rev. {\bf 38}, 1758 (1931).

\bibitem{TTT2}   G. Lema\^itre, Ann. Soc. Sci. Bruxelles A {\bf 53}, 51 (1933).

\bibitem{TTT3}  M.~Giovannini, Class. Quant. Grav. \textbf{41},  105010 (2024).

\bibitem{ST8} B.~L.~Hu and H.~E.~Kandrup, Phys. Rev. D \textbf{35}, 1776 (1987).

\bibitem{ST9} H. E. Kandrup, Phys. Rev. D \textbf{37}, 3505 (1988).

\bibitem{ST9a} M.~Gasperini and M.~Giovannini, Phys. Lett. B \textbf{301}, 334 (1993).

\bibitem{ST9b} M.~Gasperini and M.~Giovannini, Class. Quant. Grav. \textbf{10}, L133 (1993).

\bibitem{ST10} B. L. Hu, G. Kang, and A. Matacz, Int. J. Mod. Phys. A {\bf 9}, 991 (1994).

\bibitem{ST11} J.~T.~Hsiang and B.~L.~Hu, Universe \textbf{8}, 27 (2022).

\bibitem{CL1} S. Weinberg, {\it Cosmology} (Oxford University Press, Oxford, UK, 2008).

\bibitem{SAK} A. D. Sakharov,  Sov. Phys. JETP {\bf 22}, 241 (1966) [Zh. Eksp. Teor. Fiz. {\bf 49}, 345 (1965)].

\bibitem{CL2} E. M. Lifshitz and I. M. Khalatnikov, Sov. Phys. Usp. {\bf 6},  495 (1964).

\bibitem{CL2a} V. A. Belinskii, E. M. Lifshitz, and I. M.  Khalatnikov, Sov. Phys. Usp. {\bf 13}, 745 (1971).

\bibitem{CL3}  A.~A.~Starobinsky,  JETP Lett.\  {\bf 37}, 66 (1983); R.~M.~Wald,  Phys.\ Rev.\ D {\bf 28}, 2118 (1983).

\bibitem{CL4} M.~Giovannini, Phys. Lett. B {\bf 746}, 159 (2015).

\bibitem{PARKTH} L. Parker, Nature {\bf 261}, 20 (1976).

\bibitem{BIRREL}  N. D. Birrel and P. C. W. Davies, {\it Quantum fields in curved spaces} (Cambridge Univ. Press, Cambridge, England, 1982).

\bibitem{PTOMS}  L. Parker and D. Toms, {\it Quantum Field Theory in Curved Space-time}, (Cambridge University Press, Cambridge 2009).

\bibitem{MANDL} L. Mandel and E. Wolf, {\it Optical Coherence and Quantum Optics} (Cambridge University Press, Cambridge, 1995).

\bibitem{HBT4}  R. Loudon, {\it The Quantum Theory of Light} (Clarendon Press, Oxford, 1983).

\bibitem{louisell} W. H. Louisell, A. Yariv, and A. E. Siegman Phys. Rev. {\bf 124}, 1646 (1961).

\bibitem{mollow1}  B. L. Mollow and R. J. Glauber, Phys. Rev. {\bf 160}, 1076 (1967).

\bibitem{mollow2}   B. L. Mollow and R. J. Glauber, Phys. Rev. {\bf 160}, 1097 (1967).

\bibitem{RGS0} L. P. Grishchuk and Y. V. Sidorov, Phys. Rev. D {\bf 42}, 3413 (1990).

\bibitem{HBTG1} M.~Giovannini, Phys. Rev. D \textbf{83}, 023515 (2011).

\bibitem{mg0} M.~Giovannini, Phys. Lett. B \textbf{668}, 44 (2008).

\bibitem{mg1} M.~Giovannini, Class. Quant. Grav. \textbf{26}, 045004 (2009).

\bibitem{BBN1}  V.F. Schwartzman, Pis'ma Zh. Eksp. Teor. Fiz. {\bf 9}, 315 (1969) [JETP Lett. {\bf 9}, 184 (1969)].
  
\bibitem{BBN2} M.~Giovannini, H.~Kurki-Suonio and E.~Sihvola, Phys.\ Rev.\  D {\bf 66}, 043504 (2002).

\bibitem{BBN3} R. Cyburt, B.~D.~Fields, K.~A.~Olive, and E.~Skillman, Astropart.\ Phys.\ {\bf 23}, 313 (2005).

\bibitem{FGF0} F. Pegoraro, L.  Radicati, Ph. Bernard, and E. Picasso, Phys. Lett. A {\bf 68}, 165 (1978).

\bibitem{FGF1} F. Pegoraro, E. Picasso, and L. Radicati, J. Phys. A {\bf 11}, 1949 (1978).

\bibitem{FGF2}  E. Iacopini, F. Pegoraro, E. Picasso, and L. Radicati, Phys. Lett. B {\bf 73}, 140 (1979).

\bibitem{FGF3} C. M. Caves, Phys. Lett. B {\bf 80}, 323 (1979).

\bibitem{FGF4}  C.  Reece, P. Reiner, and A.  Melissinos, Phys. Lett. A {\bf 104}, 341 (1984).

\bibitem{FGF5} C.  Reece, P. Reiner, and A.  Melissinos, Nucl. Inst. and Methods, A {\bf 245}, 299 (1986).

\bibitem{FGF6} Ph. Bernard, G. Gemme, R. Parodi and E. Picasso, Rev. Sci. Instrum. {\bf 72}, 2428 (2001).

\bibitem{FGF7} R. Ballantini, P. Bernard, A. Chincarini, G. Gemme, R. Parodi and E. Picasso, Class. Quant. Grav. {\bf 21}, S1241 (2004).

\bibitem{EK1} J.~Khoury, B.~A.~Ovrut, P.~J.~Steinhardt and N.~Turok,  Phys.\ Rev.\ D {\bf 64}, 123522 (2001).
  
\bibitem{EK2} L.~A.~Boyle, P.~J.~Steinhardt and N.~Turok,  Phys.\ Rev.\ D {\bf 69}, 127302 (2004).  

\bibitem{EK3}  M.~Gasperini and M.~Giovannini,  Phys.\ Rev.\ D {\bf 47}, 1519 (1993);
 R.~Brustein, M.~Gasperini, M.~Giovannini and G.~Veneziano,  Phys.\ Lett.\ B {\bf 361}, 45 (1995).

\bibitem{sch1} C. Caves and B. L. Schumaker,  Phys.  Rev.  A {\bf 31}, 3068 (1985); 

\bibitem{sch2} C. Caves and B. L. Schumaker,  Phys.  Rev.  A {\bf 31}, 3093 (1985); B. L. Schumaker, Phys. Rept. {\bf 135}, 317 (1986).
    
\bibitem{pere} A. Perelomov, {\it Generalized coherent states and their applications}, (Springer-Verlag, Berlin, 1986).

\bibitem{instate1} M.~Giovannini, Phys. Rev. D \textbf{88},  021301 (2013); 
Class. Quant. Grav. \textbf{30}, 015009 (2013).

\bibitem{REC3} S.~Weinberg, Phys.\ Rev.\ D {\bf 77}, 123541 (2008).

\bibitem{REC3a} M.~Giovannini, Phys. Rev. D \textbf{99}, 083501 (2019).

\bibitem{REC3b} M.~Giovannini, Phys. Lett. B \textbf{819}, 136444 (2021).

\bibitem{mg3} M.~Giovannini, Phys. Rev. D \textbf{105}, 103524 (2022).

\bibitem{osc1} M. S. Turner, Phys. Rev. D \textbf{28}, 1243 (1983).

\bibitem{osc2} C. Pathinayake and L. H. Ford, Phys. Rev. D \textbf{35}, 3709 (1987).

\bibitem{osc3}  L. H. Ford, Phys. Rev. D \textbf{35}, 2955 (1987).

\bibitem{osc4} J. D. Barrow, Phys. Rev. D \textbf{48}, 1585 (1993).

\bibitem{osc5} E. Masso, F. Rota, and G. Zsembinszki, Phys. Rev. D \textbf{72}, 084007 (2005).

\bibitem{REFR1} P.~Szekeres,  Annals Phys.\  {\bf 64}, 599 (1971).

\bibitem{REFR2} P.~C.~Peters,  Phys.\ Rev.\ D {\bf 9}, 2207 (1974).

\bibitem{REFR3}  M.~Giovannini, Class.\ Quant.\ Grav.\  {\bf 33}, 125002 (2016)  [arXiv:1507.03456 [astro-ph.CO]].

\bibitem{BBBO4} M. Novello and S. E. P. Bergliaffa, Phys. Rept. {\bf 463}, 127 (2008).

\bibitem{BBBO5} A. Ashtekar and P. Singh, Class. Quant. Grav. {\bf 28}, 213001 (2011).

\bibitem{BBBO6} R.~Brandenberger, [arXiv:2306.12458 [hep-th]].

\bibitem{BBBO7} P.~Agrawal, G.~Obied, P.~J.~Steinhardt and C.~Vafa, Phys. Lett. B \textbf{784}, 271 (2018).

\bibitem{BBBO8} A.~Bedroya, R.~Brandenberger, M.~Loverde and C.~Vafa, Phys. Rev. D \textbf{101}, 103502 (2020).

\bibitem{REC4} M.~Giovannini, Phys. Rev. D \textbf{99},  123507 (2019).

\bibitem{HBT1} R. Hanbury Brown and R. Q. Twiss, Nature {\bf 178}, 1046 (1956).

\bibitem{HBT2} R. Hanbury Brown and R. Q. Twiss, Proc. R. Soc. A {\bf 242}, 300 (1957).

\bibitem{HBT3} R. Hanbury Brown and R. Q. Twiss, Proc. R. Soc. A  {\bf 243}, 291 (1958).

\bibitem{HBTG2}  M.~Giovannini,  Class.\ Quant.\ Grav.\  {\bf 34}, 035019 (2017).

\bibitem{HBTG3}  M.~Giovannini, Mod.\ Phys.\ Lett.\ A {\bf 32}, 1750191 (2017).

\bibitem{CMBT} D.~J.~Fixsen,  Astrophys.\ J.\  {\bf 707}, 916 (2009).

\bibitem{PDIST1} E. C. G. Sudarshan, Phys. Rev. Lett. {\bf 10}, 277  (1963).

\bibitem{PDIST2}  K. E. Cahill and R. J. Glauber,  Phys. Rev. {\bf 177} , 1857 (1969).

\bibitem{PDIST3}  K. E. Cahill and R. J. Glauber,   Phys. Rev. {\bf 177}   1882 (1969).

\bibitem{STO} D. Stoler, Phys. Rev. D {\bf 4}, 1925 (1971).

\bibitem{wig1} E. P. Wigner, Phys. Rev. {\bf 40}, 749 (1932).

\bibitem{wig2} M.~Hillery, R.~F.~O'Connell, M.~O.~Scully and E.~P.~Wigner, Phys. Rept. \textbf{106}, 121 (1984).

\bibitem{COMPP} P. D. Drummond and C. W. Gardiner, J. Phys. A {\bf 13}, 2353 (1980).

\bibitem{Qrep} D. F. Walls, Nature {\bf 306}, 141 (1983).

\bibitem{karlin} S. Karlin and H. M. Taylor, {\it A First Course in Stochastic Processes} (Academic Press, New York, 1975).

\bibitem{DDP1} D. Deutsch, Phys. Rev. Lett. {\bf 50}, 631 (1983).

\bibitem{DDP2} M. H. Partovi, Phys. Rev. Lett. {\bf 50}, 1883 (1983).

\bibitem{STO2} D. Stoler, Phys. Rev. D {\bf 1}, 3217 (1970).

\bibitem{STO3}  H. P. Yuen, Phys. Rev. A {\bf 13}, 2226 (1976).

\bibitem{GILMORE}  R. Gilmore, {\it Lie groups, Lie Algebras, and Some of Their Applications}, (Wiley, New York, 1974).

\bibitem{FISHER} R. A. Fisher, M.M. Neto, and Y.D.Sandberg, Phys Rev. D {\bf 29}, 1107 (1984).

\bibitem{truax1} D. R. Truax, Phys. Rev. D {\bf 31}, 1988 (1985).

\bibitem{truax2} M. M. Nieto and D. R. Truax, Phys. Rev. Lett. {\bf 71}, 2843 (1993).

\end{thebibliography}
\end{document}